\documentclass[10pt,twocolumn,prx,superscriptaddress]{revtex4-2}

\usepackage{graphicx}% Include figure files
\usepackage[tight]{units} %helps writing units correctly
\usepackage{textcomp} %units and stuff
\usepackage{gensymb} %units and stuff
\usepackage{dsfont}
\usepackage{amsmath,amsfonts}
\usepackage{bm}
\usepackage{hyperref}

\usepackage{color}

\begin{document}

\title{Correlation spectroscopy with multi-qubit-enhanced phase estimation}

\author{H.~Hainzer}
\affiliation{Institut f\"ur Quantenoptik und Quanteninformation, \"Osterreichische Akademie der Wissenschaften,
Technikerstra\ss{}e 21a, 6020 Innsbruck, Austria}
\affiliation{Institut f\"ur Experimentalphysik, Universit\"at Innsbruck, Technikerstra\ss{}e 25, 6020 Innsbruck, Austria}

\author{D.~Kiesenhofer}
\affiliation{Institut f\"ur Quantenoptik und Quanteninformation, \"Osterreichische Akademie der Wissenschaften,
Technikerstra\ss{}e 21a, 6020 Innsbruck, Austria}
\affiliation{Institut f\"ur Experimentalphysik, Universit\"at Innsbruck, Technikerstra\ss{}e 25, 6020 Innsbruck, Austria}

\author{T.~Ollikainen}
\affiliation{Institut f\"ur Quantenoptik und Quanteninformation, \"Osterreichische Akademie der Wissenschaften,
Technikerstra\ss{}e 21a, 6020 Innsbruck, Austria}

\author{M.~Bock}
\affiliation{Institut f\"ur Quantenoptik und Quanteninformation, \"Osterreichische Akademie der Wissenschaften,
Technikerstra\ss{}e 21a, 6020 Innsbruck, Austria}

\author{F.~Kranzl}
\affiliation{Institut f\"ur Quantenoptik und Quanteninformation, \"Osterreichische Akademie der Wissenschaften,
Technikerstra\ss{}e 21a, 6020 Innsbruck, Austria}
\affiliation{Institut f\"ur Experimentalphysik, Universit\"at Innsbruck, Technikerstra\ss{}e 25, 6020 Innsbruck, Austria}

\author{M.~K.~Joshi}
\affiliation{Institut f\"ur Quantenoptik und Quanteninformation, \"Osterreichische Akademie der Wissenschaften,
Technikerstra\ss{}e 21a, 6020 Innsbruck, Austria}

\author{G.~Yoeli}
\affiliation{Racah Institute of Physics, The Hebrew University of Jerusalem, Givat Ram, Jerusalem, 91904, Israel}

\author{R.~Blatt}
\affiliation{Institut f\"ur Quantenoptik und Quanteninformation, \"Osterreichische Akademie der Wissenschaften,
Technikerstra\ss{}e 21a, 6020 Innsbruck, Austria}
\affiliation{Institut f\"ur Experimentalphysik, Universit\"at Innsbruck, Technikerstra\ss{}e 25, 6020 Innsbruck, Austria}

\author{T.~Gefen}
\email{tgefen@caltech.edu}
\affiliation{Institute for Quantum Information and Matter, Caltech, Pasadena, CA, USA}

\author{C.~F.~Roos}
\email{christian.roos@uibk.ac.at}
\affiliation{Institut f\"ur Quantenoptik und Quanteninformation, \"Osterreichische Akademie der Wissenschaften,
Technikerstra\ss{}e 21a, 6020 Innsbruck, Austria}
\affiliation{Institut f\"ur Experimentalphysik, Universit\"at Innsbruck, Technikerstra\ss{}e 25, 6020 Innsbruck, Austria}

\date{\today}

\begin{abstract}
Ramsey interferometry is a widely used tool for precisely measuring transition frequencies between two energy levels of a quantum system, with applications in time-keeping, precision spectroscopy, quantum optics, and quantum information. Often, the coherence time of the quantum system surpasses the one of the oscillator probing the system, thereby limiting the interrogation time and associated spectral resolution. Correlation spectroscopy overcomes this limitation by probing two quantum systems with the same noisy oscillator for a measurement of their transition frequency difference; this technique has enabled very precise comparisons of atomic clocks. Here, we extend correlation spectroscopy to the case of multiple quantum systems undergoing strong correlated dephasing. We model Ramsey correlation spectroscopy with $N$ particles as a multi-parameter phase estimation problem and demonstrate that multiparticle quantum correlations can assist in reducing the measurement uncertainties even in the absence of entanglement. We derive precision limits and optimal sensing techniques for this problem and compare the performance of probe states and measurement with and without entanglement. Using one- and two-dimensional ion Coulomb crystals with up to 91 qubits, we experimentally demonstrate the advantage of measuring multi-particle quantum correlations for reducing phase uncertainties, and apply correlation spectroscopy to measure ion-ion distances, transition frequency shifts, laser-ion detunings, and path-length fluctuations. Our method can be straightforwardly implemented in experimental setups with globally-coherent qubit control and qubit-resolved single-shot read-out and is thus applicable to other physical systems such as neutral atoms in tweezer arrays.
\end{abstract}

\maketitle

\section{Introduction}
The ability to estimate the phase of a wave is key to practical applications such as time keeping with atomic clocks \cite{Ludlow:2015}, rotation and acceleration sensing \cite{Canuel:2006}, gravimetry \cite{Peters:1999}, but also to probing fundamental physics \cite{Hamilton:2015} and measuring fundamental constants of nature \cite{Bouchendira:2011}. 
Sensing techniques such as optical or matter wave interferometry rely on phase comparisons of two light waves or matter waves, respectively. In optical atomic clocks for instance, the phase of an atomic superposition state is compared to the phase of the laser having created the superposition. In most of these applications, a large number of uncorrelated photons or atoms are probed, giving rise to a measurement uncertainty governed by the standard quantum limit according to which the uncertainty decreases inversely with the square root of the number of particles being probed. If, however, quantum correlations exists between the particles, the scenario becomes much more interesting and complex.

In this context, phase estimation based on quantum measurements constitutes a subfield of quantum metrology, which aims at making sensitive measurements of physical quantities by harnessing quantum resources, in particular entanglement \cite{Giovannetti:2011}. To this end, it has been shown that entangled input states can be used to beat the standard quantum limit \cite{Wineland:1992,Bollinger:1996} and that entanglement can be a resource for achieving optimal phase sensing over a wider range of phases \cite{Marciniak:2022}. However, as entangled states easily decohere under environmental noise, the performance gain of entanglement-enhanced metrology protocols can be jeopardized by decoherence processes \cite{Huelga:1997,Demkowicz-Dobrzanski:2012}; the achievable precision bounds depend on whether the noise is Markovian or contains temporal or spatial correlations \cite{Chin:2011,Jeske:2014}.

Furthermore, from a practical point of view, entanglement-generating resources are often not readily available in precision experiments. For this reason, it is of interest to consider quantum metrology protocols using quantum correlations other than entanglement that might be easier to implement for carrying out quantum-enhanced measurements \cite{Braun:2018}.  
In this paper, we will focus on correlation spectroscopy \cite{Chwalla:2007,Olmschenk:2007}, a phase estimation technique for probing the phase difference of qubits subjected to spatially correlated noise, and extend it to networks of $N$ qubits. In the following, we provide our motivation for studying this measurement scenario.

Coherent probing of ultra-narrow atomic transitions in combination with outstanding characterization of systematic level shifts has led to the development of optical atomic clocks with unprecedented precision \cite{BACON:2021}. To verify a clock's performance, its frequency has to be compared with another clock. The uncertainty with which the frequency difference of the clocks can be determined within a given measurement time is usually not limited by the lifetime of the atomic energy levels but rather by the local oscillator's phase noise that sets an upper bound to the useful probe time. This limitation can be overcome by synchronous probing of the two clocks with the same local oscillator and correlating the measurement outcomes. In the case of ensemble-averaged signals, such as in optical lattice clocks where the excitation probability of a large number of atoms is measured, \cite{Takamoto:2011,Nicholson:2012}, the correlations are purely classical. If, however, the measurements probe the quantum state of individual atoms, the correlations can become non-classical, even in the absence of any entanglement \cite{Lanyon:2013}. 

It is in this context that correlation spectroscopy \cite{Chwalla:2007,Olmschenk:2007} has been developed, a technique for measuring transition frequency differences in the presence of correlated phase noise with probe times that can be significantly longer than the coherence time of each system with respect to the local oscillator \cite{Chou:2011,Marti:2018,Shanif:2019,Clements:2020,Young:2020}. It is based on a synchronous standard Ramsey-type interrogation of two or more atoms by the same oscillator: a first $\pi/2$ pulse rotates the Bloch vector into the equatorial plane where it precesses during the free evolution time with a rate set by the detuning of the oscillator from the atomic transition. The second $\pi/2$ pulse in conjunction with a state detection in the energy eigenbasis enables the measurement of a spin projection in the equatorial plane. However, instead of measuring expectation values of individual atoms, a parity measurement is used to correlate the measurement outcomes of pairs of atoms. By this approach, transition frequency differences can be measured by observing parity oscillations as a function of the duration of the free evolution time. While correlation spectroscopy only achieves a maximum parity oscillation contrast of 0.5 and therefore does not achieve the optimum signal-to-noise ratio obtainable by preparing maximally entangled states of the two systems \cite{Roos:2006,Megidish:2019,Manovitz:2019}, it is technically much easier to implement. 

\begin{figure}[t]
\centering
\includegraphics[width=0.45\textwidth]{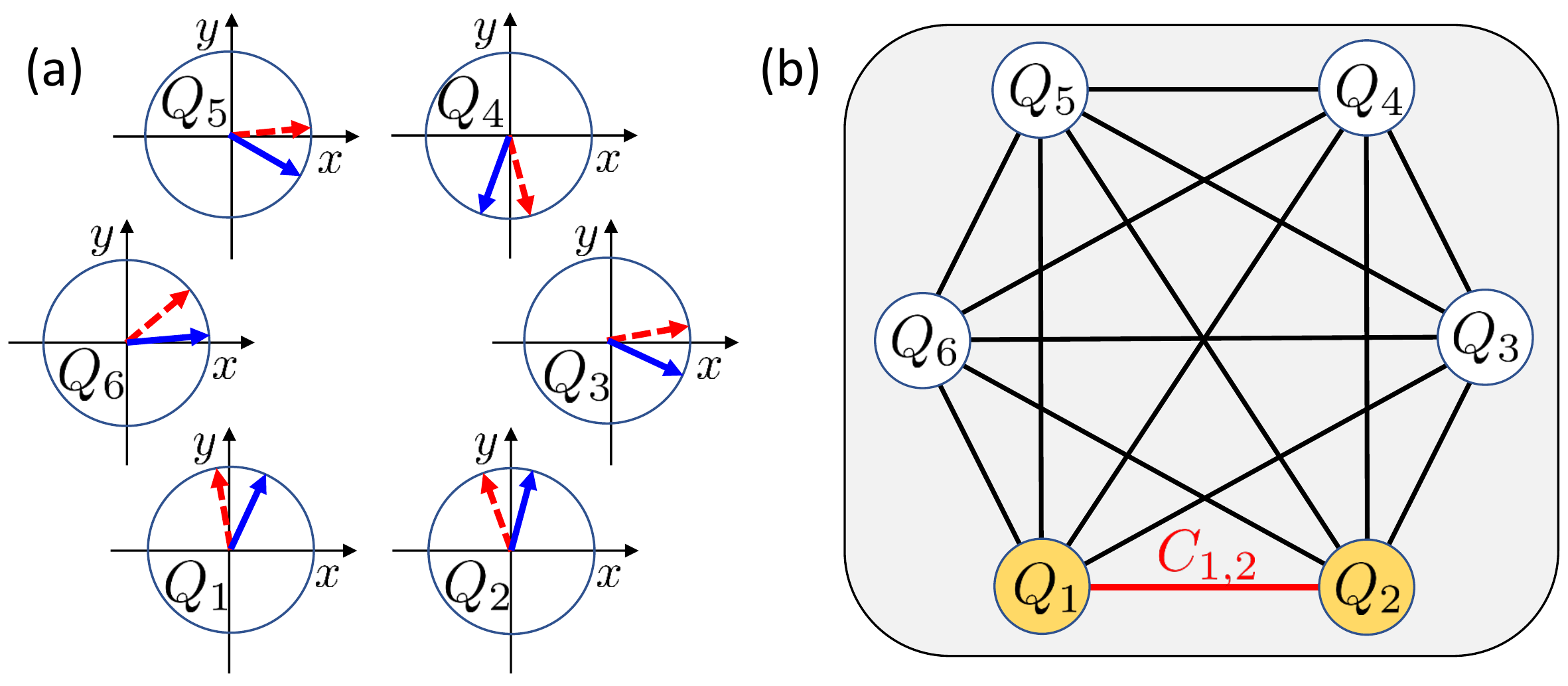}
\caption{Measurement scenario. (a) In a network of quantum sensors comprised of qubits $Q_i$, the qubits are prepared in Bloch states lying in the equatorial plane and subjected to correlated dephasing that randomly rotates all Bloch vectors by the same angle, as indicated by solid and dashed arrows. (b) Correlations $C_{i,j}$ between pairs of qubits $(i,j)$ are measured for an estimation of the angle between the respective Bloch vectors. Is it possible to reduce the measurement uncertainty of $C_{i,j}$ obtained from a finite number of experimental repetitions by taking into account all pair correlations that can be simultaneously measured, or even all $N$-qubit correlations?}
\label{fig:0}
\end{figure}

The detection of phase shifts in the presence of strong correlated phase noise is a common scenario that appears in a wide variety of sensing platforms. Apart from the example of multiple clocks probed by the same oscillator, spatially correlated noise can result from the spatial proximity of the qubits \cite{Monz:2011,Bradley:2019,Singh:2022}, instabilities of the local oscillator probing them \cite{Monz:2011,Braverman:2018}, or from the coupling of the qubits to a common bosonic mode \cite{Sawyer:2012,vonLuepke:2020}. A similar scenario appears also in interferometers and optomechanical sensors where a displacement noise of the mirrors and radiation pressure induce a correlated noise on the 
different output modes \cite{kawamura2004displacement, chen2006interferometers,gefen2022quantum}.

In this work, we investigate the parameter phase estimation scenario as sketched in Fig.~\ref{fig:0}: we consider a set of $N$ qubits all of which were prepared in states with Bloch vectors in the equatorial plane and subjected to correlated phase noise (panel (a)). We want to estimate the angle between the Bloch vectors of a pair of qubits $(i,j)$ by applying a second $\pi/2$ pulse, measuring all qubits in the energy eigenbasis and correlating the measurement outcomes ($\pm 1$) to obtain the correlation $C_{i,j}$ (panel (b)). We ask the question whether the measurement uncertainty of one of the correlations, e.~g. $C_{1,2}$, obtained from a finite number of experimental repetitions could be reduced by taking into consideration all other pair correlations that were simultaneously recorded instead of analyzing only the measurements of the particular pair, e.~g. $(1,2)$. We analyze this simple model and show that this is indeed the case. Moreover, we prove that the uncertainty can be even further reduced by considering the $N$-qubit correlations between all particles.

In this way, we generalize the notion of correlation spectroscopy to a quantum sensor network of $N$ two-level quantum systems \cite{Shanif:2019,Young:2020} and demonstrate, in theory and experiment,
an improvement compared to the traditional pair-correlations method. We derive precision bounds when all the $N \choose 2$ pair correlations of the outcomes are used and for the case where all $N$-qubit correlations are exploited. These methods are implemented in experiments with ion crystals and are used to estimate ion-ion distances and transition frequency shifts. Finally, we propose theoretical schemes for further improvement with entangled measurements and initial states. 

The manuscript is structured as follows: in section~\ref{sec:model} we describe the principle of $N$-qubit correlation spectroscopy and how the analysis of measured correlations can be used for inferring relative phase shifts between the qubits as well as tracking correlated phase shifts on all qubits in the time domain. Section~\ref{sec:implementation} demonstrates the implementation of the measurement protocol in one- and two-dimensional ion crystals with up to 91 ions. In sec.~\ref{sec:bounds}, we discuss lower bounds to the measurement uncertainties when analyzing pair correlations or $N$-qubit correlations and demonstrate that these bounds are nearly saturated in our experiments. We further discuss general quantum precision limits and show that the input states used in our experiments are near-optimal in terms of the achievable measurement precision. Section~\ref{sec:applications} discusses applications of the method in trapped-ion experiments for the determination of transition frequency differences, ion-ion distances and tracking of local oscillator noise. In sec.~\ref{sec:discussion}, we discuss the application of our measurement protocol to other experimental platforms.

%----
\begin{figure}[t]
\centering
\includegraphics[width=0.5\textwidth]{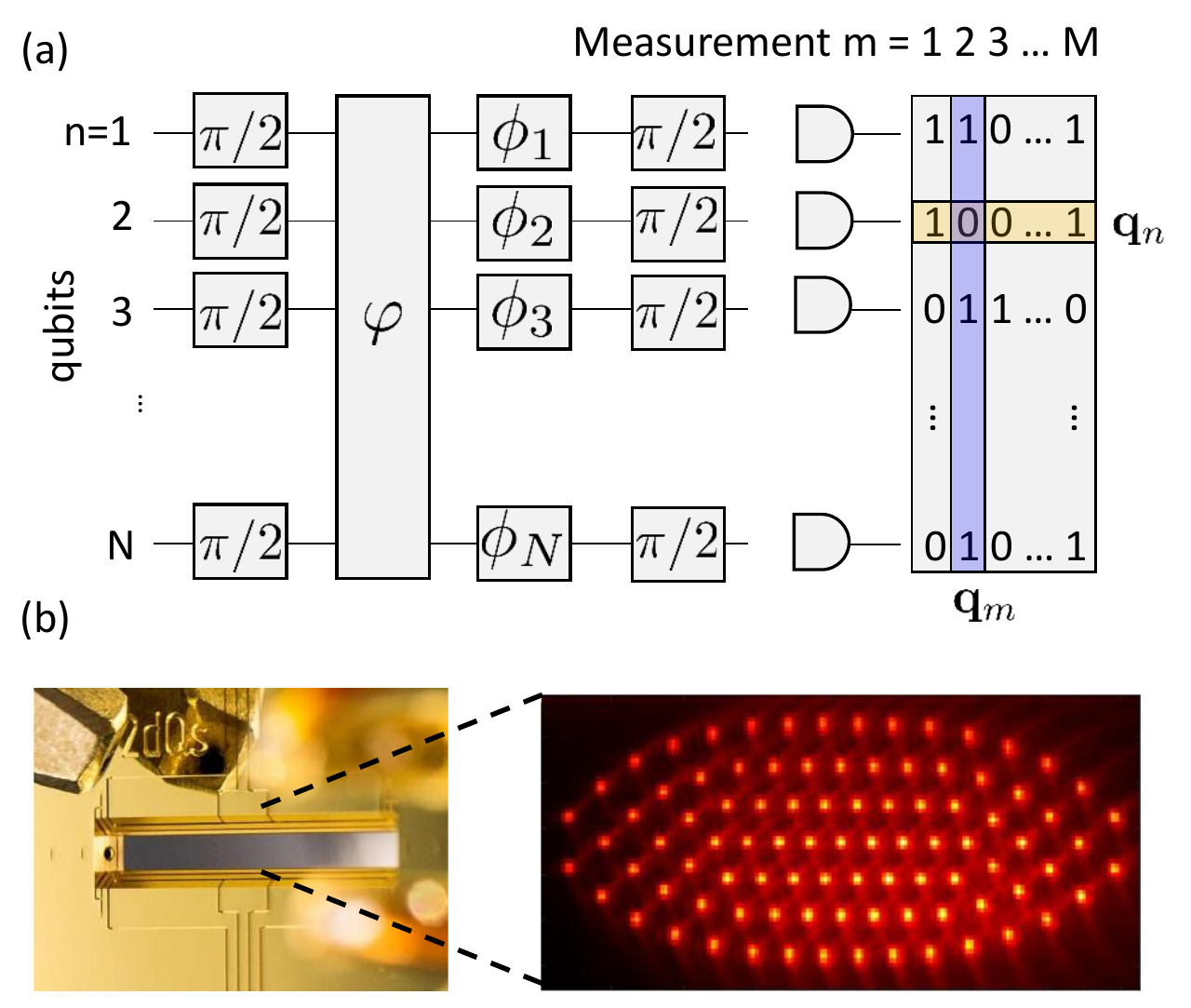}
\caption{(a) Measurement protocol: Ramsey experiments are simultaneously carried out on an ensemble of $N$ qubits subject to correlated dephasing, phase-shifting all qubits by a random phase $\varphi$, and single-qubit phase shifts $\phi_n$. The analysis of correlations between measurement outcomes on different qubits taken at the same time (column vector $\mathbf{q}_m$) enables the estimation of phase difference between qubits; similarly, the analysis of correlations between measurement outcomes taken at different times on the same qubit (row vector $\mathbf{q}_n$) provides information about the temporal evolution of phases (for details see main text). (b) We implemented correlation spectroscopy with ensembles of trapped and laser-cooled ions, such as the two-dimensional 91-ion crystal held in a monolithic ion trap shown in the picture.}
\label{fig:introduction}
\end{figure}
% data analysis: C:\Christian\tex\Papers\2021_CorrelationMeasurementsLongStrings\matlab\Figure_SpatialPositions
%----

% ------------------------------------------------------------------------
\section{Model: N-qubit correlation spectroscopy \label{sec:model}}
We consider a data set consisting of $m=1,\ldots,M$ realizations of Ramsey experiments with a free evolution time $T$, each of which is simultaneously carried out on an ensemble of $N$ qubits (see Fig.~\ref{fig:introduction}a). Prior to the second $\pi/2$ pulse, the state of the $N$ qubits during the m-th realization is
\begin{equation}
2^{-\frac{N}{2}}\underset{i=1}{\overset{N}{\Pi}}(|0\rangle+e^{i\phi_{im}}|1\rangle), \label{eq:statef}    
\end{equation}
with phases 
\begin{equation}
\phi_{im}=\phi_i+\varphi_m \label{eq:phiim}   
\end{equation}
where $\phi_i$ is a qubit-specific phase and $\varphi_m$ a random phase that is common to all qubits, i.~e. in experiments $\phi_i$ appears as a spatially varying phase whereas $\varphi_m$ encodes temporal changes. To achieve an unambiguous definition of these phases, we define $\varphi_m$ to be the phase change between the $m$th experiment and the first one. The phase $\varphi_m$ could result from a stochastic process  coupling the qubits to an environment inducing correlated dephasing; alternatively, it could be engineered in the experiment, for example by randomly shifting the phase of the first Ramsey pulse with respect to the second one. The qubit-specific phases $\phi_i=(\mathbf{k}_1-\mathbf{k}_2)\mathbf{r}_i+\Delta_iT$ arise if the qubits have different detunings $\Delta_i$ with respect to the local oscillator frequency or if the qubits are excited from different spatial directions for the first and the second $\pi/2$ pulse. Here, $\mathbf{k}_1 (\mathbf{k}_2)$ is the k-vector of the running wave inducing the first (second) $\pi/2$ pulse and $\mathbf{r}_i$ is the qubit position vector.  
We assign an outcome $q_{im}=1$ or $-1$ to the measurement, depending on whether we observe qubit $i$ in the $m$th measurement in the state $|0\rangle$  or $|1\rangle$. The probability of observing the outcome $q_{im}$ is given by $p(q_{im})=\frac{1}{2}(1+q_{im}\sin\phi_{im})$.

Here, we will study two closely related problems:

\noindent 1. We want to carry out a multi-parameter estimation of the qubit-dependent phases $\phi_i$ 
in experiments where the random phases $\varphi_m$ are uniformly distributed over the interval $[0,2\pi)$; this situation can arise if, for example, the probe time is much longer than the coherence time of the qubits. We can model this problem by preparing the qubits in
\begin{align}
\begin{split}
&\rho=\frac{1}{2\pi}\int_{0}^{2\pi}d\varphi|\Psi(\varphi)\rangle\langle\Psi(\varphi)|\;\;\;\mbox{with} \\
&|\Psi(\varphi)\rangle = 2^{-\frac{N}{2}}\prod_{i=1}^N(|0\rangle_i+e^{i\varphi}|1\rangle_i),
\end{split}
\label{eq::the state}
\end{align}
a state, which contains no entanglement but quantum correlations in the form of non-zero quantum discord \cite{Modi:2012,Lanyon:2013}. Next, the qubits are
subjected to the unitary operation $U_{\bm{\phi}}=\exp(\frac{i}{2}\sum_i\phi_i\sigma_i^z)$
followed by a global $\pi/2$ pulse around the x-axis, $U_X=\exp(-i\frac{\pi}{4}\sum_i\sigma_i^x)$, and finally a projective measurement of all qubits is carried out in the computational basis. Given a set of measurement outcomes stored in the matrix $Q=(q_{im})$, the goal is to devise a strategy for estimating all phase differences $\phi_i-\phi_j$ with optimal precision. Note that this is a special case of a quantum sensor network \cite{distributed1, distributed2, eldredge2018optimal,rubio2020quantum}, where the linear functions we wish to estimate are all the phase differences. 

\noindent 2. We are interested in characterizing the stochastic process that gives rise to temporally fluctuating random phases $\varphi_m$. Because of the symmetry of the problem in space and time as showing up in eq.~(\ref{eq:phiim}), a strategy for estimating the single-qubit phases $\phi_i$ can equally well be applied to an estimation of $\varphi_m$ by analyzing the transposed matrix $Q^t$ of measurement results.

Let us first understand the fundamental precision limits in estimating the phase differences. 
Given a pure product state, $2^{-\frac{N}{2}}\prod_{i=1}^{N}(|0\rangle_{i}+e^{i\phi_{i}}|1\rangle_{i}),$ and in the absence of noise, the precision in estimating each phase independently from $M$ measurements is $\sigma_{\phi_{i}}=\frac{1}{\sqrt{M}}.$ 
Hence the minimal uncertainty in estimating a phase difference $\Delta\phi=\phi_{2}-\phi_{1}$ is 
\begin{equation}
\sigma_{\Delta\phi}=\sqrt{\sigma_{\phi_{1}}^{2}+\sigma_{\phi_{2}}^{2}}=\sqrt{\frac{2}{M}}.\label{eq:noiselesslimit}
\end{equation}
This approach basically amounts to inferring $\Delta\phi$ from the relative phase shifts of two Ramsey fringes.
Since this is the best achievable precision with a product state, we refer to it hereafter as the noiseless precision bound.

Considering $N=2$ qubits and correlated dephasing, as in Eq.~(\ref{eq::the state}), the phase difference
$\Delta \phi$ is estimated using standard correlation spectroscopy.
Using error propagation of quantum projection noise, the uncertainty in the estimation of $\Delta\phi$
from $M$ measurements equals \begin{equation}
\sigma_{\Delta\phi}^{N=2}=\frac{\sqrt{4-\cos^{2}\left(\phi_{i}-\phi_{j}\right)}}{\sqrt{M}|\sin\left(\phi_{i}-\phi_{j}\right)|}. \label{eq:sigma_twoqubits}
\end{equation}
The uncertainty diverges when the phase difference approaches 0 or $\pi$, i.~e. the points where the parity reaches an extremum. It becomes minimal for $\Delta\phi=\pi/2$, where $\min \sigma_{\Delta\phi}^{N=2}=\frac{2}{\sqrt{M}}$, which is larger by a factor of $\sqrt{2}$ than the noiseless precision bound.
The $\sqrt{2}$ difference in the uncertainty stems from the reduced contrast ($<0.5$) in correlation spectroscopy. 

For $N>2$, one can ask whether the uncertainty of the phase difference estimation can be lowered by employing a more sophisticated analysis. Here, we provide an affirmative answer: we show that the uncertainty can be reduced by estimating the phase differences using all the $\binom{N}{2}$ pair correlations of measurement outcomes, and that in addition a further reduction is achieved by using all the multi-particle correlations.
The intuition behind this improvement is based on the following argument: An estimate of the single-qubit phase differences $\Delta\phi_{ij}=\phi_i-\phi_j$ from the observed correlations makes it possible to estimate the random phases $\varphi_m$ of each experimental realization. In the limit of a large number of qubits, the near-perfect estimation of $\varphi_m$ enables an ``unscrambling'' of the Ramsey fringes and in consequence a reconstruction of single-qubit Ramsey fringes with contrast close to $1$ instead of $1/2$ as for the two-qubit parity fringe.
This implies that we should be able to retrieve the noiseless precision bound 
of $\sigma_{\Delta\phi}^\infty=\sqrt{\frac{2}{M}}$ in the limit of $N\rightarrow\infty$. 

Here, and in the remainder of this section, we assume that in the absence of correlated dephasing Ramsey fringes would have the full contrast, i.~e. that there is no other source of decoherence. Later, this assumption will be dropped and we will also consider the influence of additional single-qubit dephasing on the measurement uncertainty. In the following, we will discuss different approaches for analyzing the multi-qubit correlations.

%------------------
\emph{Correlation spectroscopy with many qubits: pair correlations.} --- Ramsey measurements of individual qubits contain no useful information as measuring $\hat{Z}_i=|0\rangle\langle 0|-|1\rangle\langle 1|$ results in $\langle \hat{Z}_i\rangle=\mbox{Tr}(U_XU_{\bm{\phi}}\rho {U_{\bm{\phi}}}^\dagger U_X^\dagger\hat{Z}_i) 
=0$. 
Yet, information about transition frequency and position differences is obtained from correlation measurements \cite{Chwalla:2007}, 
\begin{equation}
C_{ij}\equiv\langle\hat{Z}_i\hat{Z}_j\rangle=\frac{1}{2}\cos(\phi_i-\phi_j),
\label{eq:correlation_definition}
\end{equation}
for which the correlated dephasing only reduces the maximum range of correlations by a factor of two. A fit of the correlation matrix $C=(C_{ij})$ yields estimates $\hat{\phi}_i$ of the single-qubit phases $\phi_i$ up to an irrelevant global offset phase. In experiments where $\mathbf{\Delta k}=\mathbf{k}_1-\mathbf{k}_2=0$, this approach can be used to determine differences in transition frequencies up to a global sign factor. If, on the other hand, $T=0$ and $|\mathbf{\Delta k}|\neq 0$, information about the spatial arrangement of the qubits is obtained. 

%------------------
\emph{Single-qubit phase estimates with N-particle correlations.} --- 
For an estimation of the single-qubit phases $\bm{\phi}=(\phi_1,\phi_2,\ldots)$, we calculate the likelihood of observing the single-shot measurement outcome $\mathbf{q}=(q_{1},\ldots,q_{N})$,
\begin{equation}
P(\mathbf{q}|\bm{\phi})=\frac{2^{-N}}{2\pi}\int_0^{2\pi}d\varphi\prod_{i=1}^N(1+q_i\sin(\phi_i+\varphi)). \label{eq:likelihood-of-bitstring}
\end{equation}
Given a set of measurements $Q=(q_{im})$, a maximum likelihood estimation of $\bm{\phi}$ is obtained via evaluation of the log-likelihood function
\begin{equation}
{\cal L}(Q|\bm{\phi}) = \log\prod_{m=1}^M  P(\mathbf{q}_m|\bm{\phi}). \label{eq:MLEstimation}
\end{equation}
Note that the calculation of the integral in eq.~(\ref{eq:likelihood-of-bitstring}) can be replaced by an average over $N+1$ evenly distributed phases $\varphi_m=2\pi m/N$ as the highest Fourier component of the integral kernel has a period of $2\pi/N$.

%------------------
\emph{N-particle correlations for estimating the collective random phases $\varphi_m$.} --- Once an estimate $\hat{\phi}_i$ of the single-qubit phases is available, single-shot Ramsey spectroscopy can be used for estimating the random phase $\varphi_m$ of an experimental run from the vector of outcomes $\mathbf{q}_m\equiv (q_{im})_{i=1}^N$. Towards this end, we calculate the likelihood function 
\begin{equation}
P(\mathbf{q}_m|\{\hat{\phi}_i\},\varphi)=2^{-N}\prod_{i=1}^N(1+q_{im}\sin(\hat{\phi}_i+\varphi)) \label{eq:likelihoodfunction}
\end{equation}
and use it for a Bayesian estimate of the random phase
\begin{equation}
\hat{\varphi}_m=\arg(\int_0^{2\pi}d\varphi e^{i\varphi}P(\mathbf{q}_m|\{\hat{\phi}_i\},\varphi)).\label{eq:BayesianEstimate-time-domain}
\end{equation}
This approach allows for tracking the temporal fluctuations of the local oscillator with respect to the qubit transition frequencies.

We note that the Bayesian approach can also be applied to estimating the vector of single-qubit phases $\bm{\phi}$. Given the estimates $\hat{\varphi}_m$, the single-qubit phase differences $\Delta\phi_{ij}$ can be estimated by calculating the likelihood function 
\begin{equation}
P(\mathbf{q}_i|\phi_i,\{\hat{\varphi}_m\})=2^{-N}\prod_{m=1}^M(1+q_{im}\sin(\phi_i+\hat{\varphi}_m)), \label{eq:likelihoodfunction2}
\end{equation}
where $\mathbf{q}_i\equiv (q_{im})_{m=1}^M$, in order to obtain the Bayesian estimate 
\begin{equation}
\hat{\phi}_i=\arg(\int_0^{2\pi}d\phi e^{i\phi}P(\mathbf{q}_i|\{\hat{\varphi}_m\},\phi)). \label{eq:BayesianEstimate-qubit-phases}
\end{equation}
This approach is computationally fast albeit less precise than maximum-likelihood estimation (MLE) of $\bm{\phi}$. As discussed further below, the resulting uncertainties approach the ones obtained with maximum likelihood estimation only in the limit of large number qubits whereas the performance is unsatisfactory for small numbers of qubits.
%}

%-----------------------------------------------------------------------------
\section{Experimental implementation and measurement results \label{sec:implementation}}
Measurements on linear and planar $^{40}$Ca$^+$ ion crystals are performed in two different experimental setups that will be described in the following.

%----
\begin{figure*}[t]
\centering
\includegraphics[width=1\textwidth]{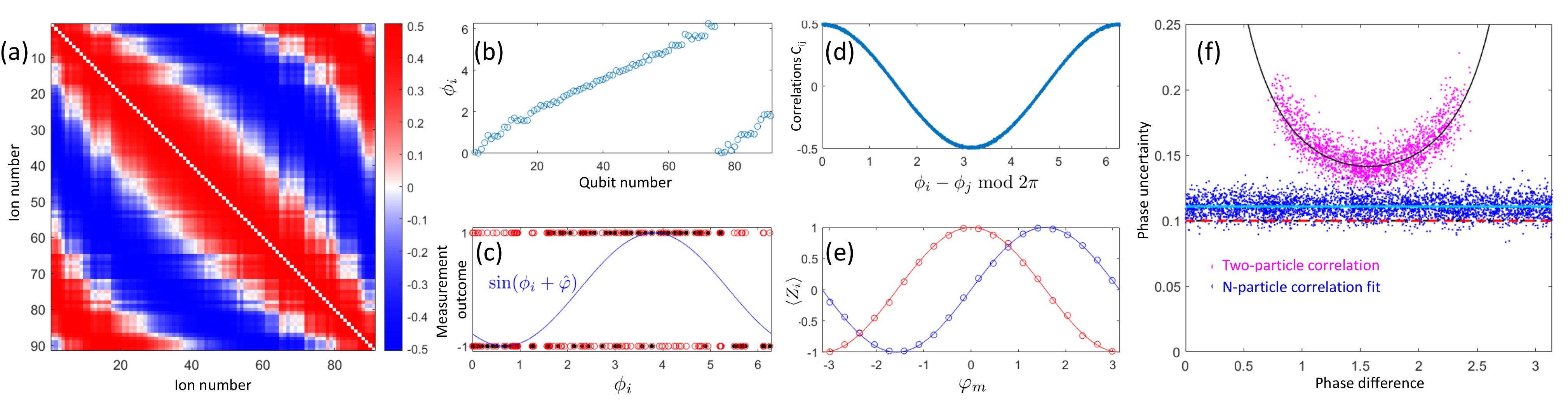}
\caption{Many-qubit correlation spectroscopy of a 91-ion planar crystal based on $M=26852$ experimental repetitions. (a) Measured correlation matrix with correlations $|C_{ij}|\leq 1/2$ limited by correlated dephasing. (b) Single-qubit phases $\hat{\phi}_i$ estimated by fitting the correlation matrix. (c) Measurement outcomes $q_{im}$ of an individual experiment (black dots) used for a Bayesian estimate of the common random phase $\hat{\varphi}_m$ (blue curve: fitted Ramsey fringe). (d) Correlation matrix elements $C_{ij}$ plotted as a function of the phase difference $\phi_i-\phi_j$ obtained by analyzing N-qubit correlations. (e) Single-qubit Ramsey fringes with nearly full contrast obtained from binning into sets of similar common random phase $\varphi_m$. The red and the blue curve are just two out of 91 measured fringes. (f) Measurement uncertainties inferred from subdividing the data into 134 data sets with 200 repetitions each. The pink dots do not cover the entire range of 0 to $\pi$ as we omit those qubit pairs (i,j) for which we measure $|C_{ij}|>0.5$ for one or several subsets.  Uncertainties obtained from individual elements $C_{ij}$ (pink dots)
and the analysis of N-qubit correlations (blue dots, solid light blue line: average over the data points). The dashed red line indicates the noiseless precision bound achievable in the limit of $N\rightarrow\infty$, the solid black line the two-qubit limit $\sigma_{\Delta\phi}^{N=2}$.}
% Scan data combined from scans on 2021.10.13 03_00_17_521386 -- 03_00_26_705498
\label{fig:ExpPhaseEstimation}
\end{figure*}
%----

The centerpiece of the apparatus for trapping planar crystals is a novel microfabricated monolithic linear Paul trap, shown in Fig.~\ref{fig:introduction}b, which allows us to create the anisotropic potentials required for trapping 2D ion crystals while simultaneously maintaining sufficient optical access perpendicular to the crystal plane for ion imaging. The trap provides a potential in which the ions are strongly confined in the direction perpendicular to the crystal plane, at an oscillation frequency of 2.196~MHz, and weakly confined along the two other directions, in which the crystal is extended, at oscillation frequencies of about 679.8~kHz and 343.0~kHz. Further details on this new ion trap apparatus can be found in Ref.~\onlinecite{Kiesenhofer:2023a}.
Ions are loaded into the trap via laser ablation and are Doppler-cooled on the $\mathrm{S}_{1/2}\leftrightarrow\mathrm{P}_{1/2}$ dipole transition. For encoding a qubit in an ion we use the two 4S$_{1/2}, m=\pm 1/2$ Zeeman ground states, coherently coupled by a magnetic radiofrequency field oscillating at approximately 11.4~MHz.  We distinguish the two qubit states by shelving the population of one of them in the long-lived 3D$_{5/2}$ Zeeman states, followed by fluorescence detection: The qubits are measured with high fidelity by exciting the ions on the $\mathrm{S}_{1/2}\leftrightarrow\mathrm{P}_{1/2}$ transition and imaging the ion fluorescence onto an electron-multiplying CCD camera. For the shelving operation, we employ $\pi$-pulses induced by a frequency-stable 729~nm laser, coming from a direction perpendicular to the crystal plane.

In contrast to the apparatus for manipulating 2D crystals, long strings of $^{40}$Ca$^+$ ions are trapped in a macroscopic linear Paul trap providing a very anisotropic trapping potential with radial oscillation frequencies of about 2.5-3~MHz and an axial oscillation frequency of about 120~kHz. After Doppler cooling, the radial modes of the ion string are cooled close to the ground state by sideband cooling and the axial modes sub-Doppler cooled by polarization-gradient cooling \cite{Joshi:2020}. The qubit is encoded in one of the two 4S$_{1/2}$ Zeeman ground states and one of the metastable 3D$_{5/2}$ Zeeman states. The ion-qubit can be coherently manipulated using 729~nm laser light resonantly exciting the S$_{1/2}\leftrightarrow$D$_{5/2}$ transition. Two laser beams with k-vectors parallel (perpendicular) to the linear ion crystal are available for collectively coupling to the qubits with the same coupling strength. Further details about this experimental setup can be found in Ref.~\onlinecite{Kranzl:2022}.

In a first measurement, we investigate multi-qubit enhanced phase estimation in a 91-ion planar crystal; the results are shown in Fig.~\ref{fig:ExpPhaseEstimation}. We probe the ground-state qubits with a Ramsey probe time of 10~ms; here, magnetic field inhomogeneities gave rise to qubit-dependent phases $\phi_i$ and correlated dephasing was the result of temporal fluctuations of the magnetic field's magnitude. 
Fig.~\ref{fig:ExpPhaseEstimation}(a) shows the measured pair correlations $C_{ij}$ used for a first estimate $\hat{\phi}_i$ of the single-qubit phases shown in panel (b). Panel (c) displays the outcomes of an individual Ramsey experiment plotted against $\hat{\phi}_i$ together with a single-shot Ramsey fringe obtained from an estimate of the collective random phase $\varphi_m$. In (d), the matrix elements $C_{ij}$ are plotted versus the improved estimate $\hat{\phi}_i-\hat{\phi}_j$ obtained by maximum-likelihood estimation based on eq.~(\ref{eq:MLEstimation}), for which we maximized the likelihood by a gradient-based optimization algorithm \cite{Schmidt:2009}. The plot shows that the contrast of the resulting fringe is close to the maximum possible value. Similarly, averaging over experiments carried out at similar values of $\varphi_m$ results in single-qubit Ramsey fringes with contrast close to 1 (panel (e)). By subdividing the data sets into subsets of 200 measurements each, it is possible to measure the uncertainty of the phase difference estimates $\hat{\phi}_i-\hat{\phi}_j$. Pink data points in Fig.~\ref{fig:ExpPhaseEstimation}(f) show the uncertainty based on estimating the phase difference from the pair correlation between two qubits, which becomes minimal for a phase difference of $\pi/2$. The measured uncertainties are in agreement with the bound provided by quantum projection noise in the presence of correlated dephasing. The lowest uncertainty is obtained by maximum likelihood estimation using $N$-qubit correlations (blue data points, blue line: average over all points). The dashed line is the lower bound in the limit of $M\rightarrow\infty$ and $N\rightarrow\infty$. All data presented in Fig.~\ref{fig:ExpPhaseEstimation} and subsequent figures is available online \cite{Hainzer:2022Zenodo}.

%----
\begin{figure}[t]
\centering
\includegraphics[width=.5\textwidth]{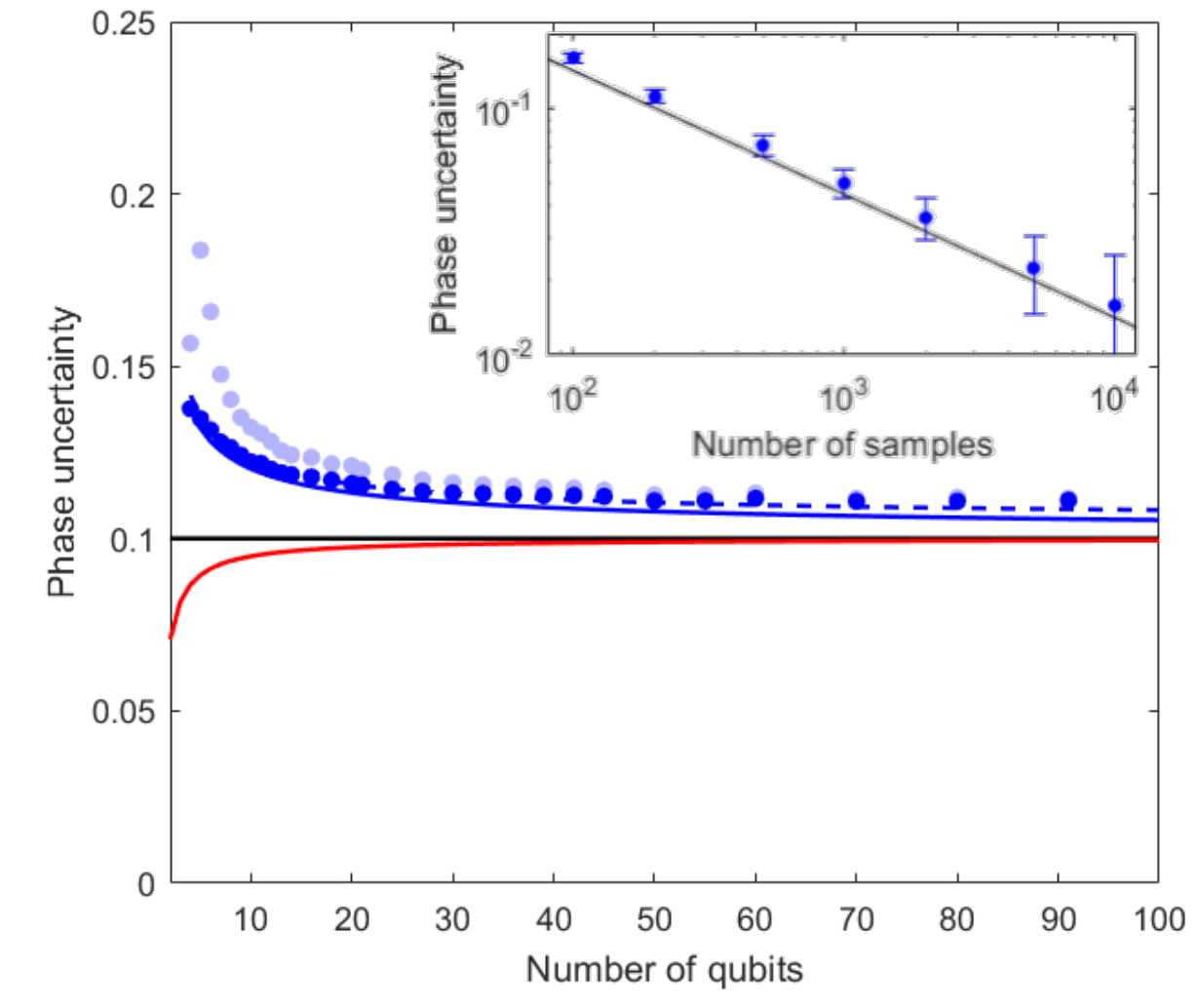}
\caption{Phase uncertainties vs number of qubits for $M=200$ samples. Dark blue dots represent uncertainties estimated from experimental data by MLE, light blue dots the uncertainties of the Bayesian estimation. The prediction of eq.~(\ref{eq:uncertainty_simplemodel}) is shown as the solid blue curve for a contrast $C_0=1$ and as a dashed blue curve for $C_0=0.995$. The latter is obtained by fitting the experimental data. The black curve represents the noiseless precision bound of $\sqrt{\frac{2}{M}}.$ The red curve represents the optimum uncertainty that is obtainable with entangled input states (see Appendix~\ref{sec:dicussion-details}). The reduction in measurement uncertainty provided by preparation of entangled input states rapidly shrinks with increasing $N$. The inset displays the measured uncertainty for $N=91$ on the number samples, with the black curve representing the noiseless precision bound.}
\label{fig:ExpPhaseUncertaintyVsNumberOfQubits}
\end{figure}
%----

The same data set can also be used for investigating the measurement uncertainties as a function of the number of qubits as shown in Fig.~\ref{fig:ExpPhaseUncertaintyVsNumberOfQubits}. Towards this end, we split the data into subsets, each containing a fixed number of qubits with single-qubit phases that are approximately evenly distributed over the interval $[0, 2\pi)$. The measurements of each of these sets is further split into subsets containing $M=200$ experimental realizations from which we reconstruct the single-qubit phases for an estimate of the measurement uncertainties. Dark blue data points represent reconstructions based on MLE (eqs.~(\ref{eq:likelihood-of-bitstring}) and (\ref{eq:MLEstimation})), light blue points are the results of the non-competitive Bayesian approach (eqs.~(\ref{eq:likelihoodfunction2}) and (\ref{eq:BayesianEstimate-qubit-phases})). As shown in section~\ref{sec:bounds}, the uncertainties of the MLE estimates can be fitted by eq.~(\ref{eq:uncertainty_simplemodel}) with a fringe contrast of $C_0=0.995$, which could result from state-assignment errors and slow drifts of trap parameters over the duration of the measurement. The inset shows that the phase uncertainty decreases inversely proportional to the number of samples in a given set and is thus still projection-noise limited at $M=10^{4}$ samples. Note that we used an unbiased estimator for the determination of the uncertainties displayed in Fig.~\ref{fig:ExpPhaseEstimation} and Fig.~\ref{fig:ExpPhaseUncertaintyVsNumberOfQubits} assuming normally distributed measurement results \cite{Holtzman:1950}.

%-----------------------------------------------------------------------------
\section{Bounds to the achievable phase estimation uncertainty\label{sec:bounds}}
In this section, we compare the experimentally measured uncertainties to the theoretically achievable minimum uncertainties for an unbiased estimator and $M$ experimental samples. 
The noiseless precision bound $\sqrt{\frac{2}{M}}$ of eq.~(\ref{eq:noiselesslimit}) cannot be experimentally achieved as it assumes noiseless dynamics, i.e. that the single-qubit Ramsey fringes (cf. Fig.~\ref{fig:ExpPhaseEstimation}e) can be measured with unity contrast. However, this assumption is unrealistic in noisy experiments affected by strong correlated dephasing and small levels of single-qubit dephasing.

The impact of these two noise sources on the precision is different. Uncorrelated dephasing reduces the fringe contrast to $C_{0}<1,$ which unavoidably degrades the precision.
In contrast, the effect of strong correlated dephasing can be overcome with a suitable data analysis for a large number of ions.  
This can be understood as follows:
given a large number of ions, the random phase in each shot, $\varphi_{m}$, can be estimated with an error that goes to zero as $N \rightarrow \infty.$ Another way to understand this is that for correlated dephasing there are decoherence-free subspaces (unlike the case of uncorrelated dephasing). The density matrix has elements inside and outside the decoherence-free subspaces, and as $N \rightarrow \infty$ the contribution of the elements outside the protected subspaces goes to zero.
Here, we will take these factors into consideration.
We derive heuristic precision bounds that depend on both the number of qubits $N$ and a finite Ramsey contrast $C_0$ (in the absence of correlated dephasing). We will start with a simple analytical model for an estimator based on $N$-particle correlations and compare the predictions to numerical simulations based on the classical Fisher information.

%----------
\subsection{Measurement uncertainties with N-particle correlations} \label{sec:simple_model}

Here we derive a simple analytic model to approximate the precision bounds given correlated and uncorrelated dephasing.

Before we proceed to the model let us first understand the effect of uncorrelated dephasing and generalize the noiseless precision bound to a finite contrast limit that takes into account this dephasing and serves as a similar benchmark. Given an uncorrelated dephasing, the product state mentioned in section \ref{sec:model} becomes a mixed state:
$\frac{1}{2^{N}}\underset{i}{\prod}\left(\mathrm{I}+C_{0}\hat{R}_{\phi_{i}}\right)$, with 
$\hat{R}_{\phi_{i}}=\cos\left(\phi_{i}\right)\hat{X}_{i}+\sin\left(\phi_{i}\right)\hat{Y}_{i}.$
Measuring the $i$-th qubit in the $\hat{Y}_{i}$ basis, the probability of $1$ is: $p=\frac{1}{2}\left(1 + C_{0}\sin \left( \phi_{i} \right) \right).$ Then the uncertainty in the determination of $p$ with $M$ repetitions is given by projection noise, $\sigma_p=\sqrt{\frac{p(1-p)}{M}}$. Because of $\sigma_p=\left|\frac{dp}{d\phi_{i}}\right|\sigma_{\phi_{i}}$ and $\frac{dp}{d\phi_{i}}=\frac{C_{0}}{2}\cos \left( \phi_{i} \right)$, we have $\sigma_{\phi_{i}}=\frac{1}{\sqrt{N_{\mathrm{eff}}(C_{0},\phi_{i})}}$
where
\begin{equation}
N_{\mathrm{eff}}(C_{0},\phi_{i}) = M C_{0}^2\frac{(1-\sin^2\phi_{i})}{1-C_{0}^2\sin^2\phi_{i}}
\label{eq:Neff_1}
\end{equation}
is an effective number of measurements. Assuming $\phi_{i}$ is drawn from a uniform distribution, the uncertainty becomes $\sigma_{\phi_{i}}=\frac{1}{\sqrt{N_{\mathrm{eff}}(C_{0})}}$ with
\begin{align}
N_{\mathrm{eff}}(C_{0})&=\frac{1}{2\pi}\int_{0}^{2\pi}d\phi_{i}\,N_{\mathrm{eff}}(C_{0},\phi_{i})\nonumber\\
&=M(1-\sqrt{1-C_{0}^2}).
\label{eq:Neff_2}
\end{align}   
The uncertainty in estimating $\Delta \phi=\phi_{i}-\phi_{j}$ is $\sqrt{\sigma_{\phi_i}^{2}+\sigma_{\phi_j}^{2}},$ and thus equal to
\begin{equation}
\sigma_{\Delta\phi}=\frac{\sqrt{2}}{\sqrt{M}\sqrt{1-\sqrt{1-C_{0}^{2}}}}. 
\label{eq:sigma_uncorrelateddephasing}
\end{equation}
Since this is the minimal obtainable uncertainty given a contrast of $C_{0}$ and assuming no 
correlated dephasing we refer to it as the finite contrast precision bound.

We derive now a simple model for precision bounds given also correlated dephasing. The idea is to first find the uncertainty in estimating the common random phase $\varphi_{m}$ and then insert this uncertainty as an uncorrelated dephasing of each qubit.

We first need to find the uncertainty in estimating $\varphi_{m}$ (given that all the other phases are known). Note that this is exactly the same calculation as performed above for $\phi_{i}$, just taking $M=N$, therefore: $\sigma_{\varphi_{m}}=\frac{1}{\sqrt{N}\sqrt{1-\sqrt{1-C_{0}^{2}}}}$.  
This limits the precision with which the random phases $\varphi_m$ can be estimated.
We can thus take the distribution of the random phase to be Gaussian with this variance. By averaging the single qubit probability over the Gaussian distribution ${\cal N}\left(\varphi_{m},\sigma^{2}\right)$,
$$\langle\cos\left(\phi_{i}+\varphi\right)\rangle_{{\cal N}\left(\varphi_{m},\sigma^{2}\right)}=\cos\left(\phi_{i}+\varphi_{m}\right)\exp\left(-\frac{\sigma^{2}}{2}\right),$$ 
we observe that the contrast of the unscrambled single-qubit Ramsey fringe (Fig.~\ref{fig:ExpPhaseEstimation}e) gets reduced to \begin{equation}
C_{unscr}=C_0\exp(-\frac{1}{2N(1-\sqrt{1-C_0^2})}),
\end{equation}
if $C_0$ was the Ramsey contrast in the absence of correlated dephasing. We can now apply the same reasoning again to estimate the uncertainty with which the shift of unscrambled Ramsey fringes can be determined in order to estimate the uncertainty of the phase difference $\phi_i-\phi_j$ which becomes
\begin{equation}
\sigma_{\Delta\phi}(N,C_0) = \frac{\sqrt{2}}{\sqrt{M(1-\sqrt{1-C_{unscr}^2})}}. \label{eq:uncertainty_simplemodel}
\end{equation}
For the case of large qubit number and high contrast $C_0$, this expression can be approximated by eq.~(\ref{eq:sigma_uncorrelateddephasing}) if the replacement $C_0\rightarrow C_0\exp(-\frac{1}{N})$ is made.

%----------
\subsection{Fisher information based bounds}
\label{sec:FI}
We use a Fisher information (FI) analysis to calculate the achievable minimum uncertainties. 
According to the Cramer-Rao bound, the Fisher information matrix sets a bound on the achievable uncertainty with any unbiased estimator, 
\begin{equation}
\text{COV}\left(\bm{\phi}\right)\geq I^{-1},    
\end{equation}
where COV is the covariance matrix of the parameters 
$\bm{\phi}=(\phi_i)$ and $I^{-1}$ 
is the inverse of the FI matrix. In case $I$ is singular, i.e. information can be obtained only about a subspace of the parameters, $I^{-1}$ is the Moore-Penrose pseudoinverse, defined only on this subspace. 
This implies that 
the variance of the phase difference $\phi_i-\phi_j$ is given by
\begin{equation}
\mbox{Var}(\phi_i-\phi_j)\ge\mathbf{v}_{ij}^tI^{-1}\mathbf{v}_{ij},
\end{equation}
where $\mathbf{v}_{ij}$ is a column vector with components $(\mathbf{v}_{ij})_n=\delta_{in}-\delta_{jn}$
and $I^{-1}$ the inverse of the relevant FI matrix.

The Fisher information matrix $I=(I_{ij})$ can be calculated by the following formula:
\begin{equation}
I_{ij}=\sum_kp_k^{-1}(\bm{\phi})\frac{\partial p_k(\bm{\phi})}{\partial\phi_i}\frac{\partial p_k(\bm{\phi})}{\partial\phi_j},
\label{eq:FImatrix}
\end{equation}
where $\bm{\phi}=(\phi_i)$ is the vector of parameters
and $p_k$ the probability distribution of the observations.
As a simple example, observe that for a single-parameter Bernoulli distribution, $p\left(\phi\right)$, the FI about $\phi$ is $I=\frac{\left(\partial_{\phi}p\right)^{2}}{p\left(1-p\right)}$ and hence $\sigma_{\phi}=\frac{\sqrt{p\left(1-p\right)}}{\partial_{\phi}p}.$ Given $M$ identical independent Bernoulli trials, the FI about $\phi$ is multiplied by a factor of $M$ and thus $\sigma_{\phi}=\frac{\sqrt{p\left(1-p\right)}}{\sqrt{M}\partial_{\phi}p}.$ This expression coincides with the uncertainty of eqs.~ (\ref{eq:noiselesslimit})-(\ref{eq:sigma_twoqubits}).
Furthermore, note that the FI is a generalization of $N_{\mathrm{eff}}$ defined in section \ref{sec:simple_model}.

%--------
\subsubsection{Fisher information bound for the pair correlations}
In pair correlation analysis, we estimate the ion phases $\left(\phi_{i}\right)$ using the pair correlations of the measurement outcomes, i.e. the correlation matrix $C_{i,j}$ defined in eq. (\ref{eq:correlation_definition}) and presented in Fig. \ref{fig:ExpPhaseEstimation}(a). More precisely we take the averages $\left\{ \frac{1}{M}\underset{m=1}{\overset{M}{\sum}}q_{i,m}q_{j,m}\right\} _{i\neq j}$ and estimate the phases according to it. According to the central limit theorem, the averages converge to a Gaussian random variable ${\cal N}\left(\left(\mathbf{\mu}_{i,j}\right)_{i\neq j},M^{-1} \Sigma\right),$ where $\mathbf{\mu}_{i,j}=\langle q_{i}q_{j}\rangle=\ensuremath{\frac{1}{2}C_{0}^{2}\cos\left(2\left(\phi_{i}-\phi_{j}\right)\right)}$ and $\Sigma$ is the covariance matrix $\Sigma_{\left(i,j\right),\left(k,m\right)}=\langle q_{i}q_{j}q_{k}q_{m}\rangle-\langle q_{i}q_{j}\rangle\langle q_{k}q_{m}\rangle.$ An explicit calculation of the covariance matrix elements is presented in Appendix~\ref{sec:details-on-bounds}. 

Since the distribution is normal the FI matrix about $\left(\phi_{i}\right)$ is given by \cite{cover1999elements}:
\begin{equation}
I=\left(\partial_{\phi}\boldsymbol{\mu}\right)^{\dagger}\Sigma^{-1}\left(\partial_{\phi}\boldsymbol{\mu}\right).
\label{eq:gaussian_FI}
\end{equation} 
$\left(\partial_{\phi}\boldsymbol{\mu}\right)_{i,\left(k,m\right)}=\partial_{\phi_{i}}\mu_{k,m}$ is the information gained due to the change in the mean values, i.e. the signal, and $\Sigma$ is the covariance matrix of the different correlations representing the noise.

Applying Eq.~(\ref{eq:gaussian_FI}) for a single pair correlation $\left(i,j\right)$ we retrieve the uncertainty in Eq.~(\ref{eq:sigma_twoqubits}):  the only linear combination of $\phi_{i},\phi_{j}$ that has a non-vanishing FI is $\phi_{i}-\phi_{j},$ for which the FI is $\frac{\sin^{2}\left(\phi_{i}-\phi_{j}\right)}{4-\cos^{2}\left(\phi_{i}-\phi_{j}\right)},$ i.e. $\sigma_{\Delta\phi}=\frac{\sqrt{4-\cos^{2}\left(\phi_{i}-\phi_{j}\right)}}{|\sin\left(\phi_{i}-\phi_{j}\right)|}.$ The minimal uncertainty per measurement is $2$, and a divergence occurs for $\phi_{i}-\phi_{j}=n\pi\;\left(n\in\mathbb{Z}\right)$ due to the vanishing derivative and non-vanishing noise. 

Since information about $\phi_{i}-\phi_{j}$ is encoded not only in the (i,j) correlations but in other pairs as well, using all pairs improves the uncertainty, and removes the divergence around $n \pi.$ We use eq.~(\ref{eq:gaussian_FI}) to perform an exact numerical calculation of the FI.
The behavior of the FI is presented in Figs. \ref{fig:2corrs}, \ref{fig:N_qubit_corrs} in Appendix~\ref{sec:details-on-bounds}. 
It can be seen from the figures that as $N \rightarrow \infty$ the FI with pair correlations does not saturate the noiseless precision bound. The reason for this is the information encoded only in higher moments. Using an analytical approximation we show in Appendix~\ref{sec:details-on-bounds} that the variance for large $N$ converges to $\frac{4-C_{0}^{2}}{C_{0}^{2}},$ while the finite contrast precision bound to the variance is $\frac{2}{1-\sqrt{1-C_{0}^{2}} }.$ As $C_{0}$ gets smaller the variance with pair correlations gets closer to the this bound since the information from higher moments becomes smaller.

%--------
\subsubsection{Fisher information bound with N-particle correlations}
When using the full counting statistics, the probability distribution entering into the calculation of the Fisher information matrix is given by eq.~(\ref{eq:likelihood-of-bitstring}) with the replacement $q_i\rightarrow C_0q_i$ in order to account for a Ramsey contrast $C_0<1$. 
In Appendix~\ref{sec:details-on-bounds}, we numerically calculate the Fisher information matrix for finding the lower limit to the achievable uncertainty as a function of qubit number $N$ and contrast $C_0$ (see Fig.~\ref{fig:N_qubit_corrs}). When $N$ becomes large, an exact evaluation of the Fisher information matrix by eq.~(\ref{eq:FImatrix}) becomes impractical as a summation over $2^N$ terms would have to be carried out. For $N>24$, we sampled bit strings from the underlying probability distribution for a  Monte-Carlo calculation of the empirical Fisher information matrix. The uncertainty achievable in experiments with a finite number of repetitions are numerically investigated in Appendix~\ref{sec:numerical-simulations}.

%---------------------------
\subsubsection{Improving precision limits using entanglement}
In our experiments, the qubits are initialized to a product state and measured in a local $X$ basis. Hence, no entanglement occurs in these experiments, and an analysis based on classical Fisher information suffices. This raises the question of whether non-classical protocols that involve entangled states or different measurement bases can yield an advantage.
It turns out that this is indeed the case: more general quantum protocols can obtain the noiseless precision bound of $\sqrt{\frac{2}{M}}$ with an initial product state for every $N$ and with an entangled initial state we can further reduce the uncertainty to $\sqrt{\frac{N-1}{N}} \sqrt{\frac{2}{M}}.$
We prove in Appendix~\ref{sec:dicussion-details} that this uncertainty is optimal.

To obtain these results we use the quantum Fisher information (QFI), which is the FI optimized over all possible measurement strategies \cite{braunstein1994statistical,liu2019quantum}. After averaging the quantum state over the random phase (Eq.~(\ref{eq::the state})), we show that the noiseless precision bound can be achieved for every $N$ with a suitable measurement strategy (see appendix \ref{sec:dicussion-details-nonlocal-measurements}).
To gain intuition, let us examine the case of $N=2$: when measuring in the local $X$ basis, eq.~(\ref{eq:sigma_twoqubits}) predicts $\sigma_{\Delta\phi}\geq \frac{2}{\sqrt{M}}$. However if we first measure the total number of excitations, i.e. $Z_{1}+Z_{2}$, and then measure in the local $X$ basis, the noiseless precision bound of $\sqrt{ \frac{2}{M}  }$ is achieved. 

Optimizing over both initial states and measurement strategies, we prove in appendix \ref{sec:dicussion-details-initial-states} that the ultimate precision limit is $\sqrt{\frac{N-1}{N}} \sqrt{\frac{2}{M}}.$ 
Several initialization strategies saturate this bound, in particular any initial pure state with $\langle Z_{j}\rangle=0, \langle Z_{j}Z_{k}\rangle=-\frac{1}{N-1}$ for all $j\neq k$ achieves it. The reason for this improvement is the minimal $\underset{j \neq k} {\sum} \langle Z_j Z_k \rangle,$ which guarantees minimal uncertainty .
The reason this limit grows with $N$ is frustration: one cannot make all pairs of spins anti-parallel. While the number of pairs is ${N \choose 2}$, the minimal $\underset{j < k}{\sum}\langle Z_{j}Z_{k}\rangle$ is $-N/2$ and thus the optimal $\langle Z_j Z_k \rangle$ is $-\frac{1}{N-1}$.

It can be immediately observed that the symmetric Dicke state with $N/2$ excitations satisfies these conditions and thus is optimal.
Another optimal strategy is to employ a probabilistic initialization to products of anti-parallel Bell states, i.e.  in each experiment different pairs are being entangled to form an anti-parallel Bell state. With these two initialization strategies, the optimal sensitivity can be achieved with local measurements in the $X$ or $Y$ basis. This bound is plotted as a red curve in Fig.~\ref{fig:ExpPhaseUncertaintyVsNumberOfQubits} along with other theoretical limits. A detailed derivation of the bound and the required initial states and measurements is presented in Appendix \ref{sec:dicussion-details-initial-states}.

These theoretical quantum limits imply that some improvement can indeed be obtained using entangled states or non-local measurements, however this improvement becomes negligible in the limit of large number of ions.
This potential improvement and a comparison between the different precision limits is presented in Fig.~\ref{fig:ExpPhaseUncertaintyVsNumberOfQubits}.

%-----------------------------------------------------------------------------
\section{Applications in trapped-ion experiments \label{sec:applications}}
In the following, different applications of correlation spectroscopy in trapped-ion experiments will be presented. 

%----------------------------------------------------------
\subsection{Measurement of ion positions\label{subsec:ionpositions}}
Very anisotropic potentials are required for confining many ions in the form of a linear string. As a consequence of the weak axial confinement, these strings have lengths that are no longer small as compared to the distance between the ions and the nearest trap electrode. Therefore, the trapping potential can no longer be modeled as being purely harmonic and anharmonicities, which might affect the ion string's normal modes of motion, have to be considered.

\begin{figure}[t]
\centering
\includegraphics[width=0.5\textwidth]{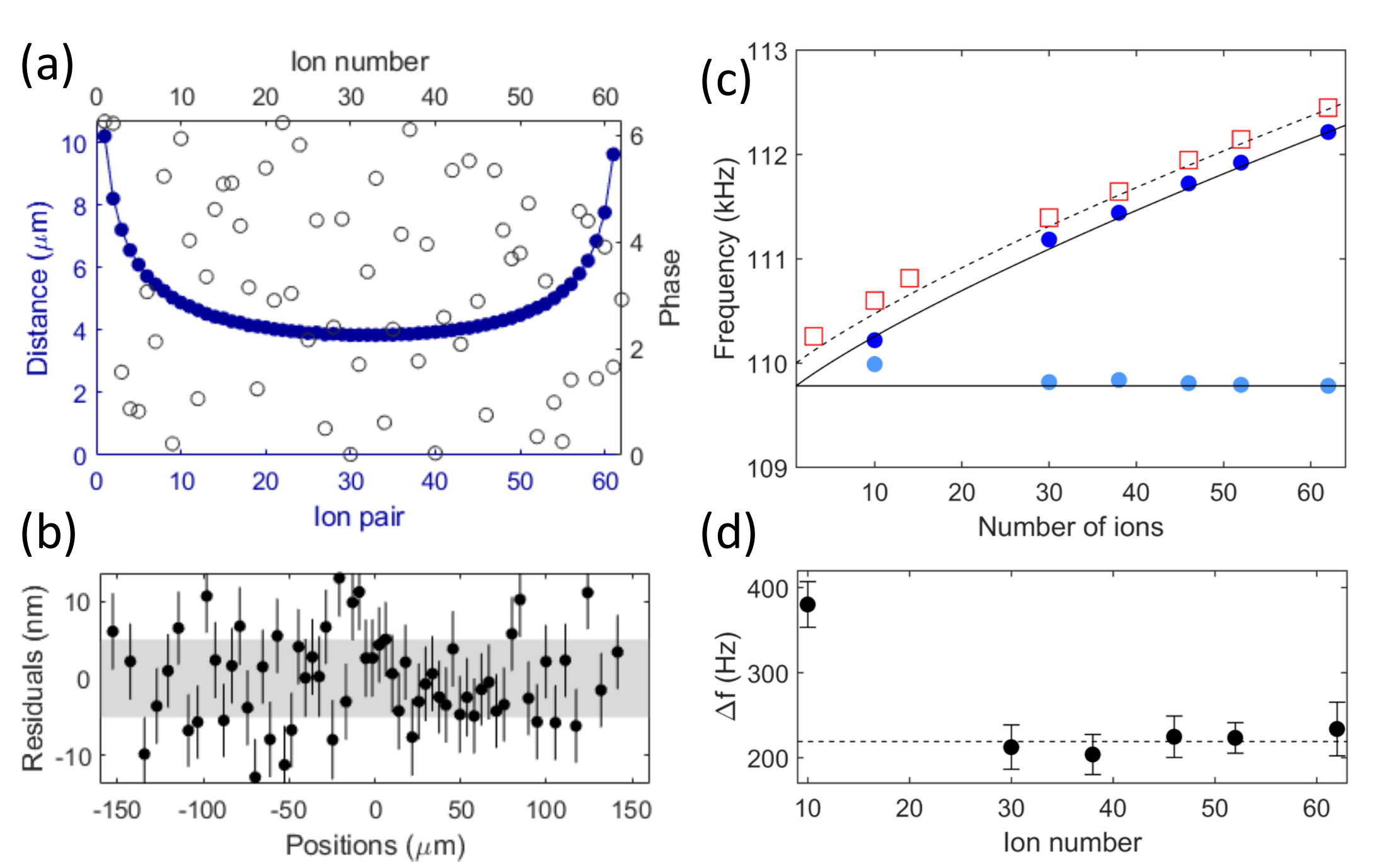}
\caption{(a) Reconstruction of the axial trapping potential. (a) Measured phases $\phi_i$ (open symbols) and nearest-neighbour distances (full symbols) obtained by fitting a model potential. (b) Residuals. Shaded area: theoretical minimum measurement uncertainty. (c) Center-of-mass mode frequency measured by correlation spectroscopy, $\omega_{c}/(2\pi)$ (dark blue circles) and by sideband spectroscopy, $\omega_{sb}/(2\pi)$, (squares) vs number of ions, together with the fitted value of $\omega_0/(2\pi)$ (light blue circles). (d) Measured $(\omega_{sb}-\omega_c)/(2\pi)$.}
\label{fig:SpatialCorrelations}
\end{figure}

We reconstruct the trapping potential in the axial direction by Ramsey experiments probing an optical qubit on the S$_{1/2}\leftrightarrow$D$_{5/2}$ transition, in which the first (second) $\pi/2$ pulse is realized by a laser beam impinging on the ions from the axial (perpendicular) direction. This setting results in qubit-specific phases $\phi_i=kx_i$ where $k$ is the wave number and $x_i$ denotes the coordinate of the $i^{th}$ ion along the direction of the ion string. To suppress energy-dependent phase contributions, we use short $\pi/2$ pulses without any free-evolution time in between.
Following the previously outlined procedure, we first reconstruct the qubit phase $\phi_i$ and the measurement contrast by fitting the correlation matrix. Next we use these phases for reconstructing the time-dependent random phases $\varphi_m$. Using N-qubit correlations, we finally use the Bayesian approach of eq.~(\ref{eq:BayesianEstimate-qubit-phases}) for an improved phase estimate of $\phi_i$, shown as open symbols in Figure~\ref{fig:SpatialCorrelations} (a) for a string of 62 ions.

%
%----
% data analysis: C:\Christian\tex\Papers\2021_CorrelationMeasurementsLongStrings\matlab\Figure_SpatialPositions
%----
%

In a second step, we extract the trapping potential from the measured correlations. We approximate the potential by Taylor-expanding it up to fourth order, $V(z)=\frac{1}{2}m\omega_0^2z^2(1+z/l_3 + (z/l_4)^2)$, where $\omega_0$ is the oscillation frequency of a single ion and $l_3$ ($l_4$) account for the cubic (quartic) anharmonicity of the potential. By calculating the ion positions in this potential, we fit the measured $\phi_i$ and find $\omega_0=(2\pi)~109.728(3)$~kHz, $l_3=2.1(7)$~mm, $l_4=0.8(6)$~mm, where the error bars are obtained from nonlinear regression assuming quantum projection noise as the only source of errors.
We compare the measured phases $\phi_i$ to the ones obtained from fitting the potential ($\phi_i^{fit}$) by calculating the residual position errors, $\delta x_i=(\phi-\phi_i^{fit})/k$. Fig.~\ref{fig:SpatialCorrelations} (b) shows that these residuals have a standard deviation of 6.0~nm, barely above the theoretically expected error $\sigma_{\Delta\phi}=5.1$~nm. Moreover, the absence of spatial correlations in the residuals demonstrates that Taylor-expanding the potential up to the fourth order is an adequate approximation to the exact potential.

To further test the method, we carried out the reconstruction of the potential for a fixed set of trap parameters but different number of ions ($10\le N\le 62$) and obtained consistent results. Fig.~\ref{fig:SpatialCorrelations} (c) shows the inferred oscillation frequency $\omega_0$ (light blue points) and the lowest collective mode frequency $\omega_c$  (dark blue points). For an independent cross-check, the latter was also measured via sideband spectroscopy on the S$_{1/2}$ to D$_{5/2}$ transition (red squares). We observe that the correlation measurement systematically underestimates the mode frequency by about 220~Hz (Fig.~\ref{fig:SpatialCorrelations} (d)). This discrepancy could be explained by the perpendicular laser beam being misaligned by about 1~mrad. Apart from this systematic error, the match between the two methods is quite good for $N>10$ ions: the inset shows the difference of the predicted mode frequencies, which have a standard deviation of only 14~Hz if the N=10 data point is excluded on the basis of the rather uneven distribution of the phases $\phi_i$ over the interval from 0 to $2\pi$. Systematic effects in the measured frequencies by imperfect laser beam misalignment could be further reduced by replacing the perpendicular beam by another axial beam that is counter-propagating to the axial beam in place, because small alignment errors of the beams with the direction of the ion string would affect the measurement outcomes only in second order.

%-----------------------------------------------------------
\subsection{Measurement of transition frequency differences}
Correlation spectroscopy with long probe times provides a tool for precisely measuring spatial transition frequency variations, which are relevant for frequency standards and quantum simulation experiments. For $^{40}$Ca$^+$ ions, the dominant frequency shifts are Zeeman and electric quadrupole shifts. We measure the spatial dependence of these shifts by probing 
the stretched S$_{1/2},m=\pm1/2\leftrightarrow$D$_{5/2},m=\pm 5/2$ transitions with a Ramsey time of $\tau=40$~ms duration with a 51-ion string. In contrast to the experiments of subsection \ref{subsec:ionpositions}, both Ramsey pulses are realized by the same laser beam. Writing the spatially resolved shifts $\Delta_i^\pm$ as $\Delta_i^Q=(\Delta_i^++\Delta_i^-)/2$ and $\Delta_i^B=(\Delta_i^+-\Delta_i^-)/2$ enables a separation of electric quadrupole and magnetic field shifts.

%----
\begin{figure}[t]
\centering
\includegraphics[width=0.5\textwidth]{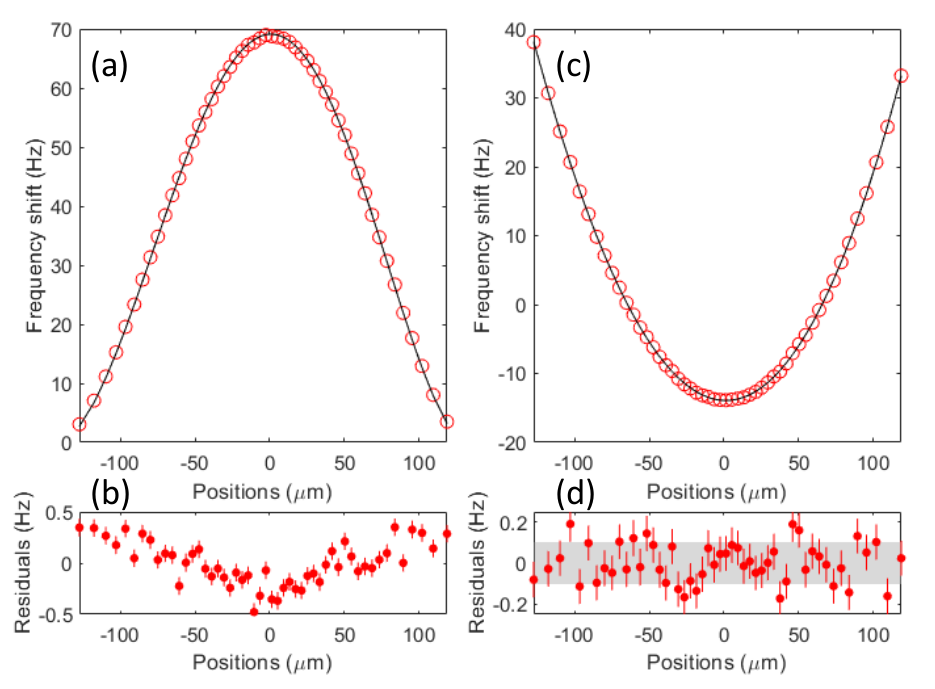}
\caption{Transition frequency shift measurement obtained by probing the quadrupole transitions between stretched states with 1500 experimental repetitions each. (a) Quadrupole shift of the D$_{5/2},m=\pm 5/2$ states (red circles: measured shift, black line: predicted shift). The frequency shift was measured with respect to the first ion; in the figure, a constant offset was added so that the averaged shift equaled the calculated average quadrupole shift. (b) Measurement residuals. (c) Differential Zeeman shift of the S$_{1/2},m=1/2\leftrightarrow$D$_{5/2},m=5/2$ transition frequency. An offset was added so that the average shift became equal to zero. The black line is a fit to the data by a third-order polynomial. (d) Residuals. The gray rectangle indicates the measurement uncertainty (1$\sigma$) predicted for quantum projection noise.}  
% data taken on 20210708
\label{fig:LevelShifts}
\end{figure}
%----
Figure~\ref{fig:LevelShifts}(a,b) displays the measured quadrupole shift together with a calculated shift obtained from a measurement of the ion positions and the known quadrupole moment $\theta(3d,5/2)$ of the $D_{5/2}$ level \cite{Roos:2006}. The systematic variation of the residuals on the scale of 0.5~Hz could be explained by a 1.5$\sigma$ error in the determination of $\theta(3d,5/2)$ or by a misalignment of the perpendicular laser beam by 3~mrad. Figure~\ref{fig:LevelShifts}(c,d) display the level shifts by the inhomogeneous magnetic field produced by the permanent magnets defining the quantization axis. We fit the Zeeman shifts with a third-order polynomial of the ion positions in order to extract the residuals. The latter have a standard deviation of 109 mHz, approaching the minimal uncertainty of 103 mHz predicted by the noiseless limit. 

%----
\begin{figure}[t]
\centering
\includegraphics[width=0.5\textwidth]{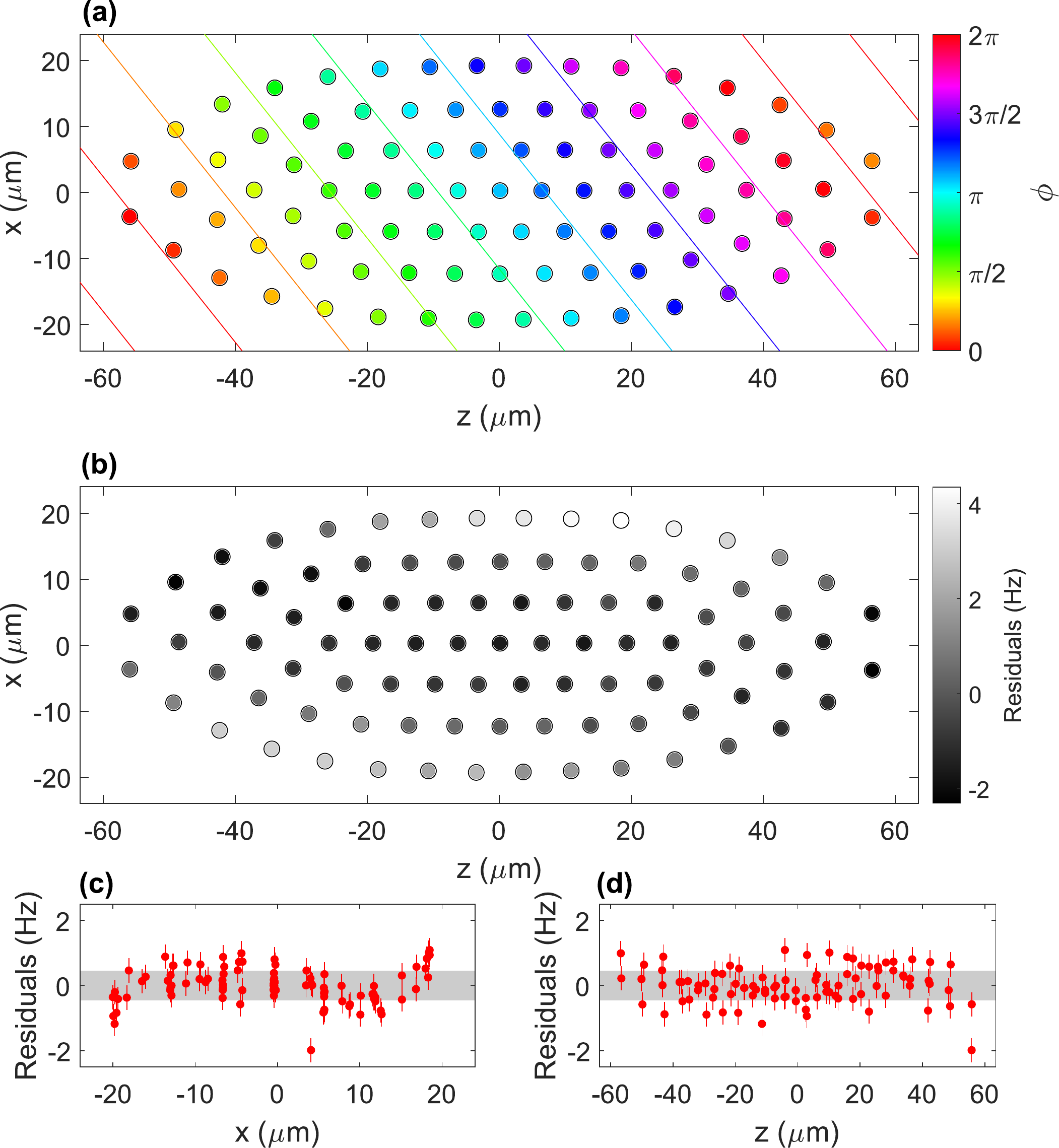}
\caption{Transition frequency shift measurement in a two-dimensional crystal obtained by probing the ground-state transition with 19736 experimental repetitions. (a) Single-ion phases together with contour lines of constant phases obtained from a linear fit. (b) Residuals from the linear fit. (c,d) Residuals from the quadratic fit along (c) $x$ and (d) $z$ axis. The gray rectangle indicates the 1$\sigma$ measurement uncertainty predicted for quantum projection noise.
}  
% Crystal data from 2021.10.13 02:55:15
% \\zidshare.uibk.ac.at\qos\spicy\Data\CorrelationSpectroscopy\211013\91_ions_correlation\91_ions_correlation_dataset3\Calibration_Data\02_55_15\Calibration_Data_Info\CrystalImage.mat
%
% Scan data combined from scans on 2021.10.13 03_00_17_521386 -- 03_00_26_705498
% \\zidshare.uibk.ac.at\qos\spicy\Data\CorrelationSpectroscopy\211013\91_ions_correlation\91_ions_correlation_dataset3\sim_data\SimData_10.mat
\label{fig:2DGradient}
\end{figure}
%----
Figure~\ref{fig:2DGradient}(a) shows the measured single-ion phases for a two-dimensional 91-ion crystal in the presence of a spatially-varying magnetic field. We probe the ground-state transition, $\mathrm{S}_{1/2}, m=-1/2 \leftrightarrow \mathrm{S}_{1/2}, m=+1/2$, by a  Ramsey experiment of $\tau=5~\mathrm{ms}$ duration. We fit a linear function to the measured phases and show the contour lines of constant phases from the fit in Fig.~\ref{fig:2DGradient}(a). We extract a magnetic-field gradient of $0.85(1)~\mathrm{G/m}$ from the linear fit with an angle of $38.6(4)$ degrees with respect to the horizontal direction. The maximal measured transition frequency difference between the ions is $218.2(8)~\mathrm{Hz}$. The spatial distribution of the residuals from the linear fit, shown in Fig.~\ref{fig:2DGradient}(b), reveal that the magnetic field contains higher-order terms in addition to the linear gradient. We further fit a quadratic function to the residuals, and show the remaining residuals in Fig.~\ref{fig:2DGradient}(c,d) along the two orthogonal directions. The majority of the spatial structure in the magnetic-field can be explained with linear and quadratic terms, as the remaining residuals show almost no systematic structure. Experimentally, it is straightforward to cancel linear variations of the magnetic field across the ion crystal with permanent magnets or coils placed outside the vacuum system.

\subsection{Single-shot Ramsey interferometry}
\begin{figure}
\centering
\includegraphics[width=0.45\textwidth]{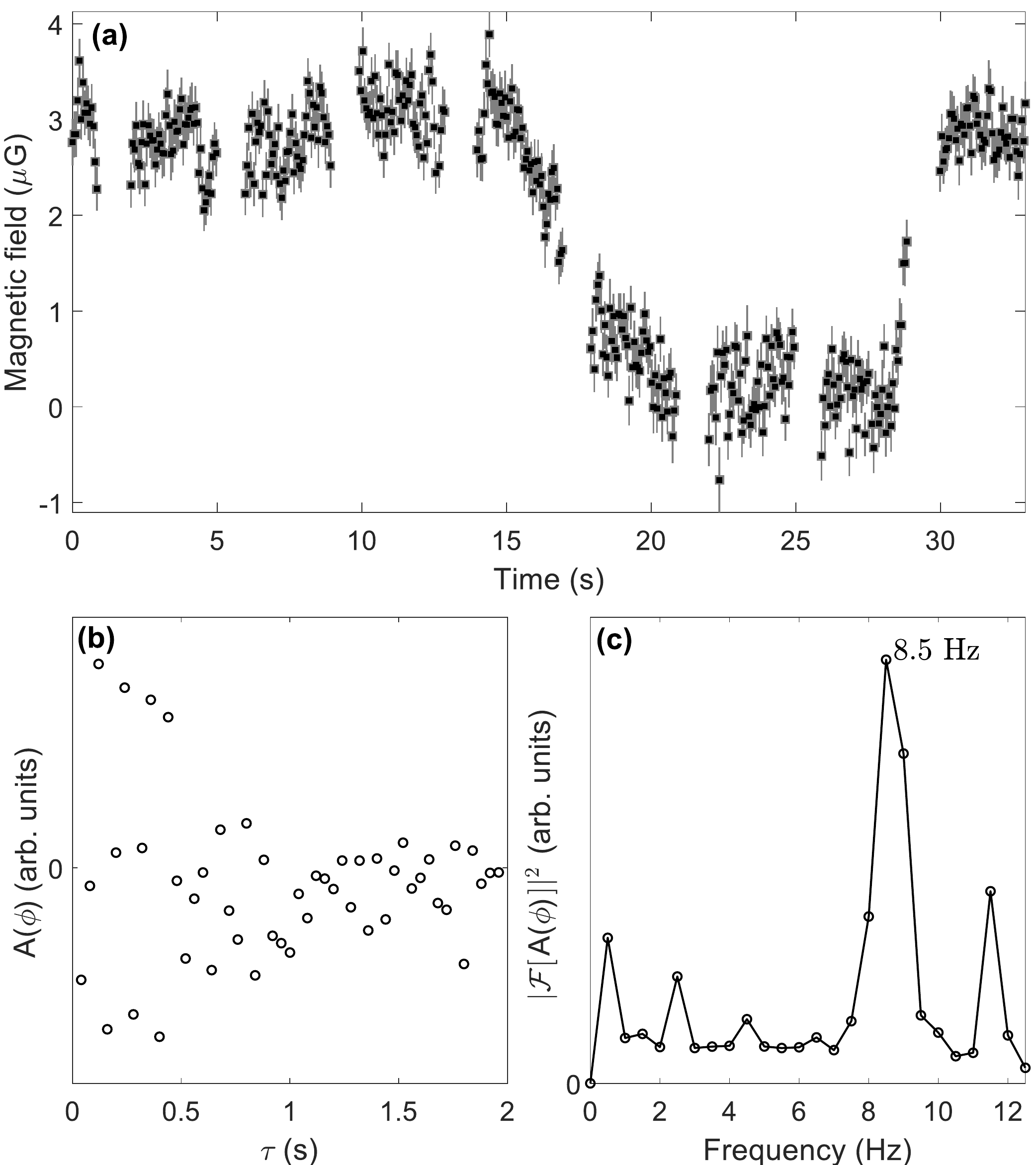}
\caption{Single-shot Ramsey interferometry tracking temporal phase fluctuations. (a) Phase fluctuations induced by a time-varying magnetic field probed with a 49-ion string. (b,c) Tracking laser frequency variations with a 51-ion string: Autocorrelation function $A(T)$ of laser phase fluctuations (left plot) together with its spectral density $|\mathcal{F}[A(T)]|^2$ (right plot).}
% data: magnetic field changes measured on 20200917-2313
% data: [Figure7_TimeDomain_LevelShifts_NewData.pdf] Scan data from 06.10.20,20:57:55 (51 ions)
\label{fig:time-domain-level-shifts}
\end{figure}

\begin{figure}
\centering
\includegraphics[width=0.5\textwidth]{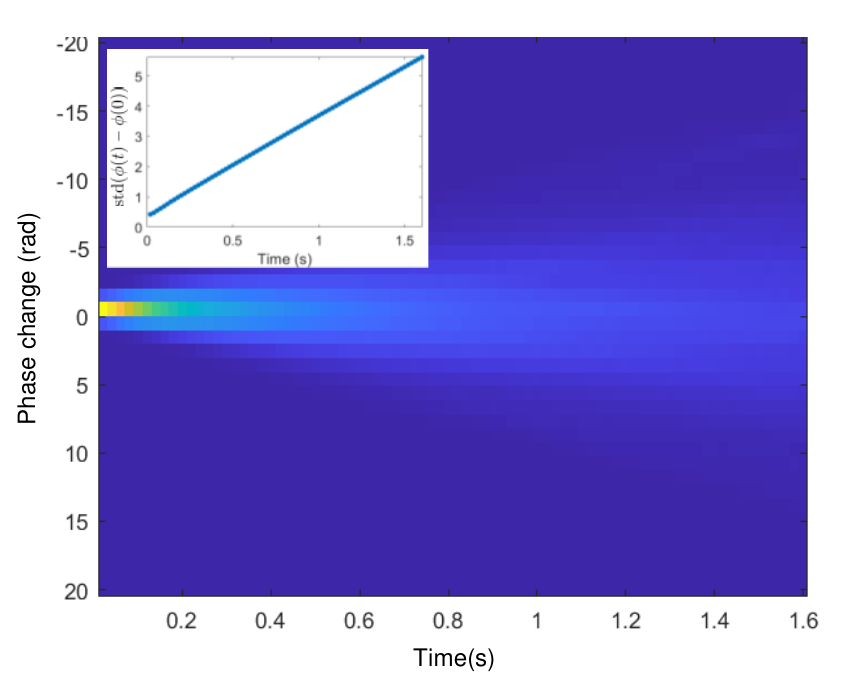}
\caption{Measurement of temporal relative optical path length fluctuations in two beam path delivering the laser pulses to a string of 40 ions.}
% data taken on 20211117.
\label{fig:time-domain-path-length}
\end{figure}

The data taken for probing the spatial dependence of phase shifts can also be analyzed in the time-domain: we probe temporal fluctuations of the local oscillator's phase at the locations of the ions by single-shot Ramsey interferometry.  Fig.~\ref{fig:time-domain-level-shifts} shows examples of such temporal phase changes that are caused by magnetic field fluctuations, laser frequency noise and optical path length fluctuations, respectively. Panel (a) shows a magnetic-field change of about $3\,\mu$G at the location of the ions induced by the arrival of an elevator at the lab floor. The magnetic field was sensed by a 49-ion string probed by a 40~ms Ramsey experiment on the Zeeman ground state qubit transition. For the data shown in (b), the S$_{1/2},m=1/2\leftrightarrow$D$_{5/2},m=3/2$ transition was probed for $\tau=20$~ms. Here, 371 data point were acquired, each containing 50 experiments that were recorded at a repetition rate of 25~Hz. Laser phase noise gave rise to phase fluctuations for which an autocorrelation was calculated. The spectral density of the autocorrelation function reveals distinct components at low frequencies contributing to the laser noise. The dominant component at $\sim8~\mathrm{Hz}$ introduces a frequency excursion on the order of $1~\mathrm{Hz}$. Figure ~\ref{fig:time-domain-path-length} shows differential path length fluctuations in the time domain, measured with short Ramsey experiments using two different laser beam paths for the two Ramsey pulses. For durations below 2\,s, the data shows phase fluctuations $\langle(\varphi(t+\tau)-\varphi(t))^2\rangle_t$ between experiments separated by a time $\tau$ that increase in proportion to $\tau$ as shown in the inset. The phase diffusion is predominantly caused by path length fluctuations in the two optical fibers delivering the light to the ion trap.

%----
%----

%----
%----

\section{Discussion and outlook \label{sec:discussion}}

We have investigated many-qubit correlation spectroscopy for probing qubits subjected to spatially correlated noise. The technique enables phase comparisons between any pair of qubits, provided that the qubit states are about evenly distributed over the equatorial plane of the Bloch sphere. The latter condition does not impose a strong restriction, as in most experimental setups it should be possible to deliberately imprint spatial phase gradients on the qubit array to satisfy this requirement.  In the limit of large qubit number and perfectly correlated noise, the quantum correlations induced by the noise enable a nearly complete restoration of the Ramsey contrast, which in the case of two-qubit correlation spectroscopy is upper-bounded to 50\%. The increased contrast gives rise to a fourfold reduction in measurement time needed to achieve the targeted measurement uncertainty. 

Many-qubit correlation spectroscopy is easy to implement as it requires only standard Ramsey spectroscopy enhanced by single-qubit read-out. The technique is therefore not limited to trapped-ion experiments but could be used in any multi-qubit physical system with high-fidelity single-shot read-out of individual qubits. In particular, it might be applicable to atomic clock experiments in tweezer arrays. Recently, experiments applying Ramsey correlation spectroscopy to subensembles of atoms held in optical lattices or tweezer arrays have demonstrated very small frequency gradients and impressive optical atomic coherence times reaching tens of seconds \cite{Young:2020, Bothwell:2022, Zheng:2022}. In one- or two-dimensional tweezer arrays, which 
feature single-atom detection of tens to hundreds of atoms \cite{Madjarov:2019,Young:2020, Singh:2022}, our method is directly applicable and could assist in reducing the measurement time required for characterizing spatially varying transition frequency shifts across the atomic array. With the further development of atomic clocks networks connected by phase-stable photonic links \cite{Komar:2014, Nichol:2022}, multi-qubit correlation spectroscopy could be applied for mutual frequency comparisons of the clocks, too. Another application of the technique might be found in quantum information processing experiments where spatially correlated noise can degrade the device performance. For example, in the atomic tweezer experiments reported in Ref.~\onlinecite{Singh:2022}, an auxialliary atomic species was employed for sensing and in-sequence correction of correlated phase noise. Here, an application of our protocol to the sensing species might increase the maximum noise level for which the correction can still be applied. 

In the context of trapped-ion experiments, many-qubit correlation spectroscopy proves to be a valuable tool for characterizing various aspects of the experimental setup with high precision. Our experiments demonstrate that experimentally observed uncertainties come close to the theoretically predicted ones. The resulting reduction in measurement time for achieving a desired uncertainty could be of interest for tracking the frequency of an unstable laser and providing feedback from individual measurements for improving its stability. Another application of multi-qubit correlation spectroscopy, which we did not explore in this paper, is to use it for thermometry and detection of structural phase transitions in ion crystals \cite{Dubin:1993, Kiethe:2021}. In this context, insufficiently cooled (low-frequency) motional modes could give rise to a reduction of fringe contrast that could be detected in correlation spectroscopy experiments.

While in our experiments both the read-out and the initial states are non-entangled, we have theoretically shown the possibility of improving the precision using an entangled initial state or non-local measurements. While being negligible for large $N$, this improvement can be considerable assuming a relatively small $N$. A possible experimental realization requires initialization to a symmetric Dicke state. In trapped-ion experiments, these  Dicke states could be engineered \cite{Linington:2008} by preparing the ions' center-of-mass mode in a Fock state with $N/2$ quanta, followed by a rapid adiabatic passage on its red-sideband transition \cite{Um:2016}, which converts motional quanta into collective electronic excitations \cite{Lechner:2016}. Finding simple protocols for generating optimal initial states and implementing these protocols in a sensing experiment is an interesting challenge for future work.

%------------------------------------------------------------------

\begin{acknowledgments}
We acknowledge useful discussions with Alex Retzker.
The project leading to this application has received funding from the European Research Council (ERC) under the European Union’s Horizon 2020 research and innovation programme (grant agreement No 741541), and from the European Union’s Horizon 2020 research and innovation programme under grant agreement No 817482. Furthermore, we acknowledge support by the Austrian Science Fund through the SFB BeyondC (F7110) and funding by the Institut f\"ur Quanteninformation GmbH. This project has received funding from the European Union’s Horizon 2020 research and innovation programme under the Marie Skłodowska‐Curie grant agreement No 801110 and the Austrian Federal Ministry of Education, Science and Research (BMBWF). TG acknowledges funding provided by the Institute for
Quantum Information and Matter and the Quantum Science and Technology Scholarship of the Israel Council for Higher Education.
\end{acknowledgments}

%----------------------------------------------------------------------------
\appendix

\section{Bounds to the achievable phase estimation uncertainty\label{sec:details-on-bounds}}
{\it{Pair correlations---}}
We analyze the Fisher information obtained using only pair correlations.
As mentioned in the main text the relevant random variables are the pair correlations
\begin{equation}
\left\{ \frac{1}{M}\underset{m=1}{\overset{M}{\sum}}q_{i,m}q_{j,m} \right\} _{i,j>i}\rightarrow {\cal N}\left(\left\{ \mu_{i,j}\right\} _{i,j>i},\Sigma\right),
\label{eq::pair_correlations_dist}
\end{equation}
which, according to the central limit theorem, converge to a Gaussian distribution, where $\mu_{i,j}$ is the average of $q_{i}q_{j}$ and $\Sigma$ is
the covariance matrix of the $\left\{ q_{i}q_{j}\right\} _{i,j>i}$.

Hence the problem boils down to calculating the FI matrix for this
Gaussian distribution. The FI matrix about $\overrightarrow{\phi}$
given this Gaussian distribution is presented in eq.~(\ref{eq:gaussian_FI}) in the main text. We write it here as 
\begin{equation}
I=D^{\dagger}\Sigma^{-1}D,
\label{eq:corr_FI_2}
\end{equation}
where: 
\[
D=\left(\begin{array}{cccc}
\partial_{\phi_{1}}\mu_{1,2} & \partial_{\phi_{2}}\mu_{1,2} & \cdots & \partial_{\phi_{N}}\mu_{1,2}\\
\vdots & \vdots & \vdots & \vdots\\
\partial_{\phi_{1}}\mu_{N-1,N} & \partial_{\phi_{2}}\mu_{N-1,N} & \cdots & \partial_{\phi_{N}}\mu_{N-1,N}
\end{array}\right).
\]

Hence we need to calculate $\left\{ \mu_{i,j}\right\} _{i,j>i}$ and
$\Sigma$ in order to get the FI matrix. Let us first assume only correlated dephasing (no uncorrelated dephasing). For the mean values, we have 
\begin{align}
\begin{split}
&\mu_{i,j} =2\left[\frac{1}{2\pi}\underset{0}{\overset{2\pi}{\int}}\cos^{2}\left(\frac{1}{2} \left( \phi_{i}+\varphi_{m} \right) \right)\cos^{2}\left(\frac{1}{2} \left(\phi_{j}+\varphi_{m} \right) \right)+\right.\\
& \left. \sin^{2}\left(\frac{1}{2} \left(\phi_{i}+\varphi_{m} \right)\right)\sin^{2}\left(\frac{1}{2}\left(\phi_{j}+\varphi_{m} \right)\right)\;d\varphi_{m} \vphantom{\frac{1}{2\pi}\underset{0}{\overset{2\pi}{\int}}\cos^{2}\left(\phi_{i}+\phi_{r}\right)\cos^{2}\left(\phi_{j}+\phi_{r}\right)} \right]-1\\
 & =\frac{1}{2}\cos\left( \phi_{i}-\phi_{j} \right).
\end{split}
 \label{FI_calc_1}
\end{align}

Let us now calculate $\Sigma$. The diagonal terms of $\Sigma$ read:
\begin{align}
\begin{split}
&\Sigma_{\left(i,j\right),\left(i,j\right)} =\langle q_{i}^{2}q_{j}^{2}\rangle-\langle q_{i}q_{j}\rangle^{2}=1-\frac{1}{4}\cos^{2}\left(\phi_{i}-\phi_{j} \right)\\
 & =\frac{7}{8}-\frac{\cos\left(2\left(\phi_{i}-\phi_{j}\right)\right)}{8}
 \end{split}
 \label{FI_calc_2}
\end{align}
 Regarding the non-diagonal terms, let us begin with non-overlapping pairs $\left(i,j\right),\left(k,n\right)$:
\begin{align*}
\begin{split}
&\langle q_{i}q_{j}q_{k}q_{n}\rangle=\frac{1}{8}\cos( \text{\ensuremath{\phi_{i}}}+\text{\ensuremath{\phi_{j}}}-\text{\ensuremath{\phi_{k}}}-\text{\ensuremath{\phi_{n}}} )\\
&+\frac{1}{4}\cos(\text{\ensuremath{\phi_{i}}}-\text{\ensuremath{\phi_{j}}})\cos( \text{\ensuremath{\phi_{k}}}-\phi_{n} ),
\end{split}
\end{align*}
\[
\langle q_{i}q_{j}\rangle\langle q_{k}q_{n}\rangle=\frac{1}{4}\cos\left(\phi_{i}-\phi_{j}\right)\cos\left(\phi_{k}-\phi_{n}\right)
\]

Hence: 
\begin{equation}
\Sigma_{\left(i,j\right),\left(k,n\right)}=\frac{1}{8}\cos\left( \text{\ensuremath{\phi_{i}}}+\text{\ensuremath{\phi_{j}}}-\text{\ensuremath{\phi_{k}}}-\text{\ensuremath{\phi_{n}}} \right)
\label{FI_calc_3}
\end{equation}
For overlapping pairs, such as $\left(i,j\right),\left(i,n\right)$
we have: 
\begin{align}
\begin{split}
&\Sigma_{\left(i,j\right),\left(i,n\right)} =\langle q_{i}^{2}q_{j}q_{n}\rangle-\langle q_{i}q_{j}\rangle\langle q_{i}q_{n}\rangle=\\
&\langle q_{j}q_{n}\rangle-\langle q_{j}q_{i}\rangle\langle q_{i}q_{n}\rangle=\\
&\frac{3}{8}\cos\left( \phi_{j}-\phi_{n}\right)-\frac{1}{8}\cos\left(\phi_{j}+\phi_{n}-2\phi_{i}\right)
\end{split}
\label{FI_calc_4}
\end{align}

The derivatives matrix, $D$, is: 
\begin{equation}
D_{\left(i,j\right),m}=\begin{cases}
-\frac{1}{2}\sin\left(\phi_{i}-\phi_{j}\right) & m=i\\
\frac{1}{2}\sin\left(\phi_{i}-\phi_{j}\right) & m=j\\
0 & m\neq i,j
\end{cases}
\label{FI_calc_5}
\end{equation}
Inserting equations (\ref{FI_calc_1}),(\ref{FI_calc_2}),(\ref{FI_calc_3}), (\ref{FI_calc_4}) into (\ref{eq:corr_FI_2}), we can perform exact numerical calculations of the FI.

Given an uncorrelated dephasing in addition to the correlated dephasing, the probabilities $p_\pm$ of observing outcomes $q_{i}=\pm 1$ are modified to $p_\pm=\frac{1}{2}(1\pm C_{0} \sin\left(\phi_{i}+\varphi_{m}\right)),$
i.e. a finite contrast of $0\leq C_{0}\leq 1$. 
It can be immediately observed that $\mu_{i,j}	=
	\frac{1}{2}C_{0}^{2}\cos\left(\phi_{i}-\phi_{j}\right).$ The covariance matrix is modified as follows:
\begin{align*}
\begin{split}
&\Sigma_{\left(i,j\right),\left(i,j\right)}=\left(1-\frac{C_{0}^{4}}{8}\right)-C_{0}^{4}\frac{\cos\left(2\left(\phi_{i}-\phi_{j}\right)\right)}{8},\\	    
&\Sigma_{\left(i,j\right),\left(k,n\right)}=\frac{1}{8}C_{0}^{4}\cos(\text{\ensuremath{\phi_{i}}}+\text{\ensuremath{\phi_{j}}}-\text{\ensuremath{\phi_{k}}}-\text{\ensuremath{\phi_{n}}}),\\    
&\Sigma_{\left(i,j\right),\left(i,n\right)}=\left(\frac{1}{2}C_{0}^{2}-\frac{1}{8}C_{0}^{4}\right)\cos\left(\phi_{j}-\phi_{n}\right)-\\
&\frac{1}{8}C_{0}^{4}\cos\left(\phi_{j}+\phi_{n}-2\phi_{i}\right).
\end{split}
\end{align*}

\begin{figure}[t]
\centering
\includegraphics[width=0.45\textwidth]{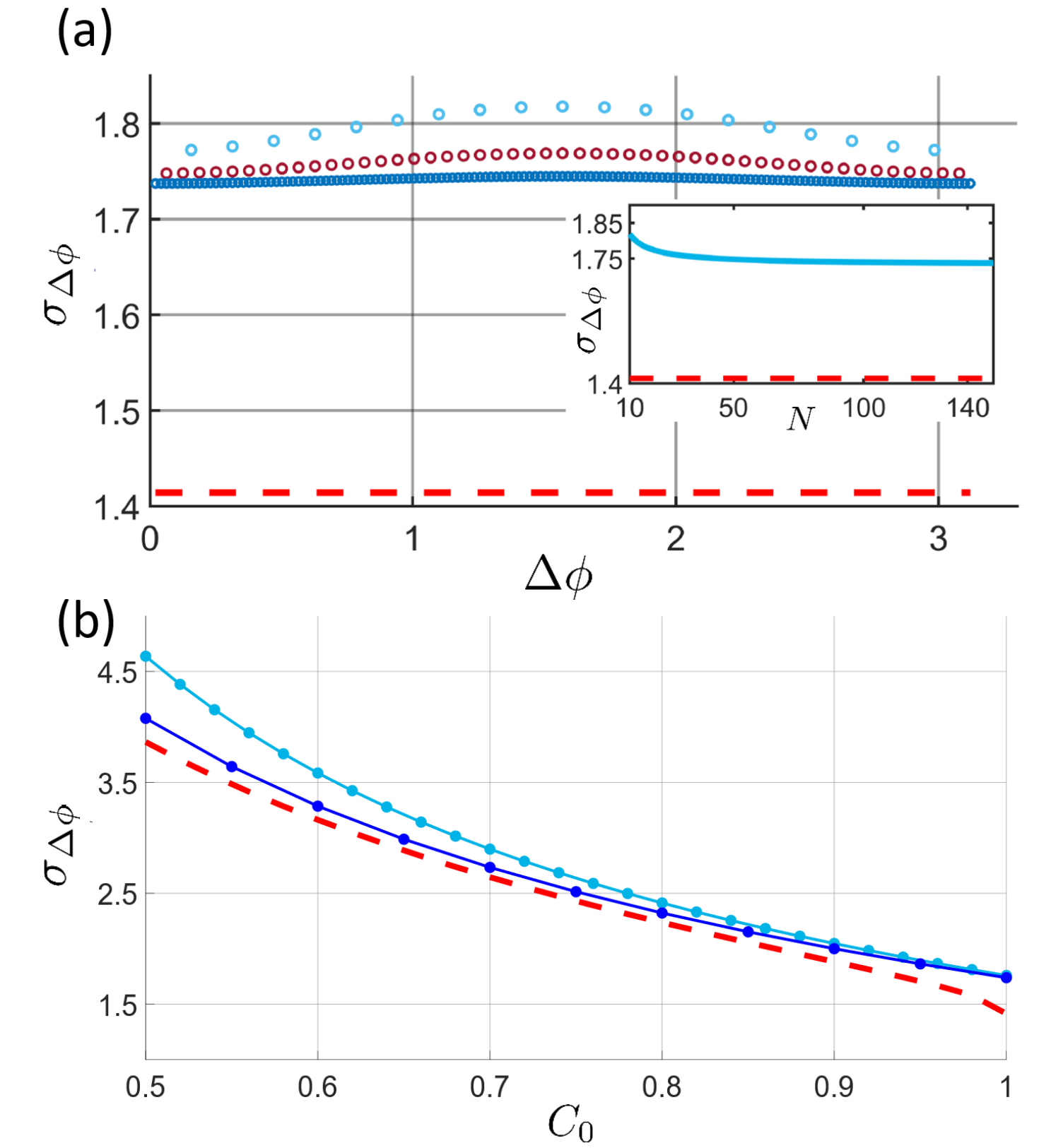}
\caption{Sensitivity (per measurement) with pair correlations compared to precision bounds (numerical analysis). (a) Uncertainty in estimating $\Delta \phi$ using all pair correlations as a function of $\Delta \phi$ for different number of ions. The dashed red line is the noiseless precision bound, pair correlations of $N=20,50,150$ correspond to light blue (top), red (middle), dark blue (bottom) points respectively. Inset: standard deviation averaged over all phases as a function of $N$, the top (bottom) lines correspond to pair correlations (noiseless precision bound).  (b) Uncertainty as a function of the contrast $C_{0}.$ The red dashed line corresponds to the finite contrast precision bound. Light blue (top) and blue (bottom) points correspond to pair correlations for $N=30, 120$ respectively.}
\label{fig:2corrs}
\end{figure}

Let us analyze the uncertainty using pair correlations as $N \rightarrow \infty.$ The behavior in the limit of large $N$ is presented in Fig. \ref{fig:2corrs}. It can be observed that for $C_0=1$ this uncertainty does not converge to the noiseless precision bound of $\sqrt{\frac{2}{M}},$ but approximately to $\sqrt{\frac{3}{M}}.$

The limit of $\sqrt{\frac{3}{M}}$ can be derived analytically, assuming that the phases are distributed evenly in $k \pi \; \left( k \in \mathbb{Z}\right).$ To obtain this result, we use the following approximation of the variance: 
\begin{align*}
\begin{split}
&\text{var}\left(\phi_{1}-\phi_{2}\right)=u_{1,2}^{\dagger}\left(D^{\dagger}\Sigma^{-1}D\right)^{-1}u_{1,2}\\
&\approx\frac{||u_{1,2}||^{4}}{||Du_{1,2}||^{4}}\left(Du_{1,2}\right)^{\dagger}\Sigma\left(Du_{1,2}\right),  
\end{split}
\end{align*}
where $u_{1,2}$ is the vector that corresponds to $\phi_{2}-\phi_{1},$ i.e. $\left(-1,1,0,...,0\right).$
This approximation is obtained by using a Cauchy-Schwarz inequality twice: $v^{\dagger}M^{-1}v\geq\frac{||v||^{4}}{v^{\dagger}Mv},$
and it can be understood as a single parameter estimation bound where the derivative of the mean is $\frac{1}{||u_{1,2}||^{2}}Du_{1,2}$ and the variance is $\frac{1}{||Du_{1,2}||^{4}}\left(Du_{1,2}\right)^{\dagger}\Sigma\left(Du_{1,2}\right).$ 
We now calculate this approximation to show that in the limit of large $N$ it converges to $3.$

Clearly $||u_{12}||^{4}=4$,
and $||Du_{1,2}||^{4}\approx\left(2\underset{k=1}{\overset{N}{\sum}}\frac{1}{4}\sin\left(2k\phi\right)^{2}\right)^{2}\approx\frac{N^{2}}{16}.$ We now need to calculate $\left(Du_{1,2}\right)^{\dagger}\Sigma\left(Du_{1,2}\right)$: note that since the denominator goes as $N^{2},$ we can omit in the calculation of this term any contributions that are smaller than $N^{2}.$ Since $Du_{1,2}$ is a real vector, this term is given by the sum (summation convention is used) $\left(Du_{1,2}\right)_{\left(i,j\right)}\Sigma_{\left(i,j\right)\left(k,n\right)}\left(Du_{1,2}\right)_{\left(k,n\right)}.$ We can neglect the $N$ terms of identical pairs. From overlapping pairs $\left(i,j \right),\left(i,n \right)$ the contribution is:
\begin{align*}
\begin{split}
&2\langle q_{i}^{2}q_{j}q_{n}\rangle\left(Du_{1,2}\right)_{\left(i,j\right)}\left(Du_{1,2}\right)_{\left(i,n\right)}\approx \\
&\frac{2}{8}\cos\left(\phi_{j}-\phi_{n}\right)\sin\left(\phi_{1}-\phi_{j}\right)\sin\left(\phi_{1}-\phi_{n}\right)+\\
&\frac{2}{8}\cos\left(\phi_{j}-\phi_{n}\right)\sin\left(\phi_{2}-\phi_{j}\right)\sin\left(\phi_{2}-\phi_{n}\right)\approx\\
&N^{2}/16.
\end{split}
\end{align*}
The contribution from the non-overlapping pairs is: 
\begin{align*}
\begin{split}
&2\langle q_{i}q_{j}q_{k}q_{n}\rangle\left(Du_{1,2}\right)_{\left(i,j\right)}\left(Du_{1,2}\right)_{\left(k,n\right)}\approx\\
&-\frac{2}{8}\cos(\text{\ensuremath{\phi_{1}}}+\text{\ensuremath{\phi_{j}}}-\text{\ensuremath{\phi_{2}}}-\text{\ensuremath{\phi_{n}}})\frac{1}{2}\sin\left(\phi_{1}-\phi_{j}\right)\frac{1}{2}\sin\left(\phi_{2}-\phi_{n}\right)\\
&\approx-N^{2}/48
\end{split}
\end{align*}
Hence $\left(Du_{1,2}\right)^{\dagger}\Sigma\left(Du_{1,2}\right) \approx \frac{N^{2}}{16}\left(1-\frac{1}{4}\right)=\frac{N^{2}}{16}\frac{3}{4}.$
Therefore altogether:
\begin{equation}
\frac{||u_{1,2}||^{4}}{||Du_{1,2}||^{4}}\left(Du_{1,2}\right){}^{\dagger}\Sigma\left(Du_{1,2}\right)\approx
3,    
\end{equation}
which matches the numerical results.

For general contrast $C_{0}$ these expressions are modified to:
\begin{align*}
\begin{split}
&||Du_{1,2}||^{4}\approx\frac{N^{2}}{16}C_{0}^{8},\\
&2\langle q_{i}^{2}q_{j}q_{n}\rangle\left(Du_{1,2}\right)_{\left(i,j\right)}\left(Du_{1,2}\right)_{\left(i,n\right)}\approx C_{0}^{8}N^{2}/16\\
& 2\langle q_{i}q_{j}q_{k}q_{n}\rangle\left(Du_{1,2}\right)_{\left(i,j\right)}\left(Du_{1,2}\right)_{\left(k,n\right)}\approx -C_{0}^{6} N^{2}/48.
\end{split}    
\end{align*}
 Hence we obtain that for a general contrast $\text{var}\left(\phi_{1}-\phi_{2}\right)$ from pair correlations converges to $\approx \frac{4-C_{0}^{2}}{C_{0}^{2}}.$
 This implies that as $C_{0}$ becomes smaller the uncertainty using pair correlations converges to the finite contrast bound of the variance $\frac{2}{1-\sqrt{1-C_{0}^{2}}}$. The reason for this convergence is that the Fisher information obtained from higher moments goes with higher powers of $C_{0},$ in general the Fisher information obtained from the $2k-$th moments goes as $C_{0}^{2k}$ and thus the contribution from the higher moment gets smaller for smaller $C_{0}.$
 In fact the FI with pair correlations coincides with the finite contrast bound up to a second order in $C_{0}^{2}$:
 $\frac{2}{1-\sqrt{1-C_{0}^{2}}}=\frac{4-C_{0}^{2}+\mathcal{O}\left(C_{0}^{4}\right)}{C_{0}^{2}}.$
 This raises a natural question: is the variance with all $m\leq k$ particle correlations equal to $\frac{2}{C_{0}^{2}}+\frac{2}{C_{0}^{2}}\left(1-\underset{l=1}{\overset{k}{\sum}}\frac{2}{l} \binom{2l-2}{l-1}
 \left(\frac{C_{0}^{2}}{4}\right)^{l}\right)$? We leave it as an open question (and conjecture) as we do not have a proof to this.

\begin{figure}
\begin{center}
\includegraphics[width=8.5 cm]{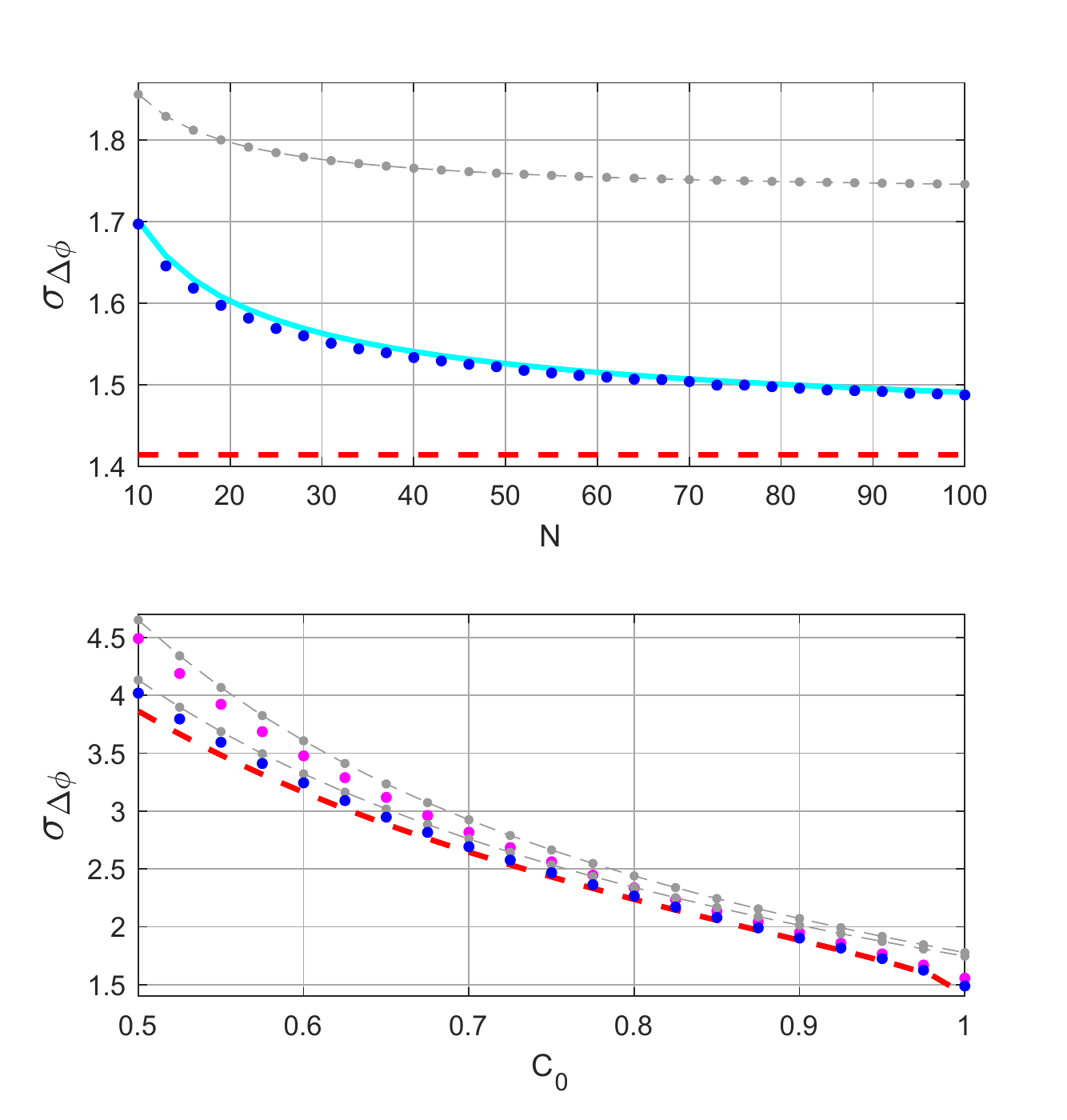}
\end{center}
\caption{Uncertainty with N-qubit correlations, pair correlations and precision limits (all values are per measurement).
(a) precision bounds as a function of $N$: grey (upper) curve and blue dots correspond to pair correlations and $N$-qubit correlations respectively.
The light blue curve correspond to the analytical approximation and the red dashed line is the fundamental noiseless limit of $\sqrt{2}.$
(b) Precision bounds as a function of $C_{0}$: Pink dots and grey (upper) curve correspond to all $N$ correlations and pair correlations respectively for $N=30.$ Same for blue dots and grey lower curve for $N=100.$ The red dashed line correspond to the finite contrast precision bound.} 

\label{fig:N_qubit_corrs}
\end{figure}

{\it{N-qubit correlations---}}
The information about the phase differences 
when taking all correlations into account is the information contained in the full distribution averaged over the random phase:
\begin{equation}
P(\mathbf{q})=\frac{2^{-N}}{2\pi}\int_{0}^{2\pi}d\varphi\prod_{i=1}^{N}(1+q_{i}\sin(\phi_{i}+\varphi)).
\end{equation}
Hence the precision bound is given by the FI matrix about $\mathbf{\phi}$ with this distribution. Since the FI matrix
involves summation over all $2^{N}$ possible $\mathbf{q}$ vectors, $I_{i,j}=\underset{\mathbf{q}}{\sum}\frac{\left(\partial_{\phi_{i}}p\left(\mathbf{q}\right)\right)\left(\partial_{\phi_{j}}p\left(\mathbf{q}\right)\right)}{p\left(\mathbf{q}\right)}$, an exact calculation becomes intractable for large $N.$
Hence to make an efficient calculation of the FI we use the fact that $I_{i,j}=\langle\frac{\left(\partial_{\phi_{i}}p\right)\left(\partial_{\phi_{j}}p\right)}{p^{2}}\rangle=\langle\partial_{\phi_{i}}\ln\left(p\right)\partial_{\phi_{j}}\ln\left(p\right)\rangle.$ This allows us to make a Monte-Carlo calculation of the FI matrix by sampling $\partial_{\phi_{i}}\ln\left(p\right)\partial_{\phi_{j}}\ln\left(p\right).$ Simulation results are shown in Fig.~\ref{fig:N_qubit_corrs} for the case of evenly distributed single-qubit phases, $\phi_j=2\pi j/N$.

\section{Numerical simulations\label{sec:numerical-simulations}}
{\it{Estimating the phases with pair correlations: Maximum likelihood and least-squares estimation---}}
We calculated precision bounds of the phases given the pair correlations; in this part we discuss estimation methods using pair correlations and the saturability of these precision limits. 
We compare between two estimation methods: simple least-squares estimation, i.e. minimizing $\mathbf{V^{\dagger}V}$ where $\mathbf{V}=\left\{ \frac{1}{M}\underset{m=1}{\overset{M}{\sum}}q_{i,m}q_{j,m}-\mu_{i,j}\right\} _{i,j},$ and maximum-likelihood estimation.
Note that since the relevant distribution, eq.~(\ref{eq::pair_correlations_dist}), is Gaussian, the maximum-likelihood estimation becomes a weighted least-squares estimation \cite{kay1993fundamentals}:
\begin{equation*}
\underset{\mathbf{\phi}}{\text{max}}\mathcal{L}\left(\mathbf{V}|\mathbf{\phi}\right)=\underset{\mathbf{\phi}}{\text{min}}\mathbf{V}^{\dagger}\Sigma^{-1}\mathbf{V}.    
\end{equation*}
The difference between the two estimation methods is thus rooted in the weights given by the inverse of the covariance matrix, $\Sigma^{-1}.$ The maximum-likelihood is in general asymptotically efficient, i.e. saturates the Fisher information, whereas the simple least square is more straightforward as it does not require evaluation of the covariance matrix.

\begin{figure}
\begin{center}
\includegraphics[width=6.5 cm]{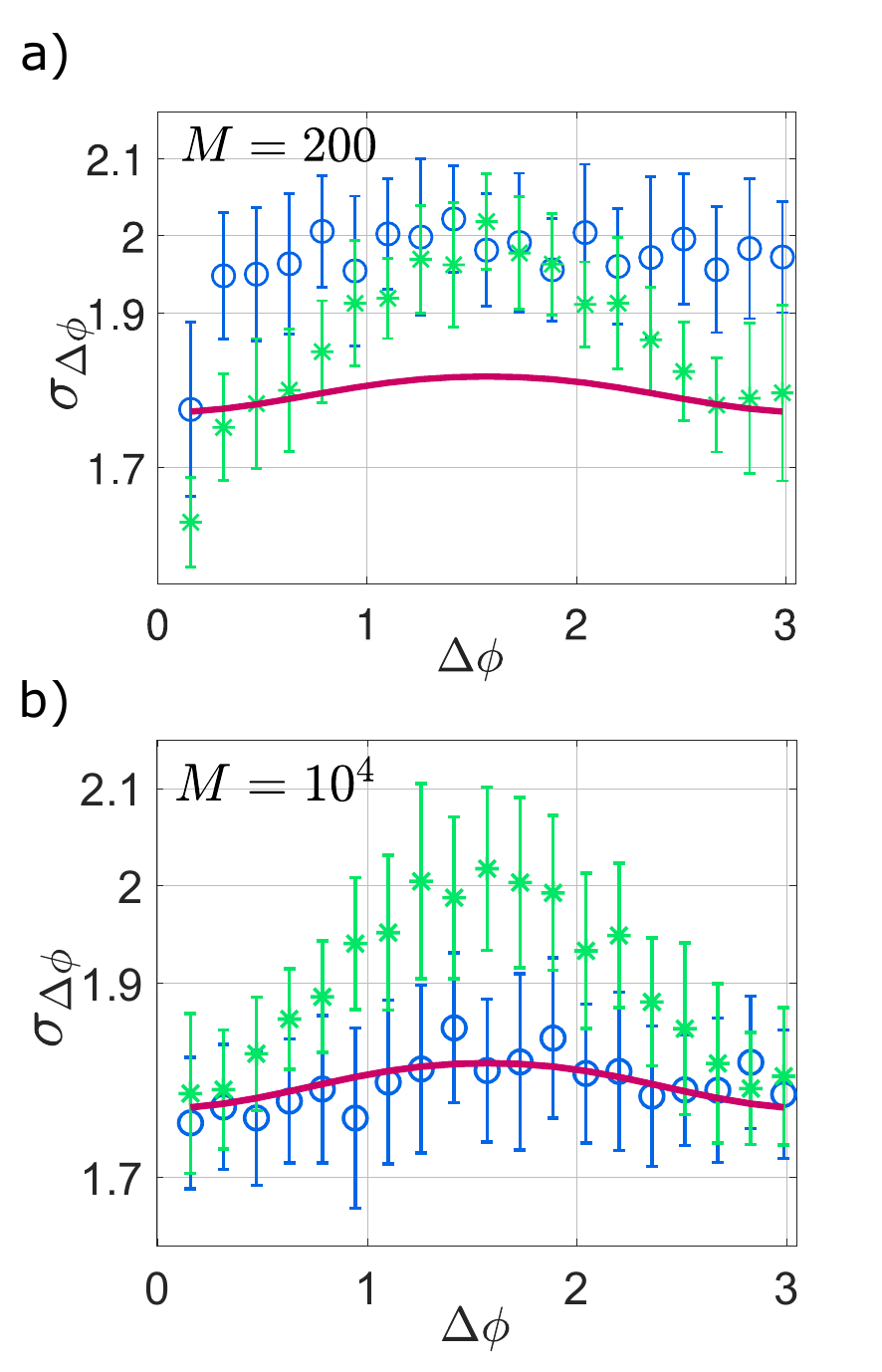}
\end{center}
\caption{Phase estimation errors using simple least squares and maximum likelihood estimation for $200$ and $10^{4}$ samples and $20$ qubits (data from numerical simulation).
Green lines and blue lines correspond to the distribution of the estimation errors with simple least squares and maximum likelihood respectively. The green circles and blue diamonds correspond to the average estimation error. The red solid line correspond to the Cramer-Rao limit.
(a) For $200$ samples the estimation errors are above the limit and simple least squares performs better than maximum likelihood. This behavior is due to the small number of samples.
(b) For $10^{4}$ samples the behavior matches the expectations: maximum likelihood coincides with the Fisher information and outperforms simple least squares.}
\label{fig::ml_vs_least_squares}
\end{figure}

A comparison of both approaches is presented in Fig. \ref{fig::ml_vs_least_squares}. It can be observed that for a large number of samples (here $M= 10^{4}$) the maximum likelihood indeed saturates the FI, while the simple least-squares method does not saturate it.
Interestingly, for a smaller number of samples (here $M=200$) maximum likelihood does not saturate the FI and a simple least-squares approximation outperforms it. In fact, for some phases simple least squares even outperform the FI (due to its bias for small number of samples).

{\it{Estimating the phases from N-qubit correlations---}}
A maximum-likelihood estimation of the phases by analysis of N-qubit correlations via eq.~(\ref{eq:MLEstimation}) satisfies the FI-based bound in the limit of infinite sample size. We carried out numerical simulations of the phase estimation process to investigate the influence of a finite number of samples on the phase uncertainties. The simulations showed the uncertainty increasing over the FI-bound with decreasing  sample size $M$; however, the effect was not very pronounced: in simulations with 20 and 100 qubits, we observed an increase by about 10\% for $M=50$ and by about 1\% for $M=500$.

\section{Improving the precision with entangled states and non-local measurements\label{sec:dicussion-details}}
Let us inquire about the optimality of our scheme by asking the following question: what is the optimal precision optimized over all possible initial states and measurement strategies? The figure of merit is $\frac{1}{\binom{N}{2}}\underset{j>k}{\sum}\text{Var}\left(\phi_{j}-\phi_{k}\right).$ We first show that the noiseless precision bound can be obtained by modifying the measurement basis to a non-local one.   
In a second step, we consider also entangled input states and find an optimal initial state for this sensing task.

Let us introduce the following notation for this part: The state $|\vec{z}\rangle$ is the product state $|z_{1}\rangle |z_{2} \rangle...|z_{N}\rangle$ where $|z_{i}\rangle$ is an eigenstate of $Z_{i}$ with eigenvalue $z_{i} \in \left\{ \pm1\right\} .$
A different notation to the same state would be $|\vec{q}\rangle$ where $q_{i}=\frac{1-z_{i}}{2},$ thus $z_{i}=1 \rightarrow q_{i}=0,$ $z_{i}=-1 \rightarrow q_{i}=1$.

\subsection{Improving precision with non-local measurements\label{sec:dicussion-details-nonlocal-measurements}}
Let us first write the quantum state after the time evolution. The initial state is a pure product state: $|+\rangle^{N}$, with $|+\rangle=\frac{1}{\sqrt{2}}\left(|0\rangle+|1\rangle\right)$ (an eigenstate of $X$ with eigenvalue $+1$). After a free evolution the state evolves into the state of Eq.~(\ref{eq:statef}) with phases $\phi_{im}=\phi_{i}+\varphi_{m}.$ $\varphi_{m}$ is a random phase that is distributed uniformly in $\left[0,2\pi\right].$ This random phase induces a correlated dephasing, i.e.  the final state, after averaging out $\varphi_{m}$, becomes a mixture of Dicke states:
\begin{equation*}
\rho_{f}=\underset{j=0}{\overset{N}{\oplus}}\rho_{j},
\end{equation*}
where $\rho_{j}$ is a Dicke state with $j$ excitation:  
\begin{equation*}
\rho_{j}=\frac{\binom{N}{j}
}{2^{N}}|\psi_{j}\rangle\langle\psi_{j}|,\;|\psi_{j}\rangle=\frac{1}{\sqrt{\binom{N}{j}
}}\underset{\vec{q},\,\underset{k}{\sum}q_{k}=j}{\sum}e^{i\vec{\phi}\cdot\vec{q}}|\vec{q}\rangle.
\end{equation*}

Given $\rho_{f} \left( \vec{\phi} \right),$ the
fundamental precision limit is set by the quantum Fisher
information matrix (QFIM) about the parameters $\phi_{1},\phi_{2},...,\phi_{N}$; this is the Fisher information matrix optimized over all possible measurement strategies.
Hence the covariance matrix
of the estimators, $\Sigma$, satisfies: 
\[
\Sigma\geq I^{-1},
\]
where $I$ is the QFIM
and thus for any $j,k$: 
\[
\text{var}\left(\phi_{j}-\phi_{k}\right)\geq u_{(j,k)}^{\dagger}I^{-1}u_{\left(j,k\right)},
\]
where $u_{\left(j,k\right)}$ is the parameter vector that corresponds
to $\phi_{j}-\phi_{k}$.

For a general mixed state $\rho,$ given its spectral decomposition $\rho=\underset{k}{\sum}p_{k}|k\rangle\langle k|,$ the QFIM is given by \cite{braunstein1994statistical}:
\begin{equation*}
I_{i,j}=2\underset{k,l}{\sum}\frac{\left(\frac{\partial\rho}{\partial\phi_{i}}\right)_{k,l}\left(\frac{\partial\rho}{\partial\phi_{j}}\right)_{l,k}}{\left(p_{l}+p_{k}\right)},
\end{equation*}
where $\left\{ p_{k} \right\}_k$ are the eigenvalues of $\rho$ and the matrix elements,$\left(\bullet\right)_{k,l}=\langle k|\bullet|l\rangle$, are with respect to the eigenbasis of $\rho.$

For pure states this expression is reduced to:
\begin{equation}
I_{i,j}=4\left(\langle\partial_{\phi_{i}}\psi|\partial_{\phi_{j}}\psi\rangle-\langle\partial_{\phi_{i}}\psi|\psi\rangle\langle\psi|\partial_{\phi_{j}}\psi\rangle\right).
\label{eq::QFI_pure}
\end{equation}

Let us calculate the QFIM of our $\rho_{f},$ which we denote as $I.$
It can be observed that in this special case $I$ 
is a weighted sum of the QFIM of each $|\psi_{j}\rangle$ \cite{demkowicz2009quantum}:
\begin{equation}
I=\underset{j=0}{\overset{N}{\sum}}\frac{\binom{N}{j}}{2^{N}}I^{\left(j\right)},
\label{eq:totalFI}
\end{equation}
where $I^{\left(j\right)}$ is the QFIM of $|\psi_{j}\rangle.$
For every $|\psi_{j}\rangle$ we have $\partial_{\phi_{i}}|\psi_{j}\rangle=-i\frac{1}{2} \left( Z_{i}+\mathds{I} \right)|\psi_{j}\rangle,$ inserting this into 
equation~(\ref{eq::QFI_pure}) we get that
the QFIM of each $|\psi_{j} \rangle$ is: 
\begin{equation*}
I_{k,l}^{\left(j\right)}=\left(\langle\psi_{j}|Z_{k}Z_{l}|\psi_{j}\rangle-\langle\psi_{j}|Z_{k}|\psi_{j}\rangle\langle\psi_{j}|Z_{l}|\psi_{j}\rangle\right).
\end{equation*}

Now $|\psi_{j}\rangle$ is a symmetric superposition of all states with
$j$ excitations, from symmetry we get: 
\[
\langle Z_{k}\rangle=\frac{1}{N}\left(N-j-j\right)=\frac{N-2j}{N}
\]
 and for $k \neq l$: 
\[
\langle Z_{k}Z_{l}\rangle=\frac{\left(N-2j\right)^{2}-N}{N\left(N-1\right)}
\]
 Hence all the non-diagonal terms of $I^{\left( j \right)}$ are: 
\begin{align*}
I_{k,l}^{\left(j\right)} & =\frac{\left(N-2j\right)^{2}-N}{N\left(N-1\right)}-\frac{\left(N-2j\right)^{2}}{N^{2}}\\
 & =\frac{4j\left(j-N\right)}{N^{2}\left(N-1\right)}.
\end{align*}
 The diagonal terms of $I^{\left( j \right)}$ read: 
\[
I_{k,k}^{\left(j\right)}=1-\left(\frac{N-2j}{N}\right)^{2}=\frac{4j\left(N-j\right)}{N^{2}}
\]
Inserting these terms into equation~(\ref{eq:totalFI}), we get that $I$ reads:
\begin{equation*}
I=\begin{cases}
\underset{j=0}{\overset{N}{\sum}}\frac{\binom{N}{j}}{2^{N}}\frac{4j\left(j-N\right)}{N^{2}\left(N-1\right)}=-\frac{1}{N} & k\neq l\\
\underset{j=0}{\overset{N}{\sum}}\frac{\binom{N}{j}}{2^{N}}\frac{4j\left(N-j\right)}{N^{2}}=\frac{N-1}{N}. & k=l
\end{cases}    
\end{equation*}
It is now simple to see that for any $k \neq m$ the vector that corresponds to
$\phi_{k}-\phi_{m}$ is an eigenvector of $I$ with an eigenvalue of $1.$
The variance per measurement is thus 
\[
\text{var}\left(\phi_{k}-\phi_{m}\right)=2
\]
and this is exactly the noiseless precision bound. 
Since the strong commutativity condition is satisfied (all Hamiltonian terms commute with each other), we know that there exists a basis that saturates this QFI \cite{pezze2017optimal}. This implies that there exists a measurement strategy such that the noiseless precision bound is obtained.

As a simple example we examine the case of two qubits.
The density matrix for two qubits reads 
\[
\rho_{f}=\frac{1}{4}\left(|11\rangle\langle11|+|00\rangle\langle00|\right)+ \frac{1}{2}|\psi_{1}\rangle\langle\psi_{1}|,
\]
with $|\psi_{1}\rangle=\frac{1}{\sqrt{2}}\left(|01\rangle+e^{i\left(\phi_{1}-\phi_{2}\right)}|10\rangle\right)$.
Measuring the local $X$ basis, we project onto the states $|\vec{x} \rangle=|x_1\rangle|x_2\rangle...|x_N\rangle$ where $|x_i\rangle$ is an eigenstate of $X_{i}$ with eigenvalues $x_i \in \left\{ \pm 1 \right \}$ and obtain the probabilities:
\begin{align*}
\begin{split}
& \text{even }\vec{x}:\frac{1}{8}+\frac{1}{4}\cos\left(\frac{\phi_{1}-\phi_{2}}{2}\right)^{2},\\
& \text{odd }\vec{x}:\frac{1}{8}+\frac{1}{4}\sin\left(\frac{\phi_{1}-\phi_{2}}{2}\right)^{2},
\end{split}
\end{align*}
where $\vec{x}$ odd (even) stands for $\# \left( x_i =-1 \right)$ odd (even).
This leads to: 
\[
\text{var}\left(\phi_{2}-\phi_{1}\right)=\frac{4-\cos\left(\phi_{1}-\phi_{2} \right)^{2}}{\sin\left(\phi_{1}-\phi_{2}\right)^{2}}\geq4,
\]
clearly this is exactly the variance with a single pair correlation and thus does not saturate the QFI. It can be observed that optimizing over all local measurement bases is equivalent to optimizing over $\phi_{1},\phi_{2}$ and thus no local measurement saturates the QFI.
There exists however a non-local measurement strategy that saturates
the QFI: consider first measuring $Z_{1}+Z_{2}$ and then measuring
the local X basis. With probability $1/2$ we get $|00\rangle,|11\rangle$ in the first
measurement and thus no information and with probability $1/2$
we collapse into $|\psi_{1}\rangle$ which yields a Fisher information
of $1$. Therefore the total Fisher information is: 
\[
\frac{1}{2}\cdot0+\frac{1}{2}\cdot1=\frac{1}{2}\rightarrow\text{var}\left(\phi_{2}-\phi_{1}\right)=2,
\]
hence the bound is saturated.

A general optimal measurement strategy would be:\\
 1. first measure $\underset{i}{\sum}Z_{i}$ (this measurement collapses the density matrix into one of the Dicke states, $|\psi_{j}\rangle$).\\  
2. Measure the Dicke state, $|\psi_{j}\rangle$, in its optimal measurement basis. 

The optimal measurement basis of $|\psi_{j}\rangle$ can be written implicitly as proven in \cite{pezze2017optimal}: projecting into a (Gram-Schmidt)
orthogonalization of $\left\{ |\psi_{j}\rangle,|\partial_{\phi_{k}}\psi_{j}\rangle\right\} _{\phi_{k}}$
would be optimal. For example, for $N=3$, given $|\psi_{1}\rangle=\frac{1}{\sqrt{3}}\left(|011\rangle+|101\rangle+|110\rangle\right)$
(we can assume for convenience $\phi_{1}=\phi_{2}=\phi_{3}=0$, this
can be achieved adaptively by local operations), an optimal measurement
basis would be: 
\begin{equation*}
\begin{split}
&\frac{1}{\sqrt{3}}\left(|011\rangle+|101\rangle+|110\rangle\right)
,\frac{1}{\sqrt{6}}(-2|011\rangle+|101\rangle+|110\rangle),\\
&\frac{1}{\sqrt{2}}\left(|101\rangle-|110\rangle\right).
\end{split}
\end{equation*}
 The construction for $|\psi_{2}\rangle$ is equivalent.

\vskip 2mm

\subsection{Optimal initial states\label{sec:dicussion-details-initial-states}} 
We show that the average variance of phase difference $\frac{1}{\binom{N}{2}}\underset{j>k}{\sum}\text{Var}\left(\phi_{j}-\phi_{k}\right)$
is lower-bounded by $2\frac{N-1}{N}$ and we find
several initialization strategies, all involve entanglement, that saturate this bound. In particular the symmetric Dicke state: 
\begin{equation*}
|\psi\rangle=\frac{1}{\sqrt{\binom{N}{N/2}}}\underset{\vec{z} \; \text{with }\underset{i}{\sum}z_{i}=0}{\sum}|\vec{z}\rangle,
\end{equation*}
and any other state that is an equal superposition of states with $\underset{i}{\sum}z_{i}=0$
saturate this optimal precision. Another strategy is a probabilistic initialization from an ensemble of products of anti-parallel Bell pairs.
It can be shown that all optimal strategies involve eigenstates of $\underset{i}{\sum}Z_{i}$
with eigenvalue $0$.
Since these states are robust against correlated dephasing, this optimal precision is achieved irrespective of whether there is correlated dephasing or not. In the following derivation we use techniques similar to those used in reference \cite{distributed2}.

We denote the final and initial state as 
$|\psi_{f}\rangle, |\psi\rangle$
respectively, where $|\psi_{f}\rangle=U |\psi\rangle,$ with $U=\exp\left(-i\frac{1}{2} \vec{\phi}\cdot\vec{Z}\right).$
Since $|\partial_{\phi_{j}}\psi_{f}\rangle=-\frac{i}{2} Z_{j}U|\psi\rangle$,
the QFIM of $|\psi_{f} \rangle$ reads: 
\begin{equation*}
I_{i,j}=\left(\langle\psi|Z_{i}Z_{j}|\psi\rangle-\langle\psi|Z_{i}|\psi\rangle\langle\psi|Z_{j}|\psi\rangle\right).
\end{equation*}

Using the QFIM we prove that the optimal achievable variance, $\frac{1}{\binom{N}{2}}\underset{j>k}{\sum}\text{Var}\left(\phi_{j}-\phi_{k}\right)$
, is $2\frac{N-1}{N}$.\\
Proof: By definition of QFIM 
\[
\text{var}\left(\phi_{j}-\phi_{k}\right)\geq u_{(j,k)}^{\dagger}I^{-1}u_{\left(j,k\right)}.
\]
 The Cauchy-Schwarz inequality $\left(u^{\dagger}I^{-1}u\right)\left(u^{\dagger}Iu\right)\geq|u^{\dagger}\sqrt{I^{-1}}\sqrt{I}u|^{2}$
implies: 
\[
\left(u_{\left(j,k\right)}^{\dagger}I^{-1}u_{j,k}\right)\geq\frac{||u_{\left(j,k\right)}||^{4}}{\left(u_{\left(j,k\right)}^{\dagger}Iu_{\left(j,k\right)}\right)}
\]

Hence: 
\begin{align*}
\begin{split}
& \text{var}\left(\phi_{j}-\phi_{k}\right)\geq\frac{||u_{\left(j,k\right)}||^{4}}{\left(u_{\left(j,k\right)}^{\dagger}Iu_{\left(j,k\right)}\right)}=\frac{4}{I_{jj}+I_{kk}-2I_{jk}} \\ 
& \geq\frac{2}{\left(1-\langle Z_{j}Z_{k}\rangle\right)}.
\end{split}
\end{align*}
 Therefore we seek to lower bound $\underset{j<k}{\sum}\frac{2}{\left(1-\langle Z_{j}Z_{k}\rangle\right)}$.
The minimal possible $\text{var}\left(\phi_{j}-\phi_{k}\right)$ is
therefore obtained when $\langle Z_{j}Z_{k}\rangle=-1$, however there is no state that satisfies $\langle Z_{j}Z_{k}\rangle=-1$ for all $j,k$.
To lower bound $\underset{j<k}{\sum}\frac{2}{\left(1-\langle Z_{j}Z_{k}\rangle\right)}$
we use the Cauchy-Schwarz inequality: 
\begin{align*}
\begin{split}
&2 \left(\underset{j<k}{\sum} \frac{1}{\left(1-\langle Z_{j}Z_{k}\rangle\right)}\right)\left(\underset{j<k}{\sum}\left(1-\langle Z_{j}Z_{k}\rangle\right)\right) \\
&\geq  2 \left(\underset{j<k}{\sum}1\right)^{2}=2 \underset{j<k}{\sum}{\binom{N}{2}}
\end{split}
\end{align*}
The first inequality is due to $\left(\underset{i}{\sum}\frac{1}{x_{i}}\right)\left(\underset{i}{\sum}x_{i}\right)\geq\left(\underset{i}{\sum}\frac{1}{\sqrt{x_{i}}}\cdot\sqrt{x_{i}}\right)^{2}=\left(\underset{i}{\sum}1\right)^{2}$
, which is just the Cauchy-Schwarz inequality.\\
Hence: 
\[
\underset{j<k}{\sum}\text{Var}\left(\phi_{j}-\phi_{k}\right)\geq 2 \underset{j<k}{\sum}\frac{\binom{N}{2}}{\underset{j<k}{\sum}\left(1-\langle Z_{j}Z_{k}\rangle\right)}.
\]

Note that $\langle\underset{i}{\sum}Z_{i}\rangle^{2}=2\underset{j<k}{\sum}\langle Z_{j}Z_{k}\rangle+N$,
therefore $2\underset{j<k}{\sum}\langle Z_{j}Z_{k}\rangle\geq-N$.
Hence: 
\begin{align*}
\begin{split}
&\underset{j<k}{\sum}\text{Var}\left(\phi_{j}-\phi_{k}\right) \geq\underset{j<k}{\sum}\frac{ 4 \binom{N}{2}}{2\binom{N}{2}+N}\\
&=2 \underset{j<k}{\sum}\frac{N-1}{N}.
\end{split}
\end{align*}

This basically proves that $\frac{1}{\binom{N}{2}}\underset{j<k}{\sum}\text{Var}\left(\phi_{j}-\phi_{k}\right)\geq 2 \frac{N-1}{N}$.
To show that this lower bound is saturable we need to find an initial
state $|\psi\rangle$, for which all these inequalities are saturated,
namely: 
\[
\underset{j<k}{\sum}u_{(j,k)}^{\dagger}I^{-1}u_{\left(j,k\right)}= 2 \underset{j<k}{\sum}\frac{N-1}{N},
\]
where $I$ is the QFIM given this $|\psi\rangle$. We observed that
a necessary condition is $\langle Z_{i} \rangle=0$
and identical $\langle Z_{j}Z_{k}\rangle=-\frac{1}{N-1}$.
Let us show that this is also 
a sufficient condition: given that this condition is satisfied the QFIM is
\[
I_{i,j}=\begin{cases}
1 & i=j\\
-\frac{1}{N-1} & i\neq j
\end{cases}.
\]
 It can be now observed that any $u_{\left(j,k\right)}$ is an eigenvector
of this matrix with eigenvalue $\frac{N}{N-1}$ and thus for any $j,k$~: $u_{(j,k)}^{\dagger}I^{-1}u_{\left(j,k\right)}=2 \frac{N-1}{N}.$

Hence any initial pure state that satisfies the conditions:
\begin{align}
\forall i \; \langle Z_{i} \rangle=0 \; \text{and} \; \forall j,k \; \langle Z_{j} Z_{k}  \rangle=-\frac{1}{N-1},
\label{eq:optimality_conditions}
\end{align}
saturates this QFIM.
We can immediately observe that the symmetric Dicke state: 
\begin{equation*}
|\psi\rangle=\frac{1}{\sqrt{\binom{N}{N/2}}}\underset{\vec{z} \; \text{with }\underset{i}{\sum}z_{i}=0}{\sum}|\vec{z}\rangle
\label{eq:symmetric_Dicke}
\end{equation*}
satisfies these conditions and thus saturates this bound. Other strategies exist, such as preparing a classical ensemble of products of anti-parallel Bell states, and they will be discussed later. For now Let us focus on the symmetric Dicke state. 
\\
To show that indeed $\text{Var}\left(\phi_{j}-\phi_{k}\right)=2 \frac{N-1}{N}$ can be achieved with $|\psi\rangle$,
we need to find a read-out strategy that achieves this bound, i.e. a measurement with a classical FI matrix that equals the QFIM. We show that local measurements in $X$ saturate this optimal variance.

To show this let us first write $|\psi_{f}\rangle,$ the final probe state given the initial symmetric Dicke state:
\begin{align}
&|\psi_{f}\rangle=\frac{1}{\sqrt{\binom{N}{N/2}}}\underset{\underset{i}{\sum}z_{i}=0,z_{1}=1 }{\sum}\left(e^{-i\frac{1}{2}\vec{\phi}\cdot\vec{z}}|\vec{z}\rangle+e^{i\frac{1}{2}\vec{\phi}\cdot\vec{z}}|-\vec{z}\rangle\right)=\\
&\frac{\sqrt{2}}{\sqrt{\binom{N}{N/2}}}\underset{\underset{i}{\sum}z_{i}=0,z_{1}=1}{\sum}\cos\left( \frac{1}{2}\vec{\phi}\cdot\vec{z}\right)|+_{\vec{z}}\rangle-i\sin\left(\frac{1}{2}\vec{\phi}\cdot\vec{z}\right)|-_{\vec{z}}\rangle,
\label{eq:psif}
\end{align} 
where $| \pm_{\vec{z}}\rangle= \frac{1}{\sqrt{2}}\left(|\vec{z}\rangle \pm |-\vec{z}\rangle\right).$

Let us now use theorem $2$ of reference \cite{pezze2017optimal}:
Given a pure probe state $|\Psi\left(\vec{\phi}\right)\rangle,$ then a projective measurement that consists of rank $1$ projectors $\left\{ \Pi_{k}\right\} _{k}$
saturates the QFIM if and only if for
every $k$, $j$: 
\begin{equation}
\text{Im}\left(\langle\Psi|\Pi_{k}|\partial_{\phi_j}\Psi_{\perp}\rangle\right)=0,
\label{condition_for_optimality}
\end{equation}
where $|\partial_{\phi_j}\Psi_{\perp}\rangle:=|\partial_{\phi_j}\Psi\rangle-|\Psi\rangle\langle\Psi|\partial_{\phi_j}\Psi\rangle,$ i.e. it is the projection of $|\partial_{\phi_j}\Psi\rangle$ onto the orthogonal subspace of $|\Psi\rangle.$
The full proof of this theorem is presented in ref. \cite{pezze2017optimal}. Let us briefly explain the intuition behind this theorem: given $|\Psi\rangle$ the probability of detecting the $k$-th result is $p_{k}=\langle\Psi|\Pi_{k}|\Psi\rangle.$ The derivative of this probability with respect to $\phi_j$ is
$\partial_{\phi_{j}}p_{k}=2\text{Re}\langle\Psi|\Pi_{k}|\partial_{\phi_{j}}\Psi\rangle.$
The parallel part of $|\partial_{\phi_{j}}\Psi\rangle,$ i.e. $|\partial_{\phi_{j}}\Psi_{\parallel}\rangle:=\langle\Psi|\partial_{\phi_{j}}\Psi\rangle|\Psi\rangle,$ does not contribute to the derivative because $\text{Re}\langle\Psi|\Pi_{k}|\partial_{\phi_{j}}\Psi_{\parallel}\rangle=0.$
The derivative can therefore be written as $\partial_{\phi_{j}}p_{k}=2\text{Re}\langle\Psi|\Pi_{k}|\partial_{\phi_{j}}\Psi_{\perp}\rangle.$
Hence if $\text{Im}\langle\Psi|\Pi_{k}|\partial_{\phi_{j}}\Psi_{\perp}\rangle\neq0$ then some of the information about $\phi_j$ is being lost when measuring in this basis, i.e. a change $\phi_{j}$ is being translated to a change in the phase and not the probability. If $\text{Im}\langle\Psi|\Pi_{k}|\partial_{\phi_{j}}\Psi_{\perp}\rangle=0$ for every $j,k$ then no information about $\vec{\phi}$ is lost and thus the QFIM is being saturated. We remark that this intuitive argument is correct only for pure states.

 Let us apply this theorem to our case:  we need to show that the condition of eq.~(\ref{condition_for_optimality})
is satisfied for our $|\psi_{f}\rangle$ and local $X$ measurements. The rank $1$ projectors in our case are thus $\left\{ \Pi_{\vec{x}}= |\vec{x}\rangle\langle\vec{x}|\right\} _{\vec{x}}$.
Hence we need to show that for every $\vec{x}$: $\text{Im}\left(\langle\psi_{f}|\Pi_{\vec{x}}|\partial_{\phi_{j}}\psi_{f\perp}\rangle\right)=0.$

Let us use the following identity:
\[
\langle\vec{x}|\frac{1}{\sqrt{2}}\left(|\vec{z}\rangle+|-\vec{z}\rangle\right)=\frac{1}{\sqrt{2^{N+1}}}\left[\left(-1\right)^{\vec{q}_{\vec{x}}\cdot\vec{q}_{\vec{z}}}+\left(-1\right)^{\vec{q}_{\vec{x}}\cdot\vec{q}_{-\vec{z}}}\right],
\]
with $\left(\vec{q}_{\vec{z}}\right)_{i}=\frac{1}{2}\left(1-z_{i}\right)$ and $\vec{q}_{\vec{x}}$ and  analogously $\left(\vec{q}_{\vec{x}}\right)_{i}=\frac{1}{2}\left(1-x_{i}\right)$.
 Note that $\vec{q}_{\vec{x}}\cdot\vec{q}_{\vec{z}}+\vec{q}_{\vec{x}}\cdot\vec{q}_{-\vec{z}}=\#\left(x_{i}=-1\right)$,
hence if $\#\left(x_{i}=-1\right)$ is even then $\left(-1\right)^{\vec{q}_{\vec{x}}\cdot\vec{q}_{\vec{z}}}=\left(-1\right)^{\vec{q}_{\vec{x}}\cdot\vec{q}_{-\vec{z}}}$,
and if it is odd $\left(-1\right)^{\vec{q}_{\vec{x}}\cdot\vec{q}_{\vec{z}}}=-\left(-1\right)^{\vec{q}_{\vec{x}}\cdot\vec{q}_{-\vec{z}}}$.
Therefore:
\begin{equation}
\langle\vec{x}|\frac{1}{\sqrt{2}}\left(|\vec{z}\rangle+|-\vec{z}\rangle\right)=\begin{cases}
0 & \vec{x} \; \text{odd}\\
\pm\frac{1}{\sqrt{2^{N}}} & \vec{x} \; \text{even},
\end{cases}
\label{inner1}
\end{equation}
and:
\begin{equation}
\langle\vec{x}|\frac{1}{\sqrt{2}}\left(|\vec{z}\rangle-|-\vec{z}\rangle\right)=\begin{cases}
\pm\frac{1}{\sqrt{2^{N}}} & \vec{x}\;\text{odd}\\
0 & \vec{x}\;\text{even},
\end{cases}
\label{inner2}
\end{equation}
Inserting equations (\ref{inner1})-(\ref{inner2}) into equation~(\ref{eq:psif}) we can observe that:
\begin{align*}
&\langle\vec{x}|\psi_{f}\rangle=\frac{1}{\sqrt{2^{N-1}{\binom{N}{N/2}}}} \cdot \\
&\begin{cases}
\left(-i\right)\underset{\underset{i}{\sum}z_{i}=0,z_{1}=1}{\sum}\left(-1\right)^{\underset{i}{\sum}\left(\vec{q}_{\vec{x}}\right)_{i}\left(\vec{q}_{\vec{z}}\right)_{i}}\sin\left(\frac{1}{2}\vec{\phi}\cdot\vec{z}\right) & \vec{x}\;\text{odd}\\
\underset{\underset{i}{\sum}z_{i}=0,z_{1}=1}{\sum}\left(-1\right)^{\underset{i}{\sum}\left(\vec{q}_{\vec{x}}\right)_{i}\left(\vec{q}_{\vec{z}}\right)_{i}}\cos\left(\frac{1}{2}\vec{\phi}\cdot\vec{z}\right) & \vec{x}\;\text{even}.
\end{cases}
\end{align*}
Therefore for any value of $\vec{\phi},$ $\langle\vec{x}|\psi_{f}\rangle$ is either real (for even $\vec{x}$) or imaginary (for odd $\vec{x}$). Similarly it is simple to observe that $\langle\vec{x}|\partial_{\phi_{j}}\psi_{f\perp}\rangle$ is real (imaginary) for even $\vec{x}$ (odd $\vec{x}$). To sum up:   
\begin{align*}
\begin{cases}
\vec{x}\;\text{odd}:\\
\langle\vec{x}|\psi_{f}\rangle,\langle\vec{x}|\partial_{\phi_{j}}\psi_{f\perp}\rangle\;\text{imaginary}\Rightarrow\langle\psi_{f}|\Pi_{\vec{x}}|\partial_{\phi_{j}}\psi_{f\perp}\rangle\;\text{real}  \\
\vec{x}\;\text{even}:\\
\langle\vec{x}|\psi_{f}\rangle,\langle\vec{x}|\partial_{\phi_{j}}\psi_{f\perp}\rangle\;\text{real}\Rightarrow\langle\psi_{f}|\Pi_{\vec{x}}|\partial_{\phi_{j}}\psi_{f\perp}\rangle\;\text{real} \;\; .
\end{cases}    
\end{align*}
Hence the condition in eq. (\ref{condition_for_optimality}) is satisfied, and thus the local $X$ basis indeed saturates the QFIM.

We remark that another strategy to saturate the QFIM is to choose the initial state of each experiment  from a classical ensemble of products of anti-parallel Bell states. An anti-parallel Bell state is defined as $|i,j\rangle=|0\rangle_{i}|1\rangle_{j}+|0\rangle_{j}|1\rangle_{i}.$ A Bell-product state is then a product of N/2 such anti-parallel Bell pairs, we denote any such state as $\underset{k=1}{\overset{N/2}{\prod}}|i_{k},j_{k}\rangle.$ It can be observed that the QFIM of any $\underset{k=1}{\overset{N/2}{\prod}}|i_{k},j_{k}\rangle$ is $\mathds{1}-\underset{k}{\prod}X_{i_{k},j_{k}}$ , where $X_{i_{k},j_{k}}=|i_{k}\rangle\langle j_{k}|+h.c.$. The total number of these states is $\left(N-1\right)\left(N-3\right)...1$, the total number of states with a specific pair $\left(i_{k},j_{k}\right)$ is $\left(N-3\right)...1$.
Hence by sampling from a uniform distribution of these Bell-product states the QFIM becomes the optimal one: $I_{i,j}=\begin{cases}
-\frac{1}{N-1} & i\neq j\\
1 & i=j
\end{cases}.$
This QFIM is saturated with local $X$ (or $Y$) measurements since these measurements saturate the QFIM of each Bell-product state individually.

Finally, we remark about the set of optimal initial pure states. These are the states that satisfy the conditions in equation \ref{eq:optimality_conditions}. 
The problem of finding optimal states (other than the symmetric Dicke state) is then basically solving a system of linear equations.
First observe that the conditions imply $\langle(\underset{i}{\sum}Z_{i})^{2}\rangle=0,$ hence any optimal pure state is an eigenstate of $\underset{i}{\sum}Z_{i}$ with eigenvalue $0$. We can therefore write the states as $\underset{\underset{i}{\sum}z_{i}=0}{\sum}\sqrt{p}_{\vec{z}}|\vec{z}\rangle.$
The conditions then become a system of ${N \choose 2}+N+1$ linear equations for the distribution $\left\{ p_{\vec{z}}\right\}$ : $\forall j,k,i \underset{\underset{i}{\sum}z_{i}=0}{\sum}p_{\vec{z}}z_{j}z_{k}=-\frac{1}{N-1} , \underset{\underset{i}{\sum}z_{i}=0}{\sum}p_{\vec{z}}z_{i}=0$ and $\underset{l}{\sum}p_{l}=1,$ with a constraint of $0 \leq p_{l}\leq 1$ for all $l$.
The number of equations, ${N \choose 2}+N+1,$ is smaller than the number of variables, ${N \choose N/2}$ , which implies that there exist solutions other than the symmetric one. Finding solutions that are simple to prepare, in terms of entanglement or circuit complexity, is an interesting problem and we leave it as an open question.

%--- bibliography --------------------------------------------------------

%\bibliography{bibliography}% Produces the bibliography via BibTeX.

\begin{thebibliography}{64}%
\makeatletter
\providecommand \@ifxundefined [1]{%
 \@ifx{#1\undefined}
}%
\providecommand \@ifnum [1]{%
 \ifnum #1\expandafter \@firstoftwo
 \else \expandafter \@secondoftwo
 \fi
}%
\providecommand \@ifx [1]{%
 \ifx #1\expandafter \@firstoftwo
 \else \expandafter \@secondoftwo
 \fi
}%
\providecommand \natexlab [1]{#1}%
\providecommand \enquote  [1]{``#1''}%
\providecommand \bibnamefont  [1]{#1}%
\providecommand \bibfnamefont [1]{#1}%
\providecommand \citenamefont [1]{#1}%
\providecommand \href@noop [0]{\@secondoftwo}%
\providecommand \href [0]{\begingroup \@sanitize@url \@href}%
\providecommand \@href[1]{\@@startlink{#1}\@@href}%
\providecommand \@@href[1]{\endgroup#1\@@endlink}%
\providecommand \@sanitize@url [0]{\catcode `\\12\catcode `\$12\catcode
  `\&12\catcode `\#12\catcode `\^12\catcode `\_12\catcode `\%12\relax}%
\providecommand \@@startlink[1]{}%
\providecommand \@@endlink[0]{}%
\providecommand \url  [0]{\begingroup\@sanitize@url \@url }%
\providecommand \@url [1]{\endgroup\@href {#1}{\urlprefix }}%
\providecommand \urlprefix  [0]{URL }%
\providecommand \Eprint [0]{\href }%
\providecommand \doibase [0]{https://doi.org/}%
\providecommand \selectlanguage [0]{\@gobble}%
\providecommand \bibinfo  [0]{\@secondoftwo}%
\providecommand \bibfield  [0]{\@secondoftwo}%
\providecommand \translation [1]{[#1]}%
\providecommand \BibitemOpen [0]{}%
\providecommand \bibitemStop [0]{}%
\providecommand \bibitemNoStop [0]{.\EOS\space}%
\providecommand \EOS [0]{\spacefactor3000\relax}%
\providecommand \BibitemShut  [1]{\csname bibitem#1\endcsname}%
\let\auto@bib@innerbib\@empty
%</preamble>
\bibitem [{\citenamefont {Ludlow}\ \emph {et~al.}(2015)\citenamefont {Ludlow},
  \citenamefont {Boyd}, \citenamefont {Ye}, \citenamefont {Peik},\ and\
  \citenamefont {Schmidt}}]{Ludlow:2015}%
  \BibitemOpen
  \bibfield  {author} {\bibinfo {author} {\bibfnamefont {A.~D.}\ \bibnamefont
  {Ludlow}}, \bibinfo {author} {\bibfnamefont {M.~M.}\ \bibnamefont {Boyd}},
  \bibinfo {author} {\bibfnamefont {J.}~\bibnamefont {Ye}}, \bibinfo {author}
  {\bibfnamefont {E.}~\bibnamefont {Peik}},\ and\ \bibinfo {author}
  {\bibfnamefont {P.~O.}\ \bibnamefont {Schmidt}},\ }\bibfield  {title}
  {\bibinfo {title} {Optical atomic clocks},\ }\href
  {https://doi.org/10.1103/RevModPhys.87.637} {\bibfield  {journal} {\bibinfo
  {journal} {Rev. Mod. Phys.}\ }\textbf {\bibinfo {volume} {87}},\ \bibinfo
  {pages} {637} (\bibinfo {year} {2015})}\BibitemShut {NoStop}%
\bibitem [{\citenamefont {Canuel}\ \emph {et~al.}(2006)\citenamefont {Canuel},
  \citenamefont {Leduc}, \citenamefont {Holleville}, \citenamefont {Gauguet},
  \citenamefont {Fils}, \citenamefont {Virdis}, \citenamefont {Clairon},
  \citenamefont {Dimarcq}, \citenamefont {Bord\'e}, \citenamefont {Landragin},\
  and\ \citenamefont {Bouyer}}]{Canuel:2006}%
  \BibitemOpen
  \bibfield  {author} {\bibinfo {author} {\bibfnamefont {B.}~\bibnamefont
  {Canuel}}, \bibinfo {author} {\bibfnamefont {F.}~\bibnamefont {Leduc}},
  \bibinfo {author} {\bibfnamefont {D.}~\bibnamefont {Holleville}}, \bibinfo
  {author} {\bibfnamefont {A.}~\bibnamefont {Gauguet}}, \bibinfo {author}
  {\bibfnamefont {J.}~\bibnamefont {Fils}}, \bibinfo {author} {\bibfnamefont
  {A.}~\bibnamefont {Virdis}}, \bibinfo {author} {\bibfnamefont
  {A.}~\bibnamefont {Clairon}}, \bibinfo {author} {\bibfnamefont
  {N.}~\bibnamefont {Dimarcq}}, \bibinfo {author} {\bibfnamefont {C.~J.}\
  \bibnamefont {Bord\'e}}, \bibinfo {author} {\bibfnamefont {A.}~\bibnamefont
  {Landragin}},\ and\ \bibinfo {author} {\bibfnamefont {P.}~\bibnamefont
  {Bouyer}},\ }\bibfield  {title} {\bibinfo {title} {Six-axis inertial sensor
  using cold-atom interferometry},\ }\href
  {https://doi.org/10.1103/PhysRevLett.97.010402} {\bibfield  {journal}
  {\bibinfo  {journal} {Phys. Rev. Lett.}\ }\textbf {\bibinfo {volume} {97}},\
  \bibinfo {pages} {010402} (\bibinfo {year} {2006})}\BibitemShut {NoStop}%
\bibitem [{\citenamefont {Peters}\ \emph {et~al.}(1999)\citenamefont {Peters},
  \citenamefont {Chung},\ and\ \citenamefont {Chu}}]{Peters:1999}%
  \BibitemOpen
  \bibfield  {author} {\bibinfo {author} {\bibfnamefont {A.}~\bibnamefont
  {Peters}}, \bibinfo {author} {\bibfnamefont {K.~Y.}\ \bibnamefont {Chung}},\
  and\ \bibinfo {author} {\bibfnamefont {S.}~\bibnamefont {Chu}},\ }\bibfield
  {title} {\bibinfo {title} {Measurement of gravitational acceleration by
  dropping atoms},\ }\href {https://doi.org/10.1038/23655} {\bibfield
  {journal} {\bibinfo  {journal} {Nature}\ }\textbf {\bibinfo {volume} {400}},\
  \bibinfo {pages} {849} (\bibinfo {year} {1999})}\BibitemShut {NoStop}%
\bibitem [{\citenamefont {Hamilton}\ \emph {et~al.}(2015)\citenamefont
  {Hamilton}, \citenamefont {Jaffe}, \citenamefont {Haslinger}, \citenamefont
  {Simmons}, \citenamefont {M\"{u}ller},\ and\ \citenamefont
  {Khoury}}]{Hamilton:2015}%
  \BibitemOpen
  \bibfield  {author} {\bibinfo {author} {\bibfnamefont {P.}~\bibnamefont
  {Hamilton}}, \bibinfo {author} {\bibfnamefont {M.}~\bibnamefont {Jaffe}},
  \bibinfo {author} {\bibfnamefont {P.}~\bibnamefont {Haslinger}}, \bibinfo
  {author} {\bibfnamefont {Q.}~\bibnamefont {Simmons}}, \bibinfo {author}
  {\bibfnamefont {H.}~\bibnamefont {M\"{u}ller}},\ and\ \bibinfo {author}
  {\bibfnamefont {J.}~\bibnamefont {Khoury}},\ }\bibfield  {title} {\bibinfo
  {title} {Atom-interferometry constraints on dark energy},\ }\href
  {https://doi.org/10.1126/science.aaa8883} {\bibfield  {journal} {\bibinfo
  {journal} {Science}\ }\textbf {\bibinfo {volume} {349}},\ \bibinfo {pages}
  {849} (\bibinfo {year} {2015})}\BibitemShut {NoStop}%
\bibitem [{\citenamefont {Bouchendira}\ \emph {et~al.}(2011)\citenamefont
  {Bouchendira}, \citenamefont {Clad\'e}, \citenamefont {Guellati-Kh\'elifa},
  \citenamefont {Nez},\ and\ \citenamefont {Biraben}}]{Bouchendira:2011}%
  \BibitemOpen
  \bibfield  {author} {\bibinfo {author} {\bibfnamefont {R.}~\bibnamefont
  {Bouchendira}}, \bibinfo {author} {\bibfnamefont {P.}~\bibnamefont
  {Clad\'e}}, \bibinfo {author} {\bibfnamefont {S.}~\bibnamefont
  {Guellati-Kh\'elifa}}, \bibinfo {author} {\bibfnamefont {F.~m.~c.}\
  \bibnamefont {Nez}},\ and\ \bibinfo {author} {\bibfnamefont {F.~m.~c.}\
  \bibnamefont {Biraben}},\ }\bibfield  {title} {\bibinfo {title} {New
  determination of the fine structure constant and test of the quantum
  electrodynamics},\ }\href {https://doi.org/10.1103/PhysRevLett.106.080801}
  {\bibfield  {journal} {\bibinfo  {journal} {Phys. Rev. Lett.}\ }\textbf
  {\bibinfo {volume} {106}},\ \bibinfo {pages} {080801} (\bibinfo {year}
  {2011})}\BibitemShut {NoStop}%
\bibitem [{\citenamefont {Giovannetti}\ \emph {et~al.}(2011)\citenamefont
  {Giovannetti}, \citenamefont {Lloyd},\ and\ \citenamefont
  {Maccone}}]{Giovannetti:2011}%
  \BibitemOpen
  \bibfield  {author} {\bibinfo {author} {\bibfnamefont {V.}~\bibnamefont
  {Giovannetti}}, \bibinfo {author} {\bibfnamefont {S.}~\bibnamefont {Lloyd}},\
  and\ \bibinfo {author} {\bibfnamefont {L.}~\bibnamefont {Maccone}},\
  }\bibfield  {title} {\bibinfo {title} {Advances in quantum metrology},\
  }\href {https://doi.org/10.1038/nphoton.2011.35} {\bibfield  {journal}
  {\bibinfo  {journal} {Nat. Photonics}\ }\textbf {\bibinfo {volume} {5}},\
  \bibinfo {pages} {222} (\bibinfo {year} {2011})}\BibitemShut {NoStop}%
\bibitem [{\citenamefont {Wineland}\ \emph {et~al.}(1992)\citenamefont
  {Wineland}, \citenamefont {Bollinger}, \citenamefont {Itano}, \citenamefont
  {Moore},\ and\ \citenamefont {Heinzen}}]{Wineland:1992}%
  \BibitemOpen
  \bibfield  {author} {\bibinfo {author} {\bibfnamefont {D.~J.}\ \bibnamefont
  {Wineland}}, \bibinfo {author} {\bibfnamefont {J.~J.}\ \bibnamefont
  {Bollinger}}, \bibinfo {author} {\bibfnamefont {W.~M.}\ \bibnamefont
  {Itano}}, \bibinfo {author} {\bibfnamefont {F.~L.}\ \bibnamefont {Moore}},\
  and\ \bibinfo {author} {\bibfnamefont {D.~J.}\ \bibnamefont {Heinzen}},\
  }\bibfield  {title} {\bibinfo {title} {{S}pin squeezing and reduced quantum
  noise in spectroscopy},\ }\href {https://doi.org/10.1103/PhysRevA.46.R6797}
  {\bibfield  {journal} {\bibinfo  {journal} {Phys.~Rev.~A}\ }\textbf {\bibinfo
  {volume} {46}},\ \bibinfo {pages} {R6797} (\bibinfo {year}
  {1992})}\BibitemShut {NoStop}%
\bibitem [{\citenamefont {Bollinger}\ \emph {et~al.}(1996)\citenamefont
  {Bollinger}, \citenamefont {Itano}, \citenamefont {Wineland},\ and\
  \citenamefont {Heinzen}}]{Bollinger:1996}%
  \BibitemOpen
  \bibfield  {author} {\bibinfo {author} {\bibfnamefont {J.~J.~.}\ \bibnamefont
  {Bollinger}}, \bibinfo {author} {\bibfnamefont {W.~M.}\ \bibnamefont
  {Itano}}, \bibinfo {author} {\bibfnamefont {D.~J.}\ \bibnamefont
  {Wineland}},\ and\ \bibinfo {author} {\bibfnamefont {D.~J.}\ \bibnamefont
  {Heinzen}},\ }\bibfield  {title} {\bibinfo {title} {Optimal frequency
  measurements with maximally correlated states},\ }\href
  {https://doi.org/10.1103/PhysRevA.54.R4649} {\bibfield  {journal} {\bibinfo
  {journal} {Phys.~Rev.~A}\ }\textbf {\bibinfo {volume} {54}},\ \bibinfo
  {pages} {R4649} (\bibinfo {year} {1996})}\BibitemShut {NoStop}%
\bibitem [{\citenamefont {Marciniak}\ \emph {et~al.}(2022)\citenamefont
  {Marciniak}, \citenamefont {Feldker}, \citenamefont {Pogorelov},
  \citenamefont {Kaubruegger}, \citenamefont {Vasilyev}, \citenamefont {van
  Bijnen}, \citenamefont {Schindler}, \citenamefont {Zoller}, \citenamefont
  {Blatt},\ and\ \citenamefont {Monz}}]{Marciniak:2022}%
  \BibitemOpen
  \bibfield  {author} {\bibinfo {author} {\bibfnamefont {C.~D.}\ \bibnamefont
  {Marciniak}}, \bibinfo {author} {\bibfnamefont {T.}~\bibnamefont {Feldker}},
  \bibinfo {author} {\bibfnamefont {I.}~\bibnamefont {Pogorelov}}, \bibinfo
  {author} {\bibfnamefont {R.}~\bibnamefont {Kaubruegger}}, \bibinfo {author}
  {\bibfnamefont {D.~V.}\ \bibnamefont {Vasilyev}}, \bibinfo {author}
  {\bibfnamefont {R.}~\bibnamefont {van Bijnen}}, \bibinfo {author}
  {\bibfnamefont {P.}~\bibnamefont {Schindler}}, \bibinfo {author}
  {\bibfnamefont {P.}~\bibnamefont {Zoller}}, \bibinfo {author} {\bibfnamefont
  {R.}~\bibnamefont {Blatt}},\ and\ \bibinfo {author} {\bibfnamefont
  {T.}~\bibnamefont {Monz}},\ }\bibfield  {title} {\bibinfo {title} {Optimal
  metrology with programmable quantum sensors},\ }\href
  {https://doi.org/https://doi.org/10.1038/s41586-022-04435-4} {\bibfield
  {journal} {\bibinfo  {journal} {Nature}\ }\textbf {\bibinfo {volume} {603}},\
  \bibinfo {pages} {604} (\bibinfo {year} {2022})}\BibitemShut {NoStop}%
\bibitem [{\citenamefont {Huelga}\ \emph {et~al.}(1997)\citenamefont {Huelga},
  \citenamefont {Macchiavello}, \citenamefont {Pellizzari}, \citenamefont
  {Ekert}, \citenamefont {Plenio},\ and\ \citenamefont {Cirac}}]{Huelga:1997}%
  \BibitemOpen
  \bibfield  {author} {\bibinfo {author} {\bibfnamefont {S.~F.}\ \bibnamefont
  {Huelga}}, \bibinfo {author} {\bibfnamefont {C.}~\bibnamefont
  {Macchiavello}}, \bibinfo {author} {\bibfnamefont {T.}~\bibnamefont
  {Pellizzari}}, \bibinfo {author} {\bibfnamefont {A.~K.}\ \bibnamefont
  {Ekert}}, \bibinfo {author} {\bibfnamefont {M.~B.}\ \bibnamefont {Plenio}},\
  and\ \bibinfo {author} {\bibfnamefont {J.~I.}\ \bibnamefont {Cirac}},\
  }\bibfield  {title} {\bibinfo {title} {Improvement of frequency standards
  with quantum entanglement},\ }\href
  {https://doi.org/10.1103/PhysRevLett.79.3865} {\bibfield  {journal} {\bibinfo
   {journal} {Phys.~Rev.~Lett.}\ }\textbf {\bibinfo {volume} {79}},\ \bibinfo
  {pages} {3865} (\bibinfo {year} {1997})}\BibitemShut {NoStop}%
\bibitem [{\citenamefont {Demkowicz-Dobrza{\'{n}}ski}\ \emph
  {et~al.}(2012)\citenamefont {Demkowicz-Dobrza{\'{n}}ski}, \citenamefont
  {Ko{\l}ody{\'{n}}ski},\ and\ \citenamefont
  {Gu{\c{t}}{\u{a}}}}]{Demkowicz-Dobrzanski:2012}%
  \BibitemOpen
  \bibfield  {author} {\bibinfo {author} {\bibfnamefont {R.}~\bibnamefont
  {Demkowicz-Dobrza{\'{n}}ski}}, \bibinfo {author} {\bibfnamefont
  {J.}~\bibnamefont {Ko{\l}ody{\'{n}}ski}},\ and\ \bibinfo {author}
  {\bibfnamefont {M.}~\bibnamefont {Gu{\c{t}}{\u{a}}}},\ }\bibfield  {title}
  {\bibinfo {title} {The elusive {H}eisenberg limit in quantum-enhanced
  metrology},\ }\href {https://doi.org/10.1038/ncomms2067} {\bibfield
  {journal} {\bibinfo  {journal} {Nat. Commun.}\ }\textbf {\bibinfo {volume}
  {3}},\ \bibinfo {pages} {1063} (\bibinfo {year} {2012})}\BibitemShut
  {NoStop}%
\bibitem [{\citenamefont {Chin}\ \emph {et~al.}(2012)\citenamefont {Chin},
  \citenamefont {Huelga},\ and\ \citenamefont {Plenio}}]{Chin:2011}%
  \BibitemOpen
  \bibfield  {author} {\bibinfo {author} {\bibfnamefont {A.~W.}\ \bibnamefont
  {Chin}}, \bibinfo {author} {\bibfnamefont {S.~F.}\ \bibnamefont {Huelga}},\
  and\ \bibinfo {author} {\bibfnamefont {M.~B.}\ \bibnamefont {Plenio}},\
  }\bibfield  {title} {\bibinfo {title} {Quantum metrology in non-markovian
  environments},\ }\href {https://doi.org/10.1103/PhysRevLett.109.233601}
  {\bibfield  {journal} {\bibinfo  {journal} {Phys. Rev. Lett.}\ }\textbf
  {\bibinfo {volume} {109}},\ \bibinfo {pages} {233601} (\bibinfo {year}
  {2012})}\BibitemShut {NoStop}%
\bibitem [{\citenamefont {Jeske}\ \emph {et~al.}(2014)\citenamefont {Jeske},
  \citenamefont {Cole},\ and\ \citenamefont {Huelga}}]{Jeske:2014}%
  \BibitemOpen
  \bibfield  {author} {\bibinfo {author} {\bibfnamefont {J.}~\bibnamefont
  {Jeske}}, \bibinfo {author} {\bibfnamefont {J.~H.}\ \bibnamefont {Cole}},\
  and\ \bibinfo {author} {\bibfnamefont {S.~F.}\ \bibnamefont {Huelga}},\
  }\bibfield  {title} {\bibinfo {title} {Quantum metrology subject to spatially
  correlated {M}arkovian noise: restoring the {H}eisenberg limit},\ }\href
  {https://doi.org/10.1088/1367-2630/16/7/073039} {\bibfield  {journal}
  {\bibinfo  {journal} {New.~J.~Phys.}\ }\textbf {\bibinfo {volume} {16}},\
  \bibinfo {pages} {073039} (\bibinfo {year} {2014})}\BibitemShut {NoStop}%
\bibitem [{\citenamefont {Braun}\ \emph {et~al.}(2018)\citenamefont {Braun},
  \citenamefont {Adesso}, \citenamefont {Benatti}, \citenamefont {Floreanini},
  \citenamefont {Marzolino}, \citenamefont {Mitchell},\ and\ \citenamefont
  {Pirandola}}]{Braun:2018}%
  \BibitemOpen
  \bibfield  {author} {\bibinfo {author} {\bibfnamefont {D.}~\bibnamefont
  {Braun}}, \bibinfo {author} {\bibfnamefont {G.}~\bibnamefont {Adesso}},
  \bibinfo {author} {\bibfnamefont {F.}~\bibnamefont {Benatti}}, \bibinfo
  {author} {\bibfnamefont {R.}~\bibnamefont {Floreanini}}, \bibinfo {author}
  {\bibfnamefont {U.}~\bibnamefont {Marzolino}}, \bibinfo {author}
  {\bibfnamefont {M.~W.}\ \bibnamefont {Mitchell}},\ and\ \bibinfo {author}
  {\bibfnamefont {S.}~\bibnamefont {Pirandola}},\ }\bibfield  {title} {\bibinfo
  {title} {Quantum-enhanced measurements without entanglement},\ }\href
  {https://doi.org/10.1103/RevModPhys.90.035006} {\bibfield  {journal}
  {\bibinfo  {journal} {Rev. Mod. Phys.}\ }\textbf {\bibinfo {volume} {90}},\
  \bibinfo {pages} {035006} (\bibinfo {year} {2018})}\BibitemShut {NoStop}%
\bibitem [{\citenamefont {Chwalla}\ \emph {et~al.}(2007)\citenamefont
  {Chwalla}, \citenamefont {Kim}, \citenamefont {Monz}, \citenamefont
  {Schindler}, \citenamefont {Riebe}, \citenamefont {Roos},\ and\ \citenamefont
  {Blatt}}]{Chwalla:2007}%
  \BibitemOpen
  \bibfield  {author} {\bibinfo {author} {\bibfnamefont {M.}~\bibnamefont
  {Chwalla}}, \bibinfo {author} {\bibfnamefont {K.}~\bibnamefont {Kim}},
  \bibinfo {author} {\bibfnamefont {T.}~\bibnamefont {Monz}}, \bibinfo {author}
  {\bibfnamefont {P.}~\bibnamefont {Schindler}}, \bibinfo {author}
  {\bibfnamefont {M.}~\bibnamefont {Riebe}}, \bibinfo {author} {\bibfnamefont
  {C.~F.}\ \bibnamefont {Roos}},\ and\ \bibinfo {author} {\bibfnamefont
  {R.}~\bibnamefont {Blatt}},\ }\bibfield  {title} {\bibinfo {title} {Precision
  spectroscopy with two correlated atoms},\ }\href
  {https://doi.org/10.1007/s00340-007-2867-4} {\bibfield  {journal} {\bibinfo
  {journal} {Appl.~Phys.~B}\ }\textbf {\bibinfo {volume} {89}},\ \bibinfo
  {pages} {483} (\bibinfo {year} {2007})}\BibitemShut {NoStop}%
\bibitem [{\citenamefont {Olmschenk}\ \emph {et~al.}(2007)\citenamefont
  {Olmschenk}, \citenamefont {Younge}, \citenamefont {Moehring}, \citenamefont
  {Matsukevich}, \citenamefont {Maunz},\ and\ \citenamefont
  {Monroe}}]{Olmschenk:2007}%
  \BibitemOpen
  \bibfield  {author} {\bibinfo {author} {\bibfnamefont {S.}~\bibnamefont
  {Olmschenk}}, \bibinfo {author} {\bibfnamefont {K.~C.}\ \bibnamefont
  {Younge}}, \bibinfo {author} {\bibfnamefont {D.~L.}\ \bibnamefont
  {Moehring}}, \bibinfo {author} {\bibfnamefont {D.~N.}\ \bibnamefont
  {Matsukevich}}, \bibinfo {author} {\bibfnamefont {P.}~\bibnamefont {Maunz}},\
  and\ \bibinfo {author} {\bibfnamefont {C.}~\bibnamefont {Monroe}},\
  }\bibfield  {title} {\bibinfo {title} {Manipulation and detection of a
  trapped {Y}b$^+$ hyperfine qubit},\ }\href
  {https://doi.org/10.1103/PhysRevA.76.052314} {\bibfield  {journal} {\bibinfo
  {journal} {Phys.~Rev.~A}\ }\textbf {\bibinfo {volume} {76}},\ \bibinfo
  {pages} {052314} (\bibinfo {year} {2007})}\BibitemShut {NoStop}%
\bibitem [{\citenamefont {{Boulder Atomic Clock Optical Network (BACON)
  Collaboration}}\ \emph {et~al.}(2021)\citenamefont {{Boulder Atomic Clock
  Optical Network (BACON) Collaboration}}, \citenamefont {Beloy},\ and\
  \citenamefont {Bodine~{\sl et al.}}}]{BACON:2021}%
  \BibitemOpen
  \bibfield  {author} {\bibinfo {author} {\bibnamefont {{Boulder Atomic Clock
  Optical Network (BACON) Collaboration}}}, \bibinfo {author} {\bibfnamefont
  {K.}~\bibnamefont {Beloy}},\ and\ \bibinfo {author} {\bibfnamefont
  {M.}~\bibnamefont {Bodine~{\sl et al.}}},\ }\bibfield  {title} {\bibinfo
  {title} {Frequency ratio measurements at 18-digit accuracy using an optical
  clock network},\ }\href {https://doi.org/10.1038/s41586-021-03253-4}
  {\bibfield  {journal} {\bibinfo  {journal} {Nature}\ }\textbf {\bibinfo
  {volume} {591}},\ \bibinfo {pages} {564} (\bibinfo {year}
  {2021})}\BibitemShut {NoStop}%
\bibitem [{\citenamefont {Takamoto}\ \emph {et~al.}(2011)\citenamefont
  {Takamoto}, \citenamefont {Takano},\ and\ \citenamefont
  {Katori}}]{Takamoto:2011}%
  \BibitemOpen
  \bibfield  {author} {\bibinfo {author} {\bibfnamefont {M.}~\bibnamefont
  {Takamoto}}, \bibinfo {author} {\bibfnamefont {T.}~\bibnamefont {Takano}},\
  and\ \bibinfo {author} {\bibfnamefont {H.}~\bibnamefont {Katori}},\
  }\bibfield  {title} {\bibinfo {title} {Frequency comparison of optical
  lattice clocks beyond the {D}ick limit},\ }\href
  {https://doi.org/10.1038/nphoton.2011.34} {\bibfield  {journal} {\bibinfo
  {journal} {Nat. Photonics}\ }\textbf {\bibinfo {volume} {5}},\ \bibinfo
  {pages} {288} (\bibinfo {year} {2011})}\BibitemShut {NoStop}%
\bibitem [{\citenamefont {Nicholson}\ \emph {et~al.}(2012)\citenamefont
  {Nicholson}, \citenamefont {Martin}, \citenamefont {Williams}, \citenamefont
  {Bloom}, \citenamefont {Bishof}, \citenamefont {Swallows}, \citenamefont
  {Campbell},\ and\ \citenamefont {Ye}}]{Nicholson:2012}%
  \BibitemOpen
  \bibfield  {author} {\bibinfo {author} {\bibfnamefont {T.~L.}\ \bibnamefont
  {Nicholson}}, \bibinfo {author} {\bibfnamefont {M.~J.}\ \bibnamefont
  {Martin}}, \bibinfo {author} {\bibfnamefont {J.~R.}\ \bibnamefont
  {Williams}}, \bibinfo {author} {\bibfnamefont {B.~J.}\ \bibnamefont {Bloom}},
  \bibinfo {author} {\bibfnamefont {M.}~\bibnamefont {Bishof}}, \bibinfo
  {author} {\bibfnamefont {M.~D.}\ \bibnamefont {Swallows}}, \bibinfo {author}
  {\bibfnamefont {S.~L.}\ \bibnamefont {Campbell}},\ and\ \bibinfo {author}
  {\bibfnamefont {J.}~\bibnamefont {Ye}},\ }\bibfield  {title} {\bibinfo
  {title} {Comparison of two independent sr optical clocks with
  $1\mathbf{\ifmmode\times\else\texttimes\fi{}}{10}^{\ensuremath{-}17}$
  stability at ${10}^{3}\text{ }\text{ }\mathbf{s}$},\ }\href
  {https://doi.org/10.1103/PhysRevLett.109.230801} {\bibfield  {journal}
  {\bibinfo  {journal} {Phys. Rev. Lett.}\ }\textbf {\bibinfo {volume} {109}},\
  \bibinfo {pages} {230801} (\bibinfo {year} {2012})}\BibitemShut {NoStop}%
\bibitem [{\citenamefont {Lanyon}\ \emph {et~al.}(2013)\citenamefont {Lanyon},
  \citenamefont {Jurcevic}, \citenamefont {Hempel}, \citenamefont {Gessner},
  \citenamefont {Vedral}, \citenamefont {Blatt},\ and\ \citenamefont
  {Roos}}]{Lanyon:2013}%
  \BibitemOpen
  \bibfield  {author} {\bibinfo {author} {\bibfnamefont {B.~P.}\ \bibnamefont
  {Lanyon}}, \bibinfo {author} {\bibfnamefont {P.}~\bibnamefont {Jurcevic}},
  \bibinfo {author} {\bibfnamefont {C.}~\bibnamefont {Hempel}}, \bibinfo
  {author} {\bibfnamefont {M.}~\bibnamefont {Gessner}}, \bibinfo {author}
  {\bibfnamefont {V.}~\bibnamefont {Vedral}}, \bibinfo {author} {\bibfnamefont
  {R.}~\bibnamefont {Blatt}},\ and\ \bibinfo {author} {\bibfnamefont {C.~F.}\
  \bibnamefont {Roos}},\ }\bibfield  {title} {\bibinfo {title} {Experimental
  generation of quantum discord via noisy processes},\ }\href
  {https://doi.org/10.1103/PhysRevLett.111.100504} {\bibfield  {journal}
  {\bibinfo  {journal} {Phys. Rev. Lett.}\ }\textbf {\bibinfo {volume} {111}},\
  \bibinfo {pages} {100504} (\bibinfo {year} {2013})}\BibitemShut {NoStop}%
\bibitem [{\citenamefont {Chou}\ \emph {et~al.}(2011)\citenamefont {Chou},
  \citenamefont {Hume}, \citenamefont {Thorpe}, \citenamefont {Wineland},\ and\
  \citenamefont {Rosenband}}]{Chou:2011}%
  \BibitemOpen
  \bibfield  {author} {\bibinfo {author} {\bibfnamefont {C.~W.}\ \bibnamefont
  {Chou}}, \bibinfo {author} {\bibfnamefont {D.~B.}\ \bibnamefont {Hume}},
  \bibinfo {author} {\bibfnamefont {M.~J.}\ \bibnamefont {Thorpe}}, \bibinfo
  {author} {\bibfnamefont {D.~J.}\ \bibnamefont {Wineland}},\ and\ \bibinfo
  {author} {\bibfnamefont {T.}~\bibnamefont {Rosenband}},\ }\bibfield  {title}
  {\bibinfo {title} {Quantum coherence between two atoms beyond
  {Q}$={10}^{15}$},\ }\href {https://doi.org/10.1103/PhysRevLett.106.160801}
  {\bibfield  {journal} {\bibinfo  {journal} {Phys. Rev. Lett.}\ }\textbf
  {\bibinfo {volume} {106}},\ \bibinfo {pages} {160801} (\bibinfo {year}
  {2011})}\BibitemShut {NoStop}%
\bibitem [{\citenamefont {Marti}\ \emph {et~al.}(2018)\citenamefont {Marti},
  \citenamefont {Hutson}, \citenamefont {Goban}, \citenamefont {Campbell},
  \citenamefont {Poli},\ and\ \citenamefont {Ye}}]{Marti:2018}%
  \BibitemOpen
  \bibfield  {author} {\bibinfo {author} {\bibfnamefont {G.~E.}\ \bibnamefont
  {Marti}}, \bibinfo {author} {\bibfnamefont {R.~B.}\ \bibnamefont {Hutson}},
  \bibinfo {author} {\bibfnamefont {A.}~\bibnamefont {Goban}}, \bibinfo
  {author} {\bibfnamefont {S.~L.}\ \bibnamefont {Campbell}}, \bibinfo {author}
  {\bibfnamefont {N.}~\bibnamefont {Poli}},\ and\ \bibinfo {author}
  {\bibfnamefont {J.}~\bibnamefont {Ye}},\ }\bibfield  {title} {\bibinfo
  {title} {Imaging optical frequencies with $100\text{ }\text{
  }\ensuremath{\mu}\mathrm{Hz}$ precision and $1.1\text{ }\text{
  }\ensuremath{\mu}\mathrm{m}$ resolution},\ }\href
  {https://doi.org/10.1103/PhysRevLett.120.103201} {\bibfield  {journal}
  {\bibinfo  {journal} {Phys. Rev. Lett.}\ }\textbf {\bibinfo {volume} {120}},\
  \bibinfo {pages} {103201} (\bibinfo {year} {2018})}\BibitemShut {NoStop}%
\bibitem [{\citenamefont {Shaniv}\ \emph {et~al.}(2019)\citenamefont {Shaniv},
  \citenamefont {Akerman}, \citenamefont {Manovitz}, \citenamefont {Shapira},\
  and\ \citenamefont {Ozeri}}]{Shanif:2019}%
  \BibitemOpen
  \bibfield  {author} {\bibinfo {author} {\bibfnamefont {R.}~\bibnamefont
  {Shaniv}}, \bibinfo {author} {\bibfnamefont {N.}~\bibnamefont {Akerman}},
  \bibinfo {author} {\bibfnamefont {T.}~\bibnamefont {Manovitz}}, \bibinfo
  {author} {\bibfnamefont {Y.}~\bibnamefont {Shapira}},\ and\ \bibinfo {author}
  {\bibfnamefont {R.}~\bibnamefont {Ozeri}},\ }\bibfield  {title} {\bibinfo
  {title} {Quadrupole shift cancellation using dynamic decoupling},\ }\href
  {https://doi.org/10.1103/PhysRevLett.122.223204} {\bibfield  {journal}
  {\bibinfo  {journal} {Phys. Rev. Lett.}\ }\textbf {\bibinfo {volume} {122}},\
  \bibinfo {pages} {223204} (\bibinfo {year} {2019})}\BibitemShut {NoStop}%
\bibitem [{\citenamefont {Clements}\ \emph {et~al.}(2020)\citenamefont
  {Clements}, \citenamefont {Kim}, \citenamefont {Cui}, \citenamefont {Hankin},
  \citenamefont {Brewer}, \citenamefont {Valencia}, \citenamefont {Chen},
  \citenamefont {Chou}, \citenamefont {Leibrandt},\ and\ \citenamefont
  {Hume}}]{Clements:2020}%
  \BibitemOpen
  \bibfield  {author} {\bibinfo {author} {\bibfnamefont {E.~R.}\ \bibnamefont
  {Clements}}, \bibinfo {author} {\bibfnamefont {M.~E.}\ \bibnamefont {Kim}},
  \bibinfo {author} {\bibfnamefont {K.}~\bibnamefont {Cui}}, \bibinfo {author}
  {\bibfnamefont {A.~M.}\ \bibnamefont {Hankin}}, \bibinfo {author}
  {\bibfnamefont {S.~M.}\ \bibnamefont {Brewer}}, \bibinfo {author}
  {\bibfnamefont {J.}~\bibnamefont {Valencia}}, \bibinfo {author}
  {\bibfnamefont {J.-S.}\ \bibnamefont {Chen}}, \bibinfo {author}
  {\bibfnamefont {C.-W.}\ \bibnamefont {Chou}}, \bibinfo {author}
  {\bibfnamefont {D.~R.}\ \bibnamefont {Leibrandt}},\ and\ \bibinfo {author}
  {\bibfnamefont {D.~B.}\ \bibnamefont {Hume}},\ }\bibfield  {title} {\bibinfo
  {title} {Lifetime-limited interrogation of two independent $^{27}${A}l$^{+}$
  clocks using correlation spectroscopy},\ }\href
  {https://doi.org/10.1103/PhysRevLett.125.243602} {\bibfield  {journal}
  {\bibinfo  {journal} {Phys. Rev. Lett.}\ }\textbf {\bibinfo {volume} {125}},\
  \bibinfo {pages} {243602} (\bibinfo {year} {2020})}\BibitemShut {NoStop}%
\bibitem [{\citenamefont {Young}\ \emph {et~al.}(2020)\citenamefont {Young},
  \citenamefont {Eckner}, \citenamefont {Milner}, \citenamefont {Kedar},
  \citenamefont {Norcia}, \citenamefont {Oelker}, \citenamefont {Schine},
  \citenamefont {Ye},\ and\ \citenamefont {Kaufman}}]{Young:2020}%
  \BibitemOpen
  \bibfield  {author} {\bibinfo {author} {\bibfnamefont {A.~W.}\ \bibnamefont
  {Young}}, \bibinfo {author} {\bibfnamefont {W.~J.}\ \bibnamefont {Eckner}},
  \bibinfo {author} {\bibfnamefont {W.~R.}\ \bibnamefont {Milner}}, \bibinfo
  {author} {\bibfnamefont {D.}~\bibnamefont {Kedar}}, \bibinfo {author}
  {\bibfnamefont {M.~A.}\ \bibnamefont {Norcia}}, \bibinfo {author}
  {\bibfnamefont {E.}~\bibnamefont {Oelker}}, \bibinfo {author} {\bibfnamefont
  {N.}~\bibnamefont {Schine}}, \bibinfo {author} {\bibfnamefont
  {J.}~\bibnamefont {Ye}},\ and\ \bibinfo {author} {\bibfnamefont {A.~M.}\
  \bibnamefont {Kaufman}},\ }\bibfield  {title} {\bibinfo {title}
  {Half-minute-scale atomic coherence and high relative stability in a tweezer
  clock},\ }\href {https://doi.org/10.1038/s41586-020-3009-y} {\bibfield
  {journal} {\bibinfo  {journal} {Nature}\ }\textbf {\bibinfo {volume} {588}},\
  \bibinfo {pages} {408} (\bibinfo {year} {2020})}\BibitemShut {NoStop}%
\bibitem [{\citenamefont {Roos}\ \emph {et~al.}(2006)\citenamefont {Roos},
  \citenamefont {Chwalla}, \citenamefont {Kim}, \citenamefont {Riebe},\ and\
  \citenamefont {Blatt}}]{Roos:2006}%
  \BibitemOpen
  \bibfield  {author} {\bibinfo {author} {\bibfnamefont {C.~F.}\ \bibnamefont
  {Roos}}, \bibinfo {author} {\bibfnamefont {M.}~\bibnamefont {Chwalla}},
  \bibinfo {author} {\bibfnamefont {K.}~\bibnamefont {Kim}}, \bibinfo {author}
  {\bibfnamefont {M.}~\bibnamefont {Riebe}},\ and\ \bibinfo {author}
  {\bibfnamefont {R.}~\bibnamefont {Blatt}},\ }\bibfield  {title} {\bibinfo
  {title} {'{D}esigner atoms' for quantum metrology},\ }\href
  {https://doi.org/10.1038/nature05101} {\bibfield  {journal} {\bibinfo
  {journal} {Nature}\ }\textbf {\bibinfo {volume} {443}},\ \bibinfo {pages}
  {316} (\bibinfo {year} {2006})}\BibitemShut {NoStop}%
\bibitem [{\citenamefont {Megidish}\ \emph {et~al.}(2019)\citenamefont
  {Megidish}, \citenamefont {Broz}, \citenamefont {Greene},\ and\ \citenamefont
  {H\"affner}}]{Megidish:2019}%
  \BibitemOpen
  \bibfield  {author} {\bibinfo {author} {\bibfnamefont {E.}~\bibnamefont
  {Megidish}}, \bibinfo {author} {\bibfnamefont {J.}~\bibnamefont {Broz}},
  \bibinfo {author} {\bibfnamefont {N.}~\bibnamefont {Greene}},\ and\ \bibinfo
  {author} {\bibfnamefont {H.}~\bibnamefont {H\"affner}},\ }\bibfield  {title}
  {\bibinfo {title} {Improved test of local {L}orentz invariance from a
  deterministic preparation of entangled states},\ }\href
  {https://doi.org/10.1103/PhysRevLett.122.123605} {\bibfield  {journal}
  {\bibinfo  {journal} {Phys. Rev. Lett.}\ }\textbf {\bibinfo {volume} {122}},\
  \bibinfo {pages} {123605} (\bibinfo {year} {2019})}\BibitemShut {NoStop}%
\bibitem [{\citenamefont {Manovitz}\ \emph {et~al.}(2019)\citenamefont
  {Manovitz}, \citenamefont {Shaniv}, \citenamefont {Shapira}, \citenamefont
  {Ozeri},\ and\ \citenamefont {Akerman}}]{Manovitz:2019}%
  \BibitemOpen
  \bibfield  {author} {\bibinfo {author} {\bibfnamefont {T.}~\bibnamefont
  {Manovitz}}, \bibinfo {author} {\bibfnamefont {R.}~\bibnamefont {Shaniv}},
  \bibinfo {author} {\bibfnamefont {Y.}~\bibnamefont {Shapira}}, \bibinfo
  {author} {\bibfnamefont {R.}~\bibnamefont {Ozeri}},\ and\ \bibinfo {author}
  {\bibfnamefont {N.}~\bibnamefont {Akerman}},\ }\bibfield  {title} {\bibinfo
  {title} {Precision measurement of atomic isotope shifts using a two-isotope
  entangled state},\ }\href {https://doi.org/10.1103/PhysRevLett.123.203001}
  {\bibfield  {journal} {\bibinfo  {journal} {Phys. Rev. Lett.}\ }\textbf
  {\bibinfo {volume} {123}},\ \bibinfo {pages} {203001} (\bibinfo {year}
  {2019})}\BibitemShut {NoStop}%
\bibitem [{\citenamefont {Monz}\ \emph {et~al.}(2011)\citenamefont {Monz},
  \citenamefont {Schindler}, \citenamefont {Barreiro}, \citenamefont {Chwalla},
  \citenamefont {Nigg}, \citenamefont {Coish}, \citenamefont {Harlander},
  \citenamefont {H\"ansel}, \citenamefont {Hennrich},\ and\ \citenamefont
  {Blatt}}]{Monz:2011}%
  \BibitemOpen
  \bibfield  {author} {\bibinfo {author} {\bibfnamefont {T.}~\bibnamefont
  {Monz}}, \bibinfo {author} {\bibfnamefont {P.}~\bibnamefont {Schindler}},
  \bibinfo {author} {\bibfnamefont {J.~T.}\ \bibnamefont {Barreiro}}, \bibinfo
  {author} {\bibfnamefont {M.}~\bibnamefont {Chwalla}}, \bibinfo {author}
  {\bibfnamefont {D.}~\bibnamefont {Nigg}}, \bibinfo {author} {\bibfnamefont
  {W.~A.}\ \bibnamefont {Coish}}, \bibinfo {author} {\bibfnamefont
  {M.}~\bibnamefont {Harlander}}, \bibinfo {author} {\bibfnamefont
  {W.}~\bibnamefont {H\"ansel}}, \bibinfo {author} {\bibfnamefont
  {M.}~\bibnamefont {Hennrich}},\ and\ \bibinfo {author} {\bibfnamefont
  {R.}~\bibnamefont {Blatt}},\ }\bibfield  {title} {\bibinfo {title} {14-qubit
  entanglement: creation and coherence},\ }\href
  {https://doi.org/10.1103/PhysRevLett.106.130506} {\bibfield  {journal}
  {\bibinfo  {journal} {Phys.~Rev.~Lett.}\ }\textbf {\bibinfo {volume} {106}},\
  \bibinfo {pages} {130506} (\bibinfo {year} {2011})}\BibitemShut {NoStop}%
\bibitem [{\citenamefont {Bradley}\ \emph {et~al.}(2019)\citenamefont
  {Bradley}, \citenamefont {Randall}, \citenamefont {Abobeih}, \citenamefont
  {Berrevoets}, \citenamefont {Degen}, \citenamefont {Bakker}, \citenamefont
  {Markham}, \citenamefont {Twitchen},\ and\ \citenamefont
  {Taminiau}}]{Bradley:2019}%
  \BibitemOpen
  \bibfield  {author} {\bibinfo {author} {\bibfnamefont {C.~E.}\ \bibnamefont
  {Bradley}}, \bibinfo {author} {\bibfnamefont {J.}~\bibnamefont {Randall}},
  \bibinfo {author} {\bibfnamefont {M.~H.}\ \bibnamefont {Abobeih}}, \bibinfo
  {author} {\bibfnamefont {R.~C.}\ \bibnamefont {Berrevoets}}, \bibinfo
  {author} {\bibfnamefont {M.~J.}\ \bibnamefont {Degen}}, \bibinfo {author}
  {\bibfnamefont {M.~A.}\ \bibnamefont {Bakker}}, \bibinfo {author}
  {\bibfnamefont {M.}~\bibnamefont {Markham}}, \bibinfo {author} {\bibfnamefont
  {D.~J.}\ \bibnamefont {Twitchen}},\ and\ \bibinfo {author} {\bibfnamefont
  {T.~H.}\ \bibnamefont {Taminiau}},\ }\bibfield  {title} {\bibinfo {title} {A
  ten-qubit solid-state spin register with quantum memory up to one minute},\
  }\href {https://doi.org/10.1103/PhysRevX.9.031045} {\bibfield  {journal}
  {\bibinfo  {journal} {Phys. Rev. X}\ }\textbf {\bibinfo {volume} {9}},\
  \bibinfo {pages} {031045} (\bibinfo {year} {2019})}\BibitemShut {NoStop}%
\bibitem [{\citenamefont {Singh}\ \emph {et~al.}(2022)\citenamefont {Singh},
  \citenamefont {Bradley}, \citenamefont {Anand}, \citenamefont {Ramesh},
  \citenamefont {White},\ and\ \citenamefont {Bernien}}]{Singh:2022}%
  \BibitemOpen
  \bibfield  {author} {\bibinfo {author} {\bibfnamefont {K.}~\bibnamefont
  {Singh}}, \bibinfo {author} {\bibfnamefont {C.~E.}\ \bibnamefont {Bradley}},
  \bibinfo {author} {\bibfnamefont {S.}~\bibnamefont {Anand}}, \bibinfo
  {author} {\bibfnamefont {V.}~\bibnamefont {Ramesh}}, \bibinfo {author}
  {\bibfnamefont {R.}~\bibnamefont {White}},\ and\ \bibinfo {author}
  {\bibfnamefont {H.}~\bibnamefont {Bernien}},\ }\bibfield  {title} {\bibinfo
  {title} {Mid-circuit correction of correlated phase errors using an array of
  spectator qubits},\ }\bibfield  {journal} {\bibinfo  {journal}
  {arXiv:2208.11716}\ }\href {https://doi.org/10.48550/ARXIV.2208.11716}
  {10.48550/ARXIV.2208.11716} (\bibinfo {year} {2022})\BibitemShut {NoStop}%
\bibitem [{\citenamefont {Braverman}\ \emph {et~al.}(2018)\citenamefont
  {Braverman}, \citenamefont {Kawasaki},\ and\ \citenamefont
  {Vuleti{\'{c}}}}]{Braverman:2018}%
  \BibitemOpen
  \bibfield  {author} {\bibinfo {author} {\bibfnamefont {B.}~\bibnamefont
  {Braverman}}, \bibinfo {author} {\bibfnamefont {A.}~\bibnamefont
  {Kawasaki}},\ and\ \bibinfo {author} {\bibfnamefont {V.}~\bibnamefont
  {Vuleti{\'{c}}}},\ }\bibfield  {title} {\bibinfo {title} {Impact of
  non-unitary spin squeezing on atomic clock performance},\ }\href
  {https://doi.org/10.1088/1367-2630/aae563} {\bibfield  {journal} {\bibinfo
  {journal} {New J. Phys.}\ }\textbf {\bibinfo {volume} {20}},\ \bibinfo
  {pages} {103019} (\bibinfo {year} {2018})}\BibitemShut {NoStop}%
\bibitem [{\citenamefont {Sawyer}\ \emph {et~al.}(2012)\citenamefont {Sawyer},
  \citenamefont {Britton}, \citenamefont {Keith}, \citenamefont {Wang},
  \citenamefont {Freericks}, \citenamefont {Uys}, \citenamefont {Biercuk},\
  and\ \citenamefont {Bollinger}}]{Sawyer:2012}%
  \BibitemOpen
  \bibfield  {author} {\bibinfo {author} {\bibfnamefont {B.~C.}\ \bibnamefont
  {Sawyer}}, \bibinfo {author} {\bibfnamefont {J.~W.}\ \bibnamefont {Britton}},
  \bibinfo {author} {\bibfnamefont {A.~C.}\ \bibnamefont {Keith}}, \bibinfo
  {author} {\bibfnamefont {C.-C.~J.}\ \bibnamefont {Wang}}, \bibinfo {author}
  {\bibfnamefont {J.~K.}\ \bibnamefont {Freericks}}, \bibinfo {author}
  {\bibfnamefont {H.}~\bibnamefont {Uys}}, \bibinfo {author} {\bibfnamefont
  {M.~J.}\ \bibnamefont {Biercuk}},\ and\ \bibinfo {author} {\bibfnamefont
  {J.~J.}\ \bibnamefont {Bollinger}},\ }\bibfield  {title} {\bibinfo {title}
  {Spectroscopy and thermometry of drumhead modes in a mesoscopic trapped-ion
  crystal using entanglement},\ }\href
  {https://doi.org/10.1103/PhysRevLett.108.213003} {\bibfield  {journal}
  {\bibinfo  {journal} {Phys. Rev. Lett.}\ }\textbf {\bibinfo {volume} {108}},\
  \bibinfo {pages} {213003} (\bibinfo {year} {2012})}\BibitemShut {NoStop}%
\bibitem [{\citenamefont {von L\"upke}\ \emph {et~al.}(2020)\citenamefont {von
  L\"upke}, \citenamefont {Beaudoin}, \citenamefont {Norris}, \citenamefont
  {Sung}, \citenamefont {Winik}, \citenamefont {Qiu}, \citenamefont
  {Kjaergaard}, \citenamefont {Kim}, \citenamefont {Yoder}, \citenamefont
  {Gustavsson}, \citenamefont {Viola},\ and\ \citenamefont
  {Oliver}}]{vonLuepke:2020}%
  \BibitemOpen
  \bibfield  {author} {\bibinfo {author} {\bibfnamefont {U.}~\bibnamefont {von
  L\"upke}}, \bibinfo {author} {\bibfnamefont {F.}~\bibnamefont {Beaudoin}},
  \bibinfo {author} {\bibfnamefont {L.~M.}\ \bibnamefont {Norris}}, \bibinfo
  {author} {\bibfnamefont {Y.}~\bibnamefont {Sung}}, \bibinfo {author}
  {\bibfnamefont {R.}~\bibnamefont {Winik}}, \bibinfo {author} {\bibfnamefont
  {J.~Y.}\ \bibnamefont {Qiu}}, \bibinfo {author} {\bibfnamefont
  {M.}~\bibnamefont {Kjaergaard}}, \bibinfo {author} {\bibfnamefont
  {D.}~\bibnamefont {Kim}}, \bibinfo {author} {\bibfnamefont {J.}~\bibnamefont
  {Yoder}}, \bibinfo {author} {\bibfnamefont {S.}~\bibnamefont {Gustavsson}},
  \bibinfo {author} {\bibfnamefont {L.}~\bibnamefont {Viola}},\ and\ \bibinfo
  {author} {\bibfnamefont {W.~D.}\ \bibnamefont {Oliver}},\ }\bibfield  {title}
  {\bibinfo {title} {Two-qubit spectroscopy of spatiotemporally correlated
  quantum noise in superconducting qubits},\ }\href
  {https://doi.org/10.1103/PRXQuantum.1.010305} {\bibfield  {journal} {\bibinfo
   {journal} {PRX Quantum}\ }\textbf {\bibinfo {volume} {1}},\ \bibinfo {pages}
  {010305} (\bibinfo {year} {2020})}\BibitemShut {NoStop}%
\bibitem [{\citenamefont {Kawamura}\ and\ \citenamefont
  {Chen}(2004)}]{kawamura2004displacement}%
  \BibitemOpen
  \bibfield  {author} {\bibinfo {author} {\bibfnamefont {S.}~\bibnamefont
  {Kawamura}}\ and\ \bibinfo {author} {\bibfnamefont {Y.}~\bibnamefont
  {Chen}},\ }\bibfield  {title} {\bibinfo {title} {Displacement-noise-free
  gravitational-wave detection},\ }\href
  {https://doi.org/10.1103/PhysRevLett.93.211103} {\bibfield  {journal}
  {\bibinfo  {journal} {Phys.~Rev.~Lett.}\ }\textbf {\bibinfo {volume} {93}},\
  \bibinfo {pages} {211103} (\bibinfo {year} {2004})}\BibitemShut {NoStop}%
\bibitem [{\citenamefont {Chen}\ \emph {et~al.}(2006)\citenamefont {Chen},
  \citenamefont {Pai}, \citenamefont {Somiya}, \citenamefont {Kawamura},
  \citenamefont {Sato}, \citenamefont {Kokeyama}, \citenamefont {Ward},
  \citenamefont {Goda},\ and\ \citenamefont
  {Mikhailov}}]{chen2006interferometers}%
  \BibitemOpen
  \bibfield  {author} {\bibinfo {author} {\bibfnamefont {Y.}~\bibnamefont
  {Chen}}, \bibinfo {author} {\bibfnamefont {A.}~\bibnamefont {Pai}}, \bibinfo
  {author} {\bibfnamefont {K.}~\bibnamefont {Somiya}}, \bibinfo {author}
  {\bibfnamefont {S.}~\bibnamefont {Kawamura}}, \bibinfo {author}
  {\bibfnamefont {S.}~\bibnamefont {Sato}}, \bibinfo {author} {\bibfnamefont
  {K.}~\bibnamefont {Kokeyama}}, \bibinfo {author} {\bibfnamefont {R.~L.}\
  \bibnamefont {Ward}}, \bibinfo {author} {\bibfnamefont {K.}~\bibnamefont
  {Goda}},\ and\ \bibinfo {author} {\bibfnamefont {E.~E.}\ \bibnamefont
  {Mikhailov}},\ }\bibfield  {title} {\bibinfo {title} {Interferometers for
  displacement-noise-free gravitational-wave detection},\ }\href
  {https://doi.org/10.1103/PhysRevLett.97.151103} {\bibfield  {journal}
  {\bibinfo  {journal} {Phys.~Rev.~Lett.}\ }\textbf {\bibinfo {volume} {97}},\
  \bibinfo {pages} {151103} (\bibinfo {year} {2006})}\BibitemShut {NoStop}%
\bibitem [{\citenamefont {Gefen}\ \emph {et~al.}(2022)\citenamefont {Gefen},
  \citenamefont {Tarafder}, \citenamefont {Adhikari},\ and\ \citenamefont
  {Chen}}]{gefen2022quantum}%
  \BibitemOpen
  \bibfield  {author} {\bibinfo {author} {\bibfnamefont {T.}~\bibnamefont
  {Gefen}}, \bibinfo {author} {\bibfnamefont {R.}~\bibnamefont {Tarafder}},
  \bibinfo {author} {\bibfnamefont {R.~X.}\ \bibnamefont {Adhikari}},\ and\
  \bibinfo {author} {\bibfnamefont {Y.}~\bibnamefont {Chen}},\ }\bibfield
  {title} {\bibinfo {title} {Quantum precision limits of displacement noise
  free interferometers},\ }\bibfield  {journal} {\bibinfo  {journal}
  {arXiv:2209.02998}\ }\href {https://doi.org/10.48550/arXiv.2209.02998}
  {10.48550/arXiv.2209.02998} (\bibinfo {year} {2022})\BibitemShut {NoStop}%
\bibitem [{\citenamefont {Modi}\ \emph {et~al.}(2012)\citenamefont {Modi},
  \citenamefont {Brodutch}, \citenamefont {Cable}, \citenamefont {Paterek},\
  and\ \citenamefont {Vedral}}]{Modi:2012}%
  \BibitemOpen
  \bibfield  {author} {\bibinfo {author} {\bibfnamefont {K.}~\bibnamefont
  {Modi}}, \bibinfo {author} {\bibfnamefont {A.}~\bibnamefont {Brodutch}},
  \bibinfo {author} {\bibfnamefont {H.}~\bibnamefont {Cable}}, \bibinfo
  {author} {\bibfnamefont {T.}~\bibnamefont {Paterek}},\ and\ \bibinfo {author}
  {\bibfnamefont {V.}~\bibnamefont {Vedral}},\ }\bibfield  {title} {\bibinfo
  {title} {The classical-quantum boundary for correlations: Discord and related
  measures},\ }\href {https://doi.org/10.1103/RevModPhys.84.1655} {\bibfield
  {journal} {\bibinfo  {journal} {Rev. Mod. Phys.}\ }\textbf {\bibinfo {volume}
  {84}},\ \bibinfo {pages} {1655} (\bibinfo {year} {2012})}\BibitemShut
  {NoStop}%
\bibitem [{\citenamefont {Proctor}\ \emph {et~al.}(2018)\citenamefont
  {Proctor}, \citenamefont {Knott},\ and\ \citenamefont
  {Dunningham}}]{distributed1}%
  \BibitemOpen
  \bibfield  {author} {\bibinfo {author} {\bibfnamefont {T.~J.}\ \bibnamefont
  {Proctor}}, \bibinfo {author} {\bibfnamefont {P.~A.}\ \bibnamefont {Knott}},\
  and\ \bibinfo {author} {\bibfnamefont {J.~A.}\ \bibnamefont {Dunningham}},\
  }\bibfield  {title} {\bibinfo {title} {Multiparameter estimation in networked
  quantum sensors},\ }\href {https://doi.org/10.1103/PhysRevLett.120.080501}
  {\bibfield  {journal} {\bibinfo  {journal} {Phys.~Rev.~Lett.}\ }\textbf
  {\bibinfo {volume} {120}},\ \bibinfo {pages} {080501} (\bibinfo {year}
  {2018})}\BibitemShut {NoStop}%
\bibitem [{\citenamefont {Ge}\ \emph {et~al.}(2018)\citenamefont {Ge},
  \citenamefont {Jacobs}, \citenamefont {Eldredge}, \citenamefont {Gorshkov},\
  and\ \citenamefont {Foss-Feig}}]{distributed2}%
  \BibitemOpen
  \bibfield  {author} {\bibinfo {author} {\bibfnamefont {W.}~\bibnamefont
  {Ge}}, \bibinfo {author} {\bibfnamefont {K.}~\bibnamefont {Jacobs}}, \bibinfo
  {author} {\bibfnamefont {Z.}~\bibnamefont {Eldredge}}, \bibinfo {author}
  {\bibfnamefont {A.~V.}\ \bibnamefont {Gorshkov}},\ and\ \bibinfo {author}
  {\bibfnamefont {M.}~\bibnamefont {Foss-Feig}},\ }\bibfield  {title} {\bibinfo
  {title} {Distributed quantum metrology with linear networks and separable
  inputs},\ }\href {https://doi.org/10.1103/PhysRevLett.121.043604} {\bibfield
  {journal} {\bibinfo  {journal} {Phys.~Rev~Lett.}\ }\textbf {\bibinfo {volume}
  {121}},\ \bibinfo {pages} {043604} (\bibinfo {year} {2018})}\BibitemShut
  {NoStop}%
\bibitem [{\citenamefont {Eldredge}\ \emph {et~al.}(2018)\citenamefont
  {Eldredge}, \citenamefont {Foss-Feig}, \citenamefont {Gross}, \citenamefont
  {Rolston},\ and\ \citenamefont {Gorshkov}}]{eldredge2018optimal}%
  \BibitemOpen
  \bibfield  {author} {\bibinfo {author} {\bibfnamefont {Z.}~\bibnamefont
  {Eldredge}}, \bibinfo {author} {\bibfnamefont {M.}~\bibnamefont {Foss-Feig}},
  \bibinfo {author} {\bibfnamefont {J.~A.}\ \bibnamefont {Gross}}, \bibinfo
  {author} {\bibfnamefont {S.~L.}\ \bibnamefont {Rolston}},\ and\ \bibinfo
  {author} {\bibfnamefont {A.~V.}\ \bibnamefont {Gorshkov}},\ }\bibfield
  {title} {\bibinfo {title} {Optimal and secure measurement protocols for
  quantum sensor networks},\ }\href
  {https://doi.org/10.1103/PhysRevA.97.042337} {\bibfield  {journal} {\bibinfo
  {journal} {Phys.~Rev.~A}\ }\textbf {\bibinfo {volume} {97}},\ \bibinfo
  {pages} {042337} (\bibinfo {year} {2018})}\BibitemShut {NoStop}%
\bibitem [{\citenamefont {Rubio}\ \emph {et~al.}(2020)\citenamefont {Rubio},
  \citenamefont {Knott}, \citenamefont {Proctor},\ and\ \citenamefont
  {Dunningham}}]{rubio2020quantum}%
  \BibitemOpen
  \bibfield  {author} {\bibinfo {author} {\bibfnamefont {J.}~\bibnamefont
  {Rubio}}, \bibinfo {author} {\bibfnamefont {P.~A.}\ \bibnamefont {Knott}},
  \bibinfo {author} {\bibfnamefont {T.~J.}\ \bibnamefont {Proctor}},\ and\
  \bibinfo {author} {\bibfnamefont {J.~A.}\ \bibnamefont {Dunningham}},\
  }\bibfield  {title} {\bibinfo {title} {Quantum sensing networks for the
  estimation of linear functions},\ }\href
  {https://doi.org/10.1088/1751-8121/ab9d46} {\bibfield  {journal} {\bibinfo
  {journal} {J.~Phys.~A~Math.~Theor.}\ }\textbf {\bibinfo {volume} {53}},\
  \bibinfo {pages} {344001} (\bibinfo {year} {2020})}\BibitemShut {NoStop}%
\bibitem [{\citenamefont {Kiesenhofer}\ \emph {et~al.}(2023)\citenamefont
  {Kiesenhofer}, \citenamefont {Hainzer}, \citenamefont {Zhdanov},
  \citenamefont {Holz}, \citenamefont {Bock}, \citenamefont {Ollikainen},\ and\
  \citenamefont {Roos}}]{Kiesenhofer:2023a}%
  \BibitemOpen
  \bibfield  {author} {\bibinfo {author} {\bibfnamefont {D.}~\bibnamefont
  {Kiesenhofer}}, \bibinfo {author} {\bibfnamefont {H.}~\bibnamefont
  {Hainzer}}, \bibinfo {author} {\bibfnamefont {A.}~\bibnamefont {Zhdanov}},
  \bibinfo {author} {\bibfnamefont {P.~C.}\ \bibnamefont {Holz}}, \bibinfo
  {author} {\bibfnamefont {M.}~\bibnamefont {Bock}}, \bibinfo {author}
  {\bibfnamefont {T.}~\bibnamefont {Ollikainen}},\ and\ \bibinfo {author}
  {\bibfnamefont {C.~F.}\ \bibnamefont {Roos}},\ }\bibfield  {title} {\bibinfo
  {title} {Controlling two-dimensional {C}oulomb crystals of more than 100 ions
  in a monolithic radio-frequency trap},\ }\bibfield  {journal} {\bibinfo
  {journal} {arXiv:2302.00565}\ }\href
  {https://doi.org/10.48550/arXiv.2302.00565} {10.48550/arXiv.2302.00565}
  (\bibinfo {year} {2023})\BibitemShut {NoStop}%
\bibitem [{\citenamefont {Joshi}\ \emph {et~al.}(2020)\citenamefont {Joshi},
  \citenamefont {Fabre}, \citenamefont {Maier}, \citenamefont {Brydges},
  \citenamefont {Kiesenhofer}, \citenamefont {Hainzer}, \citenamefont {Blatt},\
  and\ \citenamefont {Roos}}]{Joshi:2020}%
  \BibitemOpen
  \bibfield  {author} {\bibinfo {author} {\bibfnamefont {M.~K.}\ \bibnamefont
  {Joshi}}, \bibinfo {author} {\bibfnamefont {A.}~\bibnamefont {Fabre}},
  \bibinfo {author} {\bibfnamefont {C.}~\bibnamefont {Maier}}, \bibinfo
  {author} {\bibfnamefont {T.}~\bibnamefont {Brydges}}, \bibinfo {author}
  {\bibfnamefont {D.}~\bibnamefont {Kiesenhofer}}, \bibinfo {author}
  {\bibfnamefont {H.}~\bibnamefont {Hainzer}}, \bibinfo {author} {\bibfnamefont
  {R.}~\bibnamefont {Blatt}},\ and\ \bibinfo {author} {\bibfnamefont {C.~F.}\
  \bibnamefont {Roos}},\ }\bibfield  {title} {\bibinfo {title}
  {Polarization-gradient cooling of {1D} and {2D} ion {C}oulomb crystals},\
  }\href {https://doi.org/10.1088/1367-2630/abb912} {\bibfield  {journal}
  {\bibinfo  {journal} {New.~J.~Phys.}\ }\textbf {\bibinfo {volume} {22}},\
  \bibinfo {pages} {103013} (\bibinfo {year} {2020})}\BibitemShut {NoStop}%
\bibitem [{\citenamefont {Kranzl}\ \emph {et~al.}(2022)\citenamefont {Kranzl},
  \citenamefont {Joshi}, \citenamefont {Maier}, \citenamefont {Brydges},
  \citenamefont {Franke}, \citenamefont {Blatt},\ and\ \citenamefont
  {Roos}}]{Kranzl:2022}%
  \BibitemOpen
  \bibfield  {author} {\bibinfo {author} {\bibfnamefont {F.}~\bibnamefont
  {Kranzl}}, \bibinfo {author} {\bibfnamefont {M.~K.}\ \bibnamefont {Joshi}},
  \bibinfo {author} {\bibfnamefont {C.}~\bibnamefont {Maier}}, \bibinfo
  {author} {\bibfnamefont {T.}~\bibnamefont {Brydges}}, \bibinfo {author}
  {\bibfnamefont {J.}~\bibnamefont {Franke}}, \bibinfo {author} {\bibfnamefont
  {R.}~\bibnamefont {Blatt}},\ and\ \bibinfo {author} {\bibfnamefont {C.~F.}\
  \bibnamefont {Roos}},\ }\bibfield  {title} {\bibinfo {title} {Controlling
  long ion strings for quantum simulation and precision measurements},\ }\href
  {https://doi.org/10.1103/PhysRevA.105.052426} {\bibfield  {journal} {\bibinfo
   {journal} {Phys. Rev. A}\ }\textbf {\bibinfo {volume} {105}},\ \bibinfo
  {pages} {052426} (\bibinfo {year} {2022})}\BibitemShut {NoStop}%
\bibitem [{\citenamefont {Schmidt}\ \emph {et~al.}(2009)\citenamefont
  {Schmidt}, \citenamefont {van~den Berg}, \citenamefont {Friedlander},\ and\
  \citenamefont {Murphy}}]{Schmidt:2009}%
  \BibitemOpen
  \bibfield  {author} {\bibinfo {author} {\bibfnamefont {M.}~\bibnamefont
  {Schmidt}}, \bibinfo {author} {\bibfnamefont {E.}~\bibnamefont {van~den
  Berg}}, \bibinfo {author} {\bibfnamefont {M.}~\bibnamefont {Friedlander}},\
  and\ \bibinfo {author} {\bibfnamefont {K.}~\bibnamefont {Murphy}},\
  }\bibfield  {title} {\bibinfo {title} {Optimizing costly functions with
  simple constraints: A limited-memory projected quasi-newton algorithm},\ }in\
  \href@noop {} {\emph {\bibinfo {booktitle} {Proceedings of the 12$^{th}$
  International Conference on Articial Intelligence and Statistics
  (AISTATS)}}}\ (\bibinfo {year} {2009})\BibitemShut {NoStop}%
\bibitem [{\citenamefont {Hainzer}\ \emph {et~al.}(2022)\citenamefont
  {Hainzer}, \citenamefont {Kiesenhofer}, \citenamefont {Ollikainen},
  \citenamefont {Bock}, \citenamefont {Kranzl}, \citenamefont {Joshi},
  \citenamefont {Yoeli}, \citenamefont {Blatt}, \citenamefont {Gefen},\ and\
  \citenamefont {Roos}}]{Hainzer:2022Zenodo}%
  \BibitemOpen
  \bibfield  {author} {\bibinfo {author} {\bibfnamefont {H.}~\bibnamefont
  {Hainzer}}, \bibinfo {author} {\bibfnamefont {D.}~\bibnamefont
  {Kiesenhofer}}, \bibinfo {author} {\bibfnamefont {T.}~\bibnamefont
  {Ollikainen}}, \bibinfo {author} {\bibfnamefont {M.}~\bibnamefont {Bock}},
  \bibinfo {author} {\bibfnamefont {F.}~\bibnamefont {Kranzl}}, \bibinfo
  {author} {\bibfnamefont {M.~K.}\ \bibnamefont {Joshi}}, \bibinfo {author}
  {\bibfnamefont {G.}~\bibnamefont {Yoeli}}, \bibinfo {author} {\bibfnamefont
  {R.}~\bibnamefont {Blatt}}, \bibinfo {author} {\bibfnamefont
  {T.}~\bibnamefont {Gefen}},\ and\ \bibinfo {author} {\bibfnamefont {C.~F.}\
  \bibnamefont {Roos}},\ }\href {https://doi.org/10.5281/ZENODO.6396595}
  {\bibinfo {title} {Correlation spectroscopy with multi-qubit-enhanced phase
  estimation: data. {Z}enodo}} (\bibinfo {year} {2022})\BibitemShut {NoStop}%
\bibitem [{\citenamefont {Holtzman}(1950)}]{Holtzman:1950}%
  \BibitemOpen
  \bibfield  {author} {\bibinfo {author} {\bibfnamefont {W.~H.}\ \bibnamefont
  {Holtzman}},\ }\bibfield  {title} {\bibinfo {title} {The unbiased estimate of
  the population variance and standard deviation},\ }\href
  {https://doi.org/10.2307/1418879} {\bibfield  {journal} {\bibinfo  {journal}
  {The American Journal of Psychology}\ }\textbf {\bibinfo {volume} {63}},\
  \bibinfo {pages} {615} (\bibinfo {year} {1950})}\BibitemShut {NoStop}%
\bibitem [{\citenamefont {Cover}(1999)}]{cover1999elements}%
  \BibitemOpen
  \bibfield  {author} {\bibinfo {author} {\bibfnamefont {T.~M.}\ \bibnamefont
  {Cover}},\ }\href {https://doi.org/10.1002/047174882X} {\emph {\bibinfo
  {title} {Elements of information theory}}}\ (\bibinfo  {publisher} {John
  Wiley \& Sons},\ \bibinfo {year} {1999})\BibitemShut {NoStop}%
\bibitem [{\citenamefont {Braunstein}\ and\ \citenamefont
  {Caves}(1994)}]{braunstein1994statistical}%
  \BibitemOpen
  \bibfield  {author} {\bibinfo {author} {\bibfnamefont {S.~L.}\ \bibnamefont
  {Braunstein}}\ and\ \bibinfo {author} {\bibfnamefont {C.~M.}\ \bibnamefont
  {Caves}},\ }\bibfield  {title} {\bibinfo {title} {Statistical distance and
  the geometry of quantum states},\ }\href
  {https://doi.org/10.1103/PhysRevLett.72.3439} {\bibfield  {journal} {\bibinfo
   {journal} {Phys.~Rev.~Lett.}\ }\textbf {\bibinfo {volume} {72}},\ \bibinfo
  {pages} {3439} (\bibinfo {year} {1994})}\BibitemShut {NoStop}%
\bibitem [{\citenamefont {Liu}\ \emph {et~al.}(2019)\citenamefont {Liu},
  \citenamefont {Yuan}, \citenamefont {Lu},\ and\ \citenamefont
  {Wang}}]{liu2019quantum}%
  \BibitemOpen
  \bibfield  {author} {\bibinfo {author} {\bibfnamefont {J.}~\bibnamefont
  {Liu}}, \bibinfo {author} {\bibfnamefont {H.}~\bibnamefont {Yuan}}, \bibinfo
  {author} {\bibfnamefont {X.-M.}\ \bibnamefont {Lu}},\ and\ \bibinfo {author}
  {\bibfnamefont {X.}~\bibnamefont {Wang}},\ }\bibfield  {title} {\bibinfo
  {title} {Quantum fisher information matrix and multiparameter estimation},\
  }\href {https://doi.org/10.1088/1751-8121/ab5d4d} {\bibfield  {journal}
  {\bibinfo  {journal} {J.~Phys.~A~Math.~Theor.}\ }\textbf {\bibinfo {volume}
  {53}},\ \bibinfo {pages} {023001} (\bibinfo {year} {2019})}\BibitemShut
  {NoStop}%
\bibitem [{\citenamefont {Bothwell}\ \emph {et~al.}(2022)\citenamefont
  {Bothwell}, \citenamefont {Kennedy}, \citenamefont {Aeppli}, \citenamefont
  {Kedar}, \citenamefont {Robinson}, \citenamefont {Oelker}, \citenamefont
  {Staron},\ and\ \citenamefont {Ye}}]{Bothwell:2022}%
  \BibitemOpen
  \bibfield  {author} {\bibinfo {author} {\bibfnamefont {T.}~\bibnamefont
  {Bothwell}}, \bibinfo {author} {\bibfnamefont {C.~J.}\ \bibnamefont
  {Kennedy}}, \bibinfo {author} {\bibfnamefont {A.}~\bibnamefont {Aeppli}},
  \bibinfo {author} {\bibfnamefont {D.}~\bibnamefont {Kedar}}, \bibinfo
  {author} {\bibfnamefont {J.~M.}\ \bibnamefont {Robinson}}, \bibinfo {author}
  {\bibfnamefont {E.}~\bibnamefont {Oelker}}, \bibinfo {author} {\bibfnamefont
  {A.}~\bibnamefont {Staron}},\ and\ \bibinfo {author} {\bibfnamefont
  {J.}~\bibnamefont {Ye}},\ }\bibfield  {title} {\bibinfo {title} {Resolving
  the gravitational redshift across a millimetre-scale atomic sample},\ }\href
  {https://doi.org/10.1038/s41586-021-04349-7} {\bibfield  {journal} {\bibinfo
  {journal} {Nature}\ }\textbf {\bibinfo {volume} {602}},\ \bibinfo {pages}
  {420} (\bibinfo {year} {2022})}\BibitemShut {NoStop}%
\bibitem [{\citenamefont {Zheng}\ \emph {et~al.}(2022)\citenamefont {Zheng},
  \citenamefont {Dolde}, \citenamefont {Lochab}, \citenamefont {Merriman},
  \citenamefont {Li},\ and\ \citenamefont {Kolkowitz}}]{Zheng:2022}%
  \BibitemOpen
  \bibfield  {author} {\bibinfo {author} {\bibfnamefont {X.}~\bibnamefont
  {Zheng}}, \bibinfo {author} {\bibfnamefont {J.}~\bibnamefont {Dolde}},
  \bibinfo {author} {\bibfnamefont {V.}~\bibnamefont {Lochab}}, \bibinfo
  {author} {\bibfnamefont {B.~N.}\ \bibnamefont {Merriman}}, \bibinfo {author}
  {\bibfnamefont {H.}~\bibnamefont {Li}},\ and\ \bibinfo {author}
  {\bibfnamefont {S.}~\bibnamefont {Kolkowitz}},\ }\bibfield  {title} {\bibinfo
  {title} {Differential clock comparisons with a multiplexed optical lattice
  clock},\ }\href {https://doi.org/10.1038/s41586-021-04344-y} {\bibfield
  {journal} {\bibinfo  {journal} {Nature}\ }\textbf {\bibinfo {volume} {602}},\
  \bibinfo {pages} {425} (\bibinfo {year} {2022})}\BibitemShut {NoStop}%
\bibitem [{\citenamefont {Madjarov}\ \emph {et~al.}(2019)\citenamefont
  {Madjarov}, \citenamefont {Cooper}, \citenamefont {Shaw}, \citenamefont
  {Covey}, \citenamefont {Schkolnik}, \citenamefont {Yoon}, \citenamefont
  {Williams},\ and\ \citenamefont {Endres}}]{Madjarov:2019}%
  \BibitemOpen
  \bibfield  {author} {\bibinfo {author} {\bibfnamefont {I.~S.}\ \bibnamefont
  {Madjarov}}, \bibinfo {author} {\bibfnamefont {A.}~\bibnamefont {Cooper}},
  \bibinfo {author} {\bibfnamefont {A.~L.}\ \bibnamefont {Shaw}}, \bibinfo
  {author} {\bibfnamefont {J.~P.}\ \bibnamefont {Covey}}, \bibinfo {author}
  {\bibfnamefont {V.}~\bibnamefont {Schkolnik}}, \bibinfo {author}
  {\bibfnamefont {T.~H.}\ \bibnamefont {Yoon}}, \bibinfo {author}
  {\bibfnamefont {J.~R.}\ \bibnamefont {Williams}},\ and\ \bibinfo {author}
  {\bibfnamefont {M.}~\bibnamefont {Endres}},\ }\bibfield  {title} {\bibinfo
  {title} {An atomic-array optical clock with single-atom readout},\ }\href
  {https://doi.org/10.1103/PhysRevX.9.041052} {\bibfield  {journal} {\bibinfo
  {journal} {Phys. Rev. X}\ }\textbf {\bibinfo {volume} {9}},\ \bibinfo {pages}
  {041052} (\bibinfo {year} {2019})}\BibitemShut {NoStop}%
\bibitem [{\citenamefont {K{\'{o}}m{\'{a}}r}\ \emph {et~al.}(2014)\citenamefont
  {K{\'{o}}m{\'{a}}r}, \citenamefont {Kessler}, \citenamefont {Bishof},
  \citenamefont {Jiang}, \citenamefont {S{\o}rensen}, \citenamefont {Ye},\ and\
  \citenamefont {Lukin}}]{Komar:2014}%
  \BibitemOpen
  \bibfield  {author} {\bibinfo {author} {\bibfnamefont {P.}~\bibnamefont
  {K{\'{o}}m{\'{a}}r}}, \bibinfo {author} {\bibfnamefont {E.~M.}\ \bibnamefont
  {Kessler}}, \bibinfo {author} {\bibfnamefont {M.}~\bibnamefont {Bishof}},
  \bibinfo {author} {\bibfnamefont {L.}~\bibnamefont {Jiang}}, \bibinfo
  {author} {\bibfnamefont {A.~S.}\ \bibnamefont {S{\o}rensen}}, \bibinfo
  {author} {\bibfnamefont {J.}~\bibnamefont {Ye}},\ and\ \bibinfo {author}
  {\bibfnamefont {M.~D.}\ \bibnamefont {Lukin}},\ }\bibfield  {title} {\bibinfo
  {title} {A quantum network of clocks},\ }\href
  {https://doi.org/10.1038/nphys3000} {\bibfield  {journal} {\bibinfo
  {journal} {Nat.~Phys.}\ }\textbf {\bibinfo {volume} {10}},\ \bibinfo {pages}
  {582} (\bibinfo {year} {2014})}\BibitemShut {NoStop}%
\bibitem [{\citenamefont {Nichol}\ \emph {et~al.}(2022)\citenamefont {Nichol},
  \citenamefont {Srinivas}, \citenamefont {Nadlinger}, \citenamefont {Drmota},
  \citenamefont {Main}, \citenamefont {Araneda}, \citenamefont {Ballance},\
  and\ \citenamefont {Lucas}}]{Nichol:2022}%
  \BibitemOpen
  \bibfield  {author} {\bibinfo {author} {\bibfnamefont {B.~C.}\ \bibnamefont
  {Nichol}}, \bibinfo {author} {\bibfnamefont {R.}~\bibnamefont {Srinivas}},
  \bibinfo {author} {\bibfnamefont {D.~P.}\ \bibnamefont {Nadlinger}}, \bibinfo
  {author} {\bibfnamefont {P.}~\bibnamefont {Drmota}}, \bibinfo {author}
  {\bibfnamefont {D.}~\bibnamefont {Main}}, \bibinfo {author} {\bibfnamefont
  {G.}~\bibnamefont {Araneda}}, \bibinfo {author} {\bibfnamefont {C.~J.}\
  \bibnamefont {Ballance}},\ and\ \bibinfo {author} {\bibfnamefont {D.~M.}\
  \bibnamefont {Lucas}},\ }\bibfield  {title} {\bibinfo {title} {An elementary
  quantum network of entangled optical atomic clocks},\ }\href
  {https://doi.org/10.1038/s41586-022-05088-z} {\bibfield  {journal} {\bibinfo
  {journal} {Nature}\ }\textbf {\bibinfo {volume} {609}},\ \bibinfo {pages}
  {689} (\bibinfo {year} {2022})}\BibitemShut {NoStop}%
\bibitem [{\citenamefont {Dubin}(1993)}]{Dubin:1993}%
  \BibitemOpen
  \bibfield  {author} {\bibinfo {author} {\bibfnamefont {D.~H.~E.}\
  \bibnamefont {Dubin}},\ }\bibfield  {title} {\bibinfo {title} {Theory of
  structural phase transitions in a trapped {C}oulomb crystal},\ }\href
  {https://doi.org/10.1103/PhysRevLett.71.2753} {\bibfield  {journal} {\bibinfo
   {journal} {Phys.~Rev.~Lett.}\ }\textbf {\bibinfo {volume} {71}},\ \bibinfo
  {pages} {2753} (\bibinfo {year} {1993})}\BibitemShut {NoStop}%
\bibitem [{\citenamefont {Kiethe}\ \emph {et~al.}(2021)\citenamefont {Kiethe},
  \citenamefont {Timm}, \citenamefont {Landa}, \citenamefont {Kalincev},
  \citenamefont {Morigi},\ and\ \citenamefont {Mehlst\"aubler}}]{Kiethe:2021}%
  \BibitemOpen
  \bibfield  {author} {\bibinfo {author} {\bibfnamefont {J.}~\bibnamefont
  {Kiethe}}, \bibinfo {author} {\bibfnamefont {L.}~\bibnamefont {Timm}},
  \bibinfo {author} {\bibfnamefont {H.}~\bibnamefont {Landa}}, \bibinfo
  {author} {\bibfnamefont {D.}~\bibnamefont {Kalincev}}, \bibinfo {author}
  {\bibfnamefont {G.}~\bibnamefont {Morigi}},\ and\ \bibinfo {author}
  {\bibfnamefont {T.~E.}\ \bibnamefont {Mehlst\"aubler}},\ }\bibfield  {title}
  {\bibinfo {title} {Finite-temperature spectrum at the symmetry-breaking
  linear to zigzag transition},\ }\href
  {https://doi.org/10.1103/PhysRevB.103.104106} {\bibfield  {journal} {\bibinfo
   {journal} {Phys. Rev. B}\ }\textbf {\bibinfo {volume} {103}},\ \bibinfo
  {pages} {104106} (\bibinfo {year} {2021})}\BibitemShut {NoStop}%
\bibitem [{\citenamefont {Linington}\ and\ \citenamefont
  {Vitanov}(2008)}]{Linington:2008}%
  \BibitemOpen
  \bibfield  {author} {\bibinfo {author} {\bibfnamefont {I.~E.}\ \bibnamefont
  {Linington}}\ and\ \bibinfo {author} {\bibfnamefont {N.~V.}\ \bibnamefont
  {Vitanov}},\ }\bibfield  {title} {\bibinfo {title} {Robust creation of
  arbitrary-sized {D}icke states of trapped ions by global addressing},\ }\href
  {https://doi.org/10.1103/PhysRevA.77.010302} {\bibfield  {journal} {\bibinfo
  {journal} {Phys. Rev. A}\ }\textbf {\bibinfo {volume} {77}},\ \bibinfo
  {pages} {010302(R)} (\bibinfo {year} {2008})}\BibitemShut {NoStop}%
\bibitem [{\citenamefont {Um}\ \emph {et~al.}(2016)\citenamefont {Um},
  \citenamefont {Zhang}, \citenamefont {Lv}, \citenamefont {Lu}, \citenamefont
  {An}, \citenamefont {Zhang}, \citenamefont {Nha}, \citenamefont {Kim},\ and\
  \citenamefont {Kim}}]{Um:2016}%
  \BibitemOpen
  \bibfield  {author} {\bibinfo {author} {\bibfnamefont {M.}~\bibnamefont
  {Um}}, \bibinfo {author} {\bibfnamefont {J.}~\bibnamefont {Zhang}}, \bibinfo
  {author} {\bibfnamefont {D.}~\bibnamefont {Lv}}, \bibinfo {author}
  {\bibfnamefont {Y.}~\bibnamefont {Lu}}, \bibinfo {author} {\bibfnamefont
  {S.}~\bibnamefont {An}}, \bibinfo {author} {\bibfnamefont {J.-N.}\
  \bibnamefont {Zhang}}, \bibinfo {author} {\bibfnamefont {H.}~\bibnamefont
  {Nha}}, \bibinfo {author} {\bibfnamefont {M.~S.}\ \bibnamefont {Kim}},\ and\
  \bibinfo {author} {\bibfnamefont {K.}~\bibnamefont {Kim}},\ }\bibfield
  {title} {\bibinfo {title} {Phonon arithmetic in a trapped ion system},\
  }\href {https://doi.org/10.1038/ncomms11410} {\bibfield  {journal} {\bibinfo
  {journal} {Nat. Commun.}\ }\textbf {\bibinfo {volume} {7}},\ \bibinfo {pages}
  {11410} (\bibinfo {year} {2016})}\BibitemShut {NoStop}%
\bibitem [{\citenamefont {Lechner}\ \emph {et~al.}(2016)\citenamefont
  {Lechner}, \citenamefont {Maier}, \citenamefont {Hempel}, \citenamefont
  {Jurcevic}, \citenamefont {Lanyon}, \citenamefont {Monz}, \citenamefont
  {Brownnutt}, \citenamefont {Blatt},\ and\ \citenamefont
  {Roos}}]{Lechner:2016}%
  \BibitemOpen
  \bibfield  {author} {\bibinfo {author} {\bibfnamefont {R.}~\bibnamefont
  {Lechner}}, \bibinfo {author} {\bibfnamefont {C.}~\bibnamefont {Maier}},
  \bibinfo {author} {\bibfnamefont {C.}~\bibnamefont {Hempel}}, \bibinfo
  {author} {\bibfnamefont {P.}~\bibnamefont {Jurcevic}}, \bibinfo {author}
  {\bibfnamefont {B.~P.}\ \bibnamefont {Lanyon}}, \bibinfo {author}
  {\bibfnamefont {T.}~\bibnamefont {Monz}}, \bibinfo {author} {\bibfnamefont
  {M.}~\bibnamefont {Brownnutt}}, \bibinfo {author} {\bibfnamefont
  {R.}~\bibnamefont {Blatt}},\ and\ \bibinfo {author} {\bibfnamefont {C.~F.}\
  \bibnamefont {Roos}},\ }\bibfield  {title} {\bibinfo {title}
  {Electromagnetically-induced-transparency ground-state cooling of long ion
  strings},\ }\href {https://doi.org/10.1103/PhysRevA.93.053401} {\bibfield
  {journal} {\bibinfo  {journal} {Phys. Rev. A}\ }\textbf {\bibinfo {volume}
  {93}},\ \bibinfo {pages} {053401} (\bibinfo {year} {2016})}\BibitemShut
  {NoStop}%
\bibitem [{\citenamefont {Kay}(1993)}]{kay1993fundamentals}%
  \BibitemOpen
  \bibfield  {author} {\bibinfo {author} {\bibfnamefont {S.~M.}\ \bibnamefont
  {Kay}},\ }\href@noop {} {\emph {\bibinfo {title} {Fundamentals of statistical
  signal processing: estimation theory}}}\ (\bibinfo  {publisher}
  {Prentice-Hall, Inc.},\ \bibinfo {year} {1993})\BibitemShut {NoStop}%
\bibitem [{\citenamefont {Demkowicz-Dobrzanski}\ \emph
  {et~al.}(2009)\citenamefont {Demkowicz-Dobrzanski}, \citenamefont {Dorner},
  \citenamefont {Smith}, \citenamefont {Lundeen}, \citenamefont {Wasilewski},
  \citenamefont {Banaszek},\ and\ \citenamefont
  {Walmsley}}]{demkowicz2009quantum}%
  \BibitemOpen
  \bibfield  {author} {\bibinfo {author} {\bibfnamefont {R.}~\bibnamefont
  {Demkowicz-Dobrzanski}}, \bibinfo {author} {\bibfnamefont {U.}~\bibnamefont
  {Dorner}}, \bibinfo {author} {\bibfnamefont {B.~J.}\ \bibnamefont {Smith}},
  \bibinfo {author} {\bibfnamefont {J.~S.}\ \bibnamefont {Lundeen}}, \bibinfo
  {author} {\bibfnamefont {W.}~\bibnamefont {Wasilewski}}, \bibinfo {author}
  {\bibfnamefont {K.}~\bibnamefont {Banaszek}},\ and\ \bibinfo {author}
  {\bibfnamefont {I.~A.}\ \bibnamefont {Walmsley}},\ }\bibfield  {title}
  {\bibinfo {title} {Quantum phase estimation with lossy interferometers},\
  }\href {https://doi.org/10.1103/PhysRevA.80.013825} {\bibfield  {journal}
  {\bibinfo  {journal} {Phys.~Rev.~A}\ }\textbf {\bibinfo {volume} {80}},\
  \bibinfo {pages} {013825} (\bibinfo {year} {2009})}\BibitemShut {NoStop}%
\bibitem [{\citenamefont {Pezze}\ \emph {et~al.}(2017)\citenamefont {Pezze},
  \citenamefont {Ciampini}, \citenamefont {Spagnolo}, \citenamefont
  {Humphreys}, \citenamefont {Datta}, \citenamefont {Walmsley}, \citenamefont
  {Barbieri}, \citenamefont {Sciarrino},\ and\ \citenamefont
  {Smerzi}}]{pezze2017optimal}%
  \BibitemOpen
  \bibfield  {author} {\bibinfo {author} {\bibfnamefont {L.}~\bibnamefont
  {Pezze}}, \bibinfo {author} {\bibfnamefont {M.~A.}\ \bibnamefont {Ciampini}},
  \bibinfo {author} {\bibfnamefont {N.}~\bibnamefont {Spagnolo}}, \bibinfo
  {author} {\bibfnamefont {P.~C.}\ \bibnamefont {Humphreys}}, \bibinfo {author}
  {\bibfnamefont {A.}~\bibnamefont {Datta}}, \bibinfo {author} {\bibfnamefont
  {I.~A.}\ \bibnamefont {Walmsley}}, \bibinfo {author} {\bibfnamefont
  {M.}~\bibnamefont {Barbieri}}, \bibinfo {author} {\bibfnamefont
  {F.}~\bibnamefont {Sciarrino}},\ and\ \bibinfo {author} {\bibfnamefont
  {A.}~\bibnamefont {Smerzi}},\ }\bibfield  {title} {\bibinfo {title} {Optimal
  measurements for simultaneous quantum estimation of multiple phases},\ }\href
  {https://doi.org/10.1103/PhysRevLett.119.130504} {\bibfield  {journal}
  {\bibinfo  {journal} {Phys.~Rev.~Lett.}\ }\textbf {\bibinfo {volume} {119}},\
  \bibinfo {pages} {130504} (\bibinfo {year} {2017})}\BibitemShut {NoStop}%
\end{thebibliography}

%

\end{document}